\RequirePackage{silence}
\documentclass[fleqn,usenatbib,useAMS]{mnras} 
\usepackage{graphicx}
\usepackage{amsmath}
\DeclareMathOperator{\sign}{sign}
\usepackage{amsfonts}
\usepackage{float}
\usepackage{bm}
\usepackage[perpage,symbol*]{footmisc}
\usepackage{ae,aecompl}
\usepackage{array}
\usepackage{soul}
\usepackage{mathtools}
\usepackage{multirow}

\newcommand{\appropto}{\mathrel{\vcenter{
  \offinterlineskip\halign{\hfil$##$\cr 
    \propto\cr\noalign{\kern2pt}\sim\cr\noalign{\kern-2pt}}}}}

\newcommand{\ssim}{\,{\sim}\,} 

\title[Testing Gravity with wide binaries like $\alpha$ Centauri]{Testing Gravity with wide binary stars like $\alpha$ Centauri} 

\author[Indranil Banik \& Hongsheng Zhao]{Indranil Banik$^{1}$\thanks{Email: \href{mailto:ib45@st-andrews.ac.uk}{ib45@st-andrews.ac.uk} (Indranil Banik)\newline $~~~~~~~~~~~~~~$ \href{mailto:hz4@st-andrews.ac.uk}{hz4@st-andrews.ac.uk} (Hongsheng Zhao)} and Hongsheng Zhao$^{1}$\\
$^{1}$Scottish Universities Physics Alliance, University of St Andrews, North Haugh, St Andrews, Fife, KY16 9SS, UK}

\pubyear{2018}
\begin{document}
\label{firstpage}
\pagerange{\pageref{firstpage}--\pageref{lastpage}}

\maketitle

\begin{abstract}


We consider the feasibility of testing Newtonian gravity at low accelerations using wide binary (WB) stars separated by $\ga 3$ kAU. These systems probe the accelerations at which galaxy rotation curves unexpectedly flatline, possibly due to Modified Newtonian Dynamics (MOND). We conduct Newtonian and MOND simulations of WBs covering a grid of model parameters in the system mass, semi-major axis, eccentricity and orbital plane. We self-consistently include the external field (EF) from the rest of the Galaxy on the Solar neighbourhood using an axisymmetric algorithm. For a given projected separation, WB relative velocities reach larger values in MOND. The excess is ${\approx 20\%}$ adopting its simple interpolating function, as works best with a range of Galactic and extragalactic observations. This causes noticeable MOND effects in accurate observations of ${\approx 500}$ WBs, even without radial velocity measurements.

We show that the proposed Theia mission may be able to directly measure the orbital acceleration of Proxima Centauri towards the 13 kAU-distant $\alpha$ Centauri. This requires an astrometric accuracy of $\approx 1 \, \mu$as over 5 years. We also consider the long-term orbital stability of WBs with different orbital planes. As each system rotates around the Galaxy, it experiences a time-varying EF because this is directed towards the Galactic Centre. We demonstrate approximate conservation of the angular momentum component along this direction, a consequence of the WB orbit adiabatically adjusting to the much slower Galactic orbit. WBs with very little angular momentum in this direction are less stable over Gyr periods. This novel direction-dependent effect might allow for further tests of MOND.


\end{abstract}


\section{Introduction}
\label{Introduction}

The currently prevailing cosmological paradigm \citep[$\Lambda$CDM,][]{Ostriker_Steinhardt_1995} is based on the assumption that General Relativity governs the dynamics of astrophysical systems. This can be well approximated by Newtonian gravity in the non-relativistic regime, covering for instance planetary motions in the Solar System and galactic rotation curves \citep{Rowland_2015, Almeida_2016}. While the former can be well described by Newtonian gravity, this is not the case for the latter \citep[e.g.][]{Rogstad_1972}. Moreover, self-gravitating Newtonian disks are unstable both theoretically \citep{Toomre_1964} and in numerical simulations \citep{Hohl_1971}.

These apparently fatal problems with Newtonian gravity are generally explained by invoking massive halos of dark matter surrounding each galaxy \citep{Ostriker_Peebles_1973}. Constraints from gravitational microlensing experiments indicate that the Galactic dark matter can't be made of compact objects like stellar remnants \citep{MACHO_2000, EROS_2007}. Thus, it is hypothesised to be an undiscovered weakly interacting particle beyond the well-tested standard model of particle physics \citep[][and references therein]{Peebles_2017_DM_review}.

While this may be the solution, it is conceivable that Newtonian gravity does in fact break down in some astrophysical systems \citep{Zwicky_1937}. If so, this would naturally explain the remarkably tight correlation between the internal accelerations within galaxies (typically inferred from their rotation curves) and the predictions of Newtonian gravity applied to the distribution of their luminous matter \citep[e.g.][and references therein]{Famaey_McGaugh_2012}. This `radial acceleration relation' (RAR) has recently been tightened further based on near-infrared photometry taken by the Spitzer Space Telescope \citep{SPARC}, considering only the most reliable rotation curves (see their section 3.2.2) and taking advantage of reduced variability in stellar mass-to-light ratios at these wavelengths \citep{Bell_de_Jong_2001, Norris_2016}. These improvements reveal that the RAR holds with very little scatter over ${\approx 5}$ orders of magnitude in luminosity and a similar range in surface brightness \citep{McGaugh_Lelli_2016}. Fits to individual rotation curves show that any intrinsic scatter in the RAR must be ${<13\%}$ \citep{Li_2018}.

In addition to disk galaxies, the RAR also holds for ellipticals, whose internal accelerations can sometimes be measured accurately due to the presence of a thin rotation-supported gas disk \citep{Heijer_2015}. The RAR extends down to galaxies as faint as the satellites of M31 \citep{McGaugh_2013}. For a recent overview of how well the RAR works in several different types of galaxy across the Hubble sequence, we refer the reader to \citet{Lelli_2017}.

Another long-standing issue faced by $\Lambda$CDM is the highly anisotropic distribution of Milky Way (MW) satellites \citep{Kroupa_2005}. Strongly flattened satellite systems have also been identified around M31 \citep{Ibata_2013} and Centaurus A \citep{Muller_2018}. These structures are difficult to reconcile with $\Lambda$CDM \citep{Pawlowski_2018, Shao_2018}. Results from many different investigations into this issue are summarised in tables 1 and 2 of \citet{Forero_2018}. Those authors use a different way of quantifying asphericity but do not consider the particularly problematic velocity data. Even so, they find that the LG is a 3$\sigma$ outlier to $\Lambda$CDM. Their section 4.4 shows that simulations including baryonic effects have a more spherical satellite distribution, worsening the discrepancy.

The basic problem is that thin planar structures suggest some dissipative mechanism. Although this is not by itself unusual, dark matter is thought to be collisionless, with the latest results arguing against the MW possessing a dark matter disk \citep{Schutz_2018}. Thus, the only natural way to form satellite planes is out of tidal debris expelled from the baryonic disk of a galaxy that suffered an interaction with another galaxy. This phenomenon occurs in some observed galactic interactions \citep{Mirabel_1992}. Due to the way in which such tidal dwarf galaxies form out of a thin tidal tail, they would end up lying close to a plane and co-rotating within that plane \citep{Wetzstein_2007}.

Such a second-generation origin of the MW and M31 satellite planes predicts that the satellites in these planes should be free of dark matter \citep{Barnes_1992, Wetzstein_2007}. This is due to the dissipationless nature of dark matter and its initial distribution in a dispersion-supported near-spherical halo. During a tidal interaction, dark matter of this form is clearly incapable of forming into a thin dense tidal tail out of which dwarf galaxies might condense. Lacking dark matter, the MW and M31 satellite plane members should have very low internal velocity dispersions $\sigma_{_{int}}$.

This prediction is contradicted by the high observed $\sigma_{_{int}}$ of the MW satellites coherently rotating in a thin plane \citep{McGaugh_Wolf_2010}. The M31 satellite plane galaxies also have rather high $\sigma_{_{int}}$ \citep{McGaugh_Milgrom_2013}. This raises a serious objection to the idea that the anomalously strong internal accelerations within galaxies are caused by their lying within massive dark matter halos.

The leading alternative explanation for these acceleration discrepancies is Modified Newtonian Dynamics \citep[MOND,][]{Milgrom_1983}. In MOND, the dynamical effects usually attributed to dark matter are instead provided by an acceleration-dependent modification to gravity. The gravitational field strength $g$ at distance $r$ from an isolated point mass $M$ transitions from the Newtonian $\frac{GM}{r^2}$ law at short range to
\begin{eqnarray}
	g ~=~ \frac{\sqrt{GMa_{_0}}}{r} ~~~\text{for } ~ r \gg\overbrace{\sqrt{\frac{GM}{a_{_0}}}}^{r_{_M}}
	\label{Deep_MOND_limit}
\end{eqnarray}

MOND introduces $a_{_0}$ as a fundamental acceleration scale of nature below which the deviation from Newtonian dynamics becomes significant. Empirically, $a_{_0} \approx 1.2 \times {10}^{-10}$ m/s$^2$ to match galaxy rotation curves \citep{McGaugh_2011}. Remarkably, this is similar to the acceleration at which the classical energy density in a gravitational field \citep[][equation 9]{Peters_1981} becomes comparable to the dark energy density $u_{_\Lambda} \equiv \rho_{_\Lambda} c^2$ implied by the accelerating expansion of the Universe \citep{Riess_1998}.
\begin{eqnarray}
	\frac{g^2}{8\mathrm{\pi}G} ~<~ u_{_\Lambda} ~~\Leftrightarrow~~ g ~\la~ 2\mathrm{\pi}a_{_0}
	\label{MOND_quantum_link}
\end{eqnarray}

This suggests that MOND may arise from quantum gravity effects \citep[e.g.][]{Milgrom_1999, Pazy_2013, Verlinde_2016, Smolin_2017}. Regardless of its underlying microphysical explanation, it can accurately match the rotation curves of a wide variety of both spiral and elliptical galaxies across a vast range in mass, surface brightness and gas fraction \citep[][and references therein]{Lelli_2017}. It is worth emphasising that MOND does all this based solely on the distribution of luminous matter. Given that most of these rotation curves were obtained in the decades after the MOND field equation was first published \citep{Bekenstein_Milgrom_1984}, it is clear that these achievements are successful a priori predictions. These predictions work due to underlying regularities in galaxy rotation curves that are difficult to reconcile with the collisionless dark matter halos of the $\Lambda$CDM paradigm \citep{Salucci_2017, Desmond_2016, Desmond_2017}.

Although dark matter halos can be tuned to match observed rotation curves, this often requires the halo to be much more massive than the disk. The stability of galactic disks could then be rather different to a theory where the disk had all the mass. By generalising the \citet{Toomre_1964} stability condition for Newtonian disks, \citet{Milgrom_1989} showed that MOND is consistent with the stability of observed disk galaxies given reasonable velocity dispersions. This was later verified with numerical simulations, which showed that the change to the gravity law confers a similar amount of extra stability as a dark matter halo \citep{Brada_1999}. These simulations indicated a peculiarity of MOND in low surface brightness galaxies (LSBs), whose low accelerations were predicted to be associated with a large acceleration discrepancy. Though this was later verified \citep[e.g.][]{Famaey_McGaugh_2012}, the discrepancy is conventionally attributed to LSBs having a massive dark matter halo that dominates the enclosed mass down to very small radii. In MOND, all disk galaxies have self-gravitating disks, including LSBs. Thus, stability of a LSB in MOND requires a higher minimum velocity dispersion compared to $\Lambda$CDM. Observed LSBs indeed have rather high velocity dispersions compared to the very low values feasible in $\Lambda$CDM for disks which are essentially not self-gravitating \citep{Saburova_2011}.

Of course, these LSBs could be dynamically overheated as the Toomre condition only provides a lower limit to their velocity dispersion. This would make it difficult for LSBs to sustain spiral density waves, generally considered the explanation for observed spiral features in higher surface brightness galaxies \citep{Lin_1964}. Interestingly, LSBs also have spiral features \citep{McGaugh_1995}. Assuming the density wave theory applies there too, the number of spiral arms gives an idea of the critical wavelength most unstable to amplification by disk self-gravity. Indeed, \citet{Elena_2015} was able to obtain rather accurate analytic predictions for the number of spiral arms in galaxies observed as part of the DiskMass survey \citep{Bershady_2010}, though this survey `selects against LSB disks.' Using this argument, \citet{Fuchs_2003} found that LSB disks need to be much more massive than suggested by their photometry and stellar population synthesis models. A similar result was also reached by \citet{Peters_2018} using the pattern speeds of bars in LSBs, which are faster than expected in 3 of the 4 galaxies they considered. 

Bars and spiral features in galaxies can be triggered by interactions with satellites \citep{Hu_2018}. However, without disk self-gravity, any spirals formed in this way would rapidly wind up and decay due to differential rotation of the disk \citep[][page 111]{Fall_1981}. Even in a galaxy like M31, the simulations of \citet{Dubinski_2008} indicate that interactions with a realistic satellite population only cause mild disk heating in excess of that which arises in the absence of satellites.

Thus, evidence has been mounting over several decades that the gravity in a LSB generally comes from its disk. This contradicts the $\Lambda$CDM expectation that it should mostly come from its near-spherical dark matter halo given the large acceleration discrepancy at all radii in LSBs. If this discrepancy arises due to MOND, then all galaxy disks would be self-gravitating regardless of their surface brightness.

Another consequence of the MOND scenario is that it raises the expected internal velocity dispersions of \emph{purely baryonic} MW and M31 satellites enough to match observations \citep[][respectively]{McGaugh_Wolf_2010, McGaugh_Milgrom_2013}. MOND also greatly enhances the mutual attraction between the MW and M31. As a result, these galaxies must have had a close flyby ${9 \pm 2}$ Gyr ago \citep{Zhao_2013}. We conducted simulations of this flyby, treating the MW and M31 as point masses surrounded by test particle disks. The outer particles of each disk generally ended up preferentially rotating within a certain plane. If the flyby occurred in a particular orientation, then both simulated `satellite planes' matched the orientations and spatial extents of the corresponding observed structures \citep{Banik_Ryan_2018}. Their best-fitting simulation also matched several other constraints like the timing argument, the statement that the MW and M31 must have been on the Hubble flow at very early times but still end up with their presently observed separation and relative velocity \citep{Kahn_Woltjer_1959}. The calculated flyby time of 7.65 Gyr ago corresponds fairly well to the observation that the vertical velocity dispersion of the MW disk experienced a sudden jump ${\approx 7}$ Gyr ago \citep{Yu_2018}. The inner stellar halo of the MW accreted a significant proportion of its mass in a `major accretion event' around that time \citep{Belokurov_2018}. This strongly suggests that MOND can explain the Local Group satellite planes and perhaps also the Galactic thick disk \citep{Gilmore_1983} as a consequence of a past MW-M31 flyby. We are planning to test this scenario with $N$-body simulations similar to those conducted by \citet{Bilek_2018}.

As well as tidally affecting each other, the MW-M31 flyby would have dramatically affected the motion of LG dwarf galaxies caught near its spacetime location. The high MW-M31 relative velocity would allow them to gravitationally slingshot any nearby dwarf outwards at high speed, leading to some LG dwarfs having an unusually high radial velocity for their position. We did in fact find some evidence for 5 or 6 high-velocity galaxies (HVGs) like this \citep{Banik_Zhao_2016, Banik_Zhao_2017}, a result also confirmed by \citet{Peebles_2017} using his 3D $\Lambda$CDM model of the LG. We used a MOND model of the LG to demonstrate that the dwarfs reaching the fastest speeds were likely flung out almost parallel to the motion of the perturber. As a result, the HVGs ought to define the MW-M31 orbital plane \citep[][section 3]{Banik_2017_anisotropy}. Observationally, the HVGs do define a rather thin plane, with the MW-M31 line only ${16^\circ}$ out of this plane (see their table 4). Thus, we argued that the HVGs may preserve evidence of a past close MW-M31 flyby and their fast relative motion at that time.

As well as enhancing the gravity between the MW and M31, MOND should also enhance the gravity exerted by other galaxy groups. This would cause them to have a larger turnaround radius, the separation at which a galaxy has zero radial velocity with respect to the group. This turnaround radius is essentially a measure of where cosmic expansion wins the battle against the gravity of the cluster \citep{Lee_2017}. Stronger gravity would enlarge the turnaround radius, perhaps explaining why it apparently exceeds the maximum expected in $\Lambda$CDM for the NGC 5353/4 group \citep{Lee_2015} and three out of six other galaxy groups \citep{Lee_2018}.



Because MOND is an acceleration-dependent theory, its effects could become apparent in a rather small system if the system had a sufficiently low mass (Equation \ref{Deep_MOND_limit}). In fact, the MOND radius $r_{_M}$ is only 7000 astronomical units (7 kAU) for a system with $M = M_\odot$. This implies that the orbits of distant Solar System objects might be affected by MOND \citep{Pauco_2016}, perhaps accounting for certain correlations in their properties \citep{Pauco_2017}. For example, Oort cloud comets could fall into the inner Solar System more frequently as their orbits can lose their angular momentum in MOND, even without tidal effects (Section \ref{Secular_effects}). However, it is difficult to accurately constrain the dynamics of objects at such large distances.

Such constraints could be obtained more easily around other stars if they have distant binary companions. As first suggested by \citet{Hernandez_2012}, the orbital motions of these wide binaries (WBs) should be faster in MOND than in Newtonian gravity. Moreover, it is likely that many such systems would form \citep{Tokovinin_2017}, paving the way for the wide binary test (WBT) of gravity that we discuss in this contribution. Equation \ref{Deep_MOND_limit} implies that this will involve orbital velocities of $\sim \sqrt[4]{GM_\odot a_{_0}} = 0.36$ km/s.

The WBT was first attempted by \citet{Hernandez_2012} using the WB catalogue of \citet{Shaya_2011}, who analyzed Hipparcos data with Bayesian methods to identify WBs within 100 pc \citep{Leeuwen_2007}. A tentative signal was identified whereby the typical relative velocities between WB stars remained constant with increasing separation instead of following the expected Keplerian decline \citep[][figure 1]{Hernandez_2012}. However, it was later shown that their typical velocity uncertainty of 0.8 km/s was too large to draw strong conclusions about the underlying law of gravity \citep[][section 1]{Scarpa_2017}. This work obtained accurate spectra of 60 candidate WB pairs, constraining their relative radial velocity to within ${\sim 0.1}$ km/s \citep[][table 3]{Scarpa_2017}. Combined with parallaxes and proper motions, these measurements showed that a handful of the candidate systems are likely genuine WBs that may be suitable for the WBT. A few systems had a relative velocity above the Newtonian upper limit but below the MOND upper limit, though additional follow-up work will be required to confirm the nature of these systems (Section \ref{Systematics}).

Existing data from the Gaia mission \citep{Perryman_2001} strongly suggests that many more WBs will be discovered \citep{Andrews_2017}. The candidate systems they identified are mostly genuine, with a contamination rate of ${\approx 6\%}$ \citep{Andrews_2018} estimated using the second data release of the Gaia mission \citep[Gaia DR2,][]{GAIA_2018}.

The separations of WB stars are small compared to typical interstellar separations of ${\sim 1}$ pc. As a result, an individual WB system separated by 20 kAU should have a centre of mass acceleration towards the nearest star that is ${\approx 100\times}$ weaker than the internal gravity of the WB. The tidal effect would be smaller still. Moreover, the effects of stars in different directions would cancel to a large extent. For the Galaxy as a whole, the overall gravitational field is still only ${\sim a_{_0}}$ despite the Solar neighbourhood lying ${\sim 10^5 \times}$ further from the Galactic Centre than typical WB separations. This implies that WBs should not be much affected by tides from the smooth component of the Galactic potential.

However, the real Galaxy is not smooth as it contains many individual stars. Thus, one concern with the WBT is whether a sufficiently large fraction of WB systems survive encounters with passing field stars. \citet{Bahcall_1985} estimated that the survival timescale was longer than 10 Gyr for systems separated by ${< 31}$ kAU, with the survival timescale being inversely proportional to the separation. \citet{Jiang_2010} also performed a detailed study into this issue. Their figure 8 shows that a substantial fraction of WB systems should survive for 10 Gyr if we restrict to systems with separation below ${\approx 0.1}$ of their Jacobi (tidal) radius, which is 350 kAU for two Sun-like stars orbiting each other in the Solar neighbourhood (see their equation 43).

If WBs were very rare, then finding one should require us to look beyond the nearest star to the Sun, Proxima Centauri (Proxima Cen). It orbits the close (18 AU) binary $\alpha$ Cen A and B at a distance of 13 kAU \citep{Kervella_2017}. This puts the Proxima Cen orbit well within the regime where MOND would have a significant effect \citep{Beech_2009, Beech_2011}. Given the billions of stars in our Galaxy, it would be highly unusual if it did not contain a very large number of systems well suited to the WBT. This is especially true given the high (74\%) likelihood that our nearest WB was stable over the last 5 Gyr despite the effects of Galactic tides and stellar encounters \citep{Feng_2018}. 

Although these works assumed Newtonian gravity, their conclusions should also be valid in MOND as it only slightly enhances the impulse due to a stellar encounter (Section \ref{Boost_to_Newton}). The effects of the non-linear MOND gravity can cause a WB to be unstable over Gyr periods, but we find that this only affects a small proportion of WB systems in particular orientations (Section \ref{Secular_effects}).

The WBT was considered in more detail by \citet{Pittordis_2018}, who approximated MOND using their equation 21. This appears to significantly underestimate the gravitational attraction between the stars in a WB. In fact, the authors found a wide range of scenarios in which stars are expected to attract each other even less than under Newtonian gravity. Given the importance of the WBT, we revisit it using libraries of WB orbits based on more rigorous MOND force calculations. These are compared with similar orbit libraries based on Newtonian gravity. We also check our numerically determined MOND forces in very wide systems using previously derived analytic results \citep{Banik_2015}.

We then develop a statistical analysis procedure to quantify how many WB systems would be needed to conclusively distinguish between Newtonian and MOND gravity using the WBT. Our work focuses on a particular implementation of MOND with the interpolating function that works best with currently available observations. Thus, our more rigorous approach complements that of \citet{Pittordis_2018}, who considered a wider range of modified gravity formulations and free parameters.

After introducing the WBT in Section \ref{Introduction}, we explain how we determine the MOND gravitational attraction between the stars in each system and use this information to integrate the system (Section \ref{Method}). We then discuss our choice of prior distributions for the WB orbital parameters (Section \ref{Priors}). Using similar methods to obtain a Newtonian control, we compare the results using the procedure explained in Section \ref{Statistics}. This allows us to quantify how many systems are required for the WBT, the primary result of this contribution (Section \ref{Results}). We then discuss measurement uncertainties in the basic parameters of nearby WB systems (Section \ref{Measurement_uncertainties}). Using simple analytic estimates, we discuss how the WBT might or might not work with different MOND formulations and interpolating functions (Section \ref{Theoretical_uncertainties}). We also discuss which interpolating function is most appropriate in light of existing observations, especially of rotation curves (Section \ref{Interpolating_function}). The WBT can also be affected by astrophysical uncertainties regarding the properties of each system, in particular whether they contain any undetected companions (Section \ref{Systematics}). These uncertainties could be mitigated and a much more direct version of the WBT conducted if the orbital acceleration were measured directly, something that may be possible with future observations of Proxima Cen (Section \ref{Proxima_Centauri}). In this section, we also consider the long-term orbital stability of WB systems in the complicated time-dependent MOND potential. We provide our conclusions in Section \ref{Conclusions}.


\section{Method}
\label{Method}

The basic idea behind the WBT is that MOND enhances the gravitational attraction $\bm{g}$ between two widely separated stars. Currently, it is difficult to test this by directly determining their relative acceleration (though this may be possible in future, see Section \ref{Proxima_Cen_short_term}). Instead, the WBT focuses on their relative velocity $\bm{v}$, making use of the fact that stronger gravity allows systems to be bound at a higher relative speed ${v \equiv \left| \bm{v} \right|}$.

One issue with the WBT is that it necessarily requires many WB systems and thus a sufficiently large survey volume. Towards its edge, Gaia is unlikely to constrain line of sight distances accurately enough to know the true (3D) separation $r$ of each system (Section \ref{Distance_measurement}). However, the sky-projected separation $r_p$ would be known very accurately. To take advantage of this, \citet{Pittordis_2018} defined
\begin{eqnarray}
	\widetilde{v} ~\equiv ~ v \div \overbrace{\sqrt{\frac{GM}{r_p}}}^{\text{Newtonian }v_c}
	\label{v_tilde}
\end{eqnarray}

$\widetilde{v}$ is the ratio of $v$ to the Newtonian circular velocity $v_c$ of a system with total mass $M$ if its stars are separated by a distance $r_p$. Because their true (3D) separation $r > r_p$, calculating $\widetilde{v}$ in this way provides an upper limit on $v \div \sqrt{\frac{GM}{r}}$, a quantity which can't exceed $\sqrt{2}$ in Newtonian gravity. In MOND, we expect the upper limit to be somewhat higher. As a result, the probability distribution of $\widetilde{v}$ should differ between the two models, with MOND allowing for a non-zero probability that $\widetilde{v} > \sqrt{2}$. This is the basis for the WBT.

To forecast how this might work, we integrate forwards a grid of WB systems covering a range of masses and orbital parameters. At each timestep, we consider what would be seen by a distant ($\gg r$) observer at a grid of possible viewing directions. In this way, we build up a probability distribution over $r_p$ and $\widetilde{v}$. The results are compared with those of similar calculations using Newtonian gravity. We then develop a statistical procedure that quantifies how easily we could distinguish the $\widetilde{v}$ distributions of the two theories for different total numbers of WB systems (Section \ref{Statistics}). This addresses the question of how many systems would be needed for the WBT, thus helping observers plan its implementation.

In the near term, the WBT will be based on stars in the Solar neighbourhood. This means that our orbit integrations must take into account an important MOND phenomenon whereby the internal dynamics of a system is affected by any external gravitational field (EF), even if its strength $\bm{g}_{ext}$ is uniform across the system. This external field effect \citep[EFE,][]{Milgrom_1986} arises because MOND gravity is non-linear in the matter distribution (Equation \ref{QUMOND_equation}). The EFE can be understood intuitively by considering a system with low internal accelerations that would normally show strong MOND effects. However, if the system is in a high-acceleration environment (${g}_{ext} \gg a_{_0}$), then the total acceleration $g$ exceeds the $a_{_0}$ threshold, making the internal dynamics Newtonian.

For the WBT, $\bm{g}_{ext}$ is provided by the rest of the Galaxy. This leads to the force between two stars varying with their orientation relative to the EF direction $\widehat{\bm{g}}_{ext} \equiv \frac{\bm{g}_{ext}}{\left| \bm{g}_{ext} \right|}$ \citep{Banik_2015}. Thus, we need to consider WB systems with a range of different angular momentum directions $\widehat{\bm{h}}$. In general, all possible directions would need to be considered. To keep the computational cost manageable, we make the simplifying assumption that one of the stars is much less massive than the other. This makes the problem axisymmetric as the gravitational field is generated by a single point mass, with the other star treated as a test particle.

Such a dominant mass approximation is valid in the Newtonian regime as the linearity of this gravity theory means the mass ratio has no effect on the relative acceleration. MOND gravity is also linear when $\bm{g}_{ext}$ dominates the dynamics of a system \citep{Banik_2015}. In these circumstances, the mass ratio between two stars does not affect their relative acceleration (this depends only on their total mass $M$ and separation vector $\bm{r}$).

Our approximation is therefore accurate both for very close and very wide systems. At intermediate separations, the force binding a WB system would be somewhat weaker if its mass were split more equally between its components \citep[][equation 53]{QUMOND}. This would make the $\widetilde{v}$ distribution slightly more similar to the Newtonian expectation. However, we expect this to be a very small effect for reasons discussed in Section \ref{Mass_ratio_effect}, where we also perform some detailed calculations to help confirm this.

\subsection{Governing equations}
\label{Governing_equations}


We begin by describing how we advance WB systems using the quasilinear formulation of MOND \citep[QUMOND,][]{QUMOND}. Each system is treated as a single point mass $M$ plus a test particle embedded in a uniform EF $\bm{g}_{ext}$. QUMOND uses the Newtonian gravitational field $\bm{g}_{_N}$ to determine the true gravitational field $\bm{g}$ by first finding its divergence.
\begin{eqnarray}
	\label{QUMOND_equation}
	\overbrace{\nabla \cdot \bm{g}}^{\propto \rho_{_{PDM}} + \rho_{_b}} ~&=&~ \nabla \cdot \left[\nu \overbrace{\left( \frac{g_{_N}}{a_{_0}} \right)}^y \, \bm{g}_{_N} \right] ~~\text{ where} \\
	\nu \left( y \right) &=& \frac{1}{2} ~+~ \sqrt{\frac{1}{4} + \frac{1}{y}}	
\end{eqnarray}

$\nu \left(y \right)$ is the interpolating function used to transition between the Newtonian and deep-MOND regimes. We use the `simple' form of this function \citep{Famaey_Binney_2005} because it fits a wide range of data on the MW and external galaxies better than other functions with a sharper transition (Section \ref{Interpolating_function}). The source term for the gravitational field is $\nabla \cdot \left( \nu \bm{g}_{_N} \right)$, which can be thought of as an `effective' density $\rho$ composed of the baryonic density $\rho_{_b}$ and an extra contribution which we define to be the phantom dark matter density $\rho_{_{PDM}}$. This is the distribution of dark matter that would be required for Newtonian gravity to generate the same total gravitational field as QUMOND yields from the baryons alone.

The Newtonian gravity $\bm{g}_{_N}$ at position $\bm{r}$ relative to the central mass $M$ is given by
\begin{eqnarray}
	\label{Newtonian_gravity}
	\bm{g}_{_N} ~\equiv~ -\frac{GM\bm{r}}{r^3} ~+~ \bm{g}_{_{N, ext}}
\end{eqnarray}

The EF contributing to $\bm{g}_{_N}$ is not the true EF $\bm{g}_{ext}$ acting on the system. Rather, the important quantity is $\bm{g}_{_{N, ext}}$, what the EF would have been if the universe was governed by Newtonian gravity. For simplicity, we assume the spherically symmetric relation between $\bm{g}_{ext}$ and $\bm{g}_{_{N, ext}}$, reducing Equation \ref{QUMOND_equation} to
\begin{eqnarray}
	\bm{g}_{ext} ~=~ \overbrace{\nu \left(\frac{\left| \bm{g}_{_{N, ext}} \right|}{a_{_0}} \right)}^{\nu_{_{ext}}} \bm{g}_{_{N, ext}}
	\label{g_N_ext}
\end{eqnarray}

This algebraic MOND approximation should be fairly accurate given that the Solar neighbourhood is ${\approx 4}$ disk scale lengths from the Galactic Centre \citep{Bovy_2013, McMillan_2017}. Note that this does not require the gravitational field to be spherically symmetric. Instead, it requires the weaker condition that departures of $\bm{g}$ from spherical symmetry are accurately captured by applying the MOND $\nu$ function to $\bm{g}_{_{N}}$, which is itself not spherically symmetric. This may explain why \citet{Jones_2018} found that QUMOND gravitational fields in disk galaxies could be estimated rather well using the algebraic MOND approximation, justifying our use of Equation \ref{g_N_ext}. We discuss its accuracy in Section \ref{Galactic_disk_effect}, finding it should work well in the Solar neighbourhood where the WBT would be conducted.

Having found $\bm{g}_{_N}$ in this way, we use Equation \ref{QUMOND_equation} to find $\nabla \cdot \bm{g}$. We then apply a direct summation procedure to $\nabla \cdot \bm{g}$ in order to determine $\bm{g}$ itself.
\begin{eqnarray}
	\bm{g} \left( \bm{r} \right) ~=~ \int \nabla \cdot \bm{g} \left( \bm{r'}\right) \frac{\left( \bm{r} - \bm{r'} \right)}{4 \mathrm{\pi} |\bm{r} - \bm{r'}|^3} \,d^3\bm{r'}
	\label{g_direct_sum}
\end{eqnarray}

As $\bm{g}_{_N}$ is axisymmetric about $\widehat{\bm{g}}_{ext}$, the phantom dark matter distribution can be thought of as a large number of azimuthally uniform rings. At points along their symmetry axis $\widehat{\bm{g}}_{ext}$, it is thus straightforward to find $\bm{g}$ by summing the contributions from each ring. In general, the lower mass star in a system is not conveniently located along $\widehat{\bm{g}}_{ext}$ relative to the primary star. To find $\bm{g}$ at off-axis points in a computationally efficient way, we use a `ring library' that stores $\bm{g}_{_N}$ due to a unit radius ring. This saves us from having to further split each ring into a finite number of elements. Instead, we can simply interpolate within our densely allocated ring library to find the gravity exerted by any ring at the point where we wish to know its contribution to $\bm{g}$.

In this way, we can map out the gravitational field due to a point mass $M$ embedded in a uniform EF. Using a scaling trick, we only need to do this for one value of $M$. This is because the only physical lengths in the problem are the MOND radius $r_{_M}$ (Equation \ref{Deep_MOND_limit}) and the EF radius $r_{ext}$ where $g_{_N} = g_{_{N, ext}}$. If we keep $g_{_{N, ext}}$ fixed, then it is always a fixed multiple of $a_{_0}$, leading to a constant $\frac{r_{ext}}{r_{_M}}$. Thus, we construct a force library for some arbitrary mass $M = 1$ and work in units where $G = a_{_0} = 1$, causing distances to be in units of $r_{_M}$ and accelerations in units of $a_{_0}$.

Due to the finite extent of our grid, we can only consider contributions to $\bm{g}$ from the region $r < r_{_{out}}$, though we make $r_{_{out}}$ sufficiently large that $\bm{g}_{ext}$ is totally dominant beyond it. Thus, regions beyond our grid have an analytic phantom density distribution containing only a quadrupolar term \citep[][equation 24]{Banik_2015}. As explained in Appendix \ref{Potential_correction}, this leads to a correction $\Delta \Phi$ to the potential $\Phi$ in the region $r < r_{_{out}}$ covered by our grid.
\begin{eqnarray}
	\Delta \Phi \left(r, \theta \right) ~&=&~ \frac{1}{5}GM\nu_{ext}K_{_0}r^2 \left(3 \cos^2 \theta - 1 \right) \int_{r_{_{out}}}^\infty \frac{1}{{\widetilde{r}}^4} \, d\widetilde{r} \nonumber \\
	~&=&~ \frac{GM\nu_{ext}K_{_0}r^2 \left(3 \cos^2 \theta - 1 \right)}{15{r_{_{out}}}^3} ~~\text{ where} \\
	K_0 &\equiv & \frac{\partial \, Ln \, \nu_{_{ext}}}{\partial \, Ln \, g_{_{N, ext}}} = ~-\frac{1}{2} \text{ if } g_{_{N, ext}} \ll a_{_0} \\
	\cos \theta ~&\equiv &~ \widehat{\bm{r}} \cdot \widehat{\bm{g}}_{_{ext}} \nonumber
	\label{K_0}
\end{eqnarray}

This causes an adjustment to the gravitational field of
\begin{eqnarray}
	\Delta \bm{g} ~=~ \frac{2GM\nu_{_{ext}}K_0}{15 \, {r_{out}}^3} \left( \bm{r} - 3 r \cos \theta \, \widehat{\bm{g}}_{_{ext}} \right)
	\label{EXTERIOR_PDM_ADJUSTMENT}
\end{eqnarray}

When considering a point where $\bm{g}_{ext}$ is dominant, the gravity due to the star has a magnitude of $\approx \frac{GM\nu_{_{ext}}}{r^2}$, making the correction to it only $\ssim \frac{1}{15} \left(\frac{r}{r_{out}} \right)^3$ when expressed in fractional terms. Thus, the accuracy of our results should not depend much on this correction, which should in any case be very accurate as it estimates contributions from regions with $r > 66.5 \, r_{_{M}}$. There, $g_{_{_{N,ext}}}$ should be $\ga 5000 \, g_{_{N}}$, allowing $\bm{g}_{_{N}}$ to be considered perturbatively in the manner of \citet{Banik_2015}.

In the opposite extreme where the test particle gets very close to the mass, we do not need to consider the EF from the rest of the Galaxy. Thus, at distances within $0.08 \, r_{_M}$, we assume that
\begin{eqnarray}
	\bm{g} ~&=&~ \nu \bm{g}_{_N} \\
	~&=&~ -\frac{\nu GM \bm{r}}{r^3}
	\label{g_near_field}
\end{eqnarray}

At these positions, the EF (of order $a_{_0}$) should be $\ga 100 \times$ weaker than $g_{_N}$, making it reasonable to treat the situation as isolated and neglect the EF when calculating $\nu$. However, our algorithm will eventually slow down if $r$ becomes sufficiently small. Thus, we terminate the trajectory of any particle that gets within 50 AU.

\subsection{The boost to Newtonian gravity}
\label{Boost_to_Newton}

To better understand how much the gravity between two stars might be boosted by MOND effects, we determine the angle-averaged ratio $\eta$ between the MOND and Newtonian radial gravity at different separations.
\begin{eqnarray}
	\eta \left( r \right) ~\equiv ~ \frac{1}{4 \pi}\int_0^\mathrm{\pi} \frac{\bm{g}_r \left( r, \theta \right)}{\bm{g}_{_{N,r}} \left(r \right)} \, 2 \pi \sin \theta \, d\theta
	\label{eta}
\end{eqnarray}

In very widely separated systems, the total acceleration is dominated by the EF rather than self-gravity $\bm{g}$. In this limit, we can obtain $\bm{g}_r$ analytically \citep[][equation 37]{Banik_2015}.
\begin{eqnarray}
	\bm{g}_r ~=~ \bm{g}_{_{N,r}} \, \nu_{ext} \left( 1 + \frac{K_0}{2} \sin^2 \theta \right)
	\label{g_ratio_QUMOND}
\end{eqnarray}

Substituting this into Equation \ref{eta} yields
\begin{eqnarray}
	\eta ~=~ \nu_{ext} \left(1 + \frac{K_0}{3} \right)
	\label{eta_EFE}
\end{eqnarray}

\begin{figure}
	\centering
		\includegraphics[width = 8.5cm] {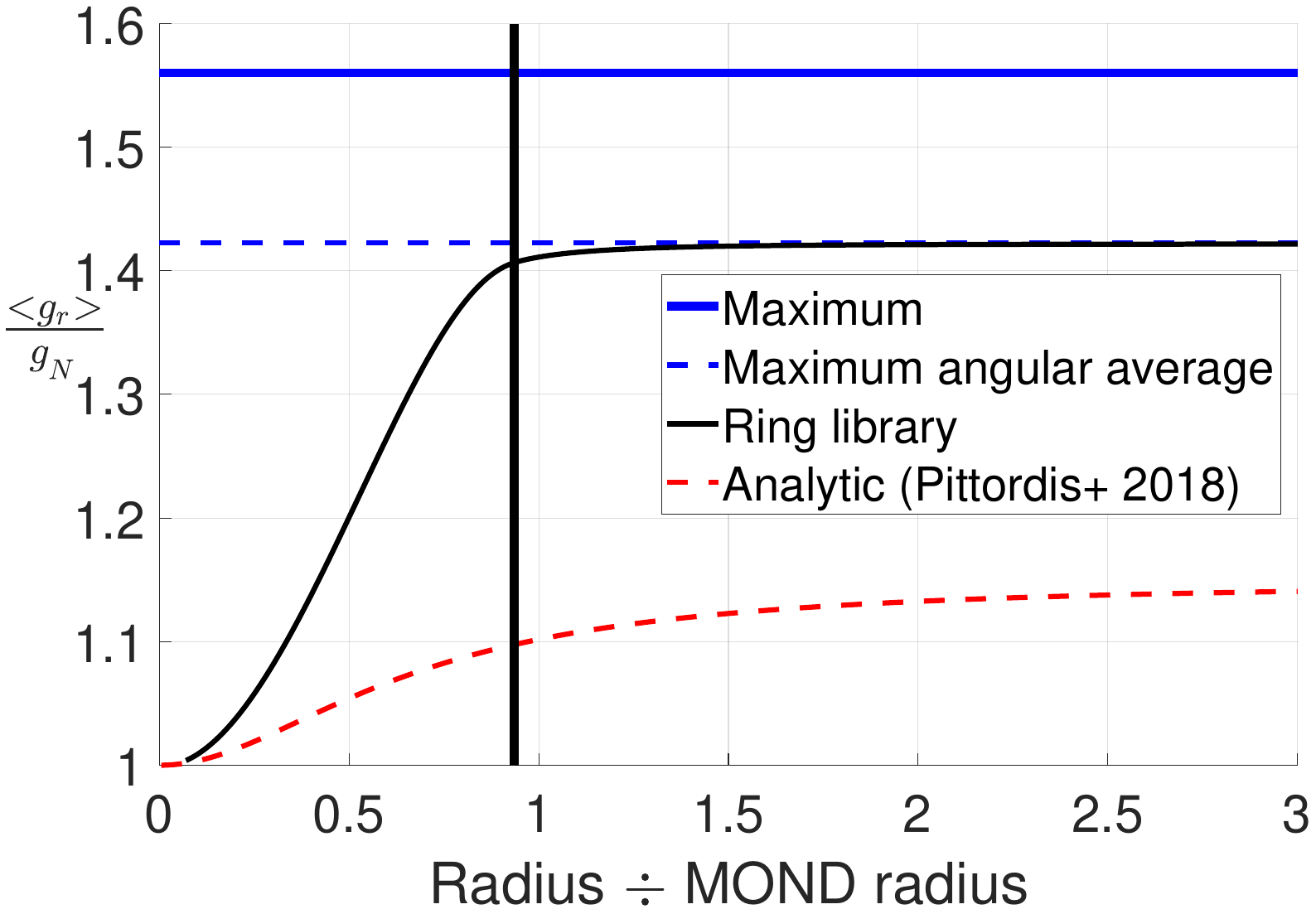}
		\caption{The azimuthally averaged quantity $\eta$ (Equation \ref{eta}) for our QUMOND force library as a function of separation between the stars in a wide binary, assuming one of the stars dominates the system (solid black curve). Our results apply to systems of any mass, as long as distances are scaled to its MOND radius $r_{_M}$ (Equation \ref{Deep_MOND_limit}) and the EF from the rest of the Galaxy has the Solar neighbourhood value (Equation \ref{g_N_ext}). We also show the result obtained by \citet{Pittordis_2018} using their equation 21 (dashed red curve). The EF dominates the dynamics of sufficiently widely separated systems (right of black vertical line). Analytic calculations in this regime show that $\eta$ asymptotically reaches the value given by Equation \ref{eta_EFE} (dashed blue line). For systems aligned with the EF, the maximum ratio of MOND and Newtonian gravity is given by Equation \ref{eta_max_EFE} (solid blue line).}
	\label{Pittordis_forces}
\end{figure}

The angle-averaging makes $\eta$ a good guide to how much gravity would be boosted by MOND effects in a system with known separation relative to its MOND radius. In Figure \ref{Pittordis_forces}, we compare the numerically determined value of $\eta$ at different radii with this EF-dominated expectation.

For completeness, we note that the maximum value of $\frac{\bm{g}_r}{\bm{g}_{_{N,r}}}$ requires not only that the EF dominate ($g_{_{N,ext}} \gg g$) but also that the angle $\theta = 0$ or $\mathrm{\pi}$ (Equation \ref{g_ratio_QUMOND}). Thus, the MOND boost to the self-gravity of the system is limited to
\begin{eqnarray}
	\frac{\bm{g}_r}{\bm{g}_{_{N,r}}} ~\leq~ \nu_{ext}
	\label{eta_max_EFE}
\end{eqnarray}

\subsection{Orbit integration}

\subsubsection{Initial conditions}
\label{Initial_conditions}

To investigate a range of WB orbital semi-major axes $a$ and eccentricities $e$, we first need to define what these quantities mean in MOND. To generalise their definitions for modified gravity theories while remaining valid in Newtonian gravity, we follow the work of \citet[][section 4.1]{Pittordis_2018}. $a$ is defined as the orbital separation $r$ at the point in the orbit where the speed $v$ satisfies
\begin{eqnarray}
	v ~=~ \sqrt{-\bm{r} \cdot \bm{g}}
	\label{a_definition}
\end{eqnarray}

There will be two points in the orbit which satisfy this equation. Either point can be used as they both have the same $r$. These points are also used to define $e$ according to
\begin{eqnarray}
	e ~\equiv ~ \left| \widehat{\bm{r}} \cdot \widehat{\bm{v}} \right|
	\label{e_definition}
\end{eqnarray}

We use the usual Galactic Cartesian co-ordinates with $\widehat{\bm{x}}$ towards the Galactic Centre, $\widehat{\bm{z}}$ towards the North Galactic Pole and $\widehat{\bm{y}} = \widehat{\bm{z}} \times \widehat{\bm{x}}$ so that the co-ordinate system is right-handed. As a result, $\widehat{\bm{y}}$ points along the direction in which the Solar neighbourhood rotates around the Galaxy.

The massive component of each WB is assumed to remain at the origin. We start the other component at the position $\left(0, a, 0 \right)$ and use Equation \ref{a_definition} to set $v$. Equation \ref{e_definition} is used to fix the component of $\bm{v}$ along the radial direction.
\begin{eqnarray}
	v_y ~=~ v \, e
\end{eqnarray}

The remaining tangential velocity $\bm{v}_{tan}$ must have a magnitude of $v\sqrt{1-e^2}$ and lie within the $xz$-plane. We adjust the direction of $\bm{v}_{tan}$ in order to change the orbital pole $\widehat{\bm{h}} \propto \bm{r} \times \bm{v}$, thereby investigating a range of possible angles between $\widehat{\bm{h}}$ and $\widehat{\bm{g}}_{ext} = \widehat{\bm{x}}$. Due to the axisymmetry of the problem, it is only necessary to consider orbital poles along a single great circle containing $\widehat{\bm{g}}_{ext}$. In our setup, this is achieved by considering all possible $\widehat{\bm{h}}$ within the $xz$-plane. As MOND orbits are not closed, we can start our simulations anywhere in the plane orthogonal to $\bm{h}$. Thus, it is always valid to start on the $y$-axis.

When running Newtonian control simulations to compare with the MOND ones, the orbit is closed. However, its conserved orientation within the orbital plane has no effect on the internal dynamics of the system as the force law is not angle-dependent. Thus, we can use the same setup for our Newtonian runs, though these benefit from a number of simplifications compared to the MOND runs.

\subsubsection{Advancing the system}

We evolve our WB systems forwards using our dimensionless force libraries (Section \ref{Governing_equations}). This requires us to scale co-ordinates down by the value of $r_{_M}$ appropriate to the mass $M$ of the system we are considering (Equation \ref{Deep_MOND_limit}). We then use interpolation to estimate the relative acceleration at the instantaneous separation $\bm{r}$ of the WB system. This is used to advance $\bm{r}$ with the fourth-order Runge-Kutta procedure. As the dynamical time should be similar to what it would be in Newtonian gravity, we use an adaptive timestep of
\begin{eqnarray}
	dt ~=~ 0.01\sqrt{\frac{r^3}{GM}}
\end{eqnarray}

We evolve each system forwards until it completes 20 revolutions, representing a rotation angle of 40$\mathrm{\pi}$ radians. To determine the rotation angle over each timestep, we use the dot product between the initial and final directions of $\widehat{\bm{r}}$. The algorithm is accelerated by using a small angle approximation at one order beyond the leading order term, thereby minimising the use of computationally expensive inverse trigonometric functions.

As the MOND potential is non-trivial, it is possible for $r$ to reach very small or very large values compared to its initial value. We therefore terminate trajectories when $r < 50$ AU or when $r > 100$ kAU. We assign zero statistical weight to the parameters which cause the system to `crash' or `escape' in this way. The upper limit is chosen based on the observed 270 kAU distance of Proxima Cen \citep{Kervella_2016}. We expect that WBs with separations exceeding about half this would be so widely separated that nearby stars could unbind the system. The lower limit is chosen to avoid spending excessive amounts of computational time on systems which would likely lose significant amounts of energy through tides, thus taking the orbital parameters outside the region of interest for the WBT. For our purposes, it is not important to know whether these systems would actually undergo a stellar collision or merely settle into a much tighter binary \citep{Kaib_2014}.

For simplicity, we neglect the fact that the Galactic orbit of a WB system will cause $\widehat{\bm{g}}_{ext}$ to gradually change. This is because the orbital timescale at 10 kAU is expected to be ${\sim 1}$ Myr, much shorter than the ${\approx 200}$ Myr taken by the Sun to orbit the Galaxy \citep{Vallee_2017}. Consequently, WB systems should gradually adjust to the changing $\bm{g}_{ext}$. In Section \ref{Proxima_Centauri}, we consider how this affects the long-term evolution of WBs. We also show that our results should not differ much if we had advanced our simulations for 5 Gyr rather than 20 revolutions and allowed $\widehat{\bm{g}}_{ext}$ to rotate (Figure \ref{Centauri_control}).

\subsubsection{Recording of results}

Due to the large number of WB parameters we explore, it is difficult to store all the information available from our trajectory calculations. Moreover, we are not interested in doing so as the observations only constrain certain features of the orbits, and even then only in a statistical sense given that we see a very small fraction of the orbit. Thus, we use our simulated trajectories to obtain the joint probability distribution of the main observable quantities $r_p$ and $\widetilde{v}$.

To do this, we create a 2D set of bins in $r_p$ and $\widetilde{v}$. At each timestep and for each viewing angle (Section \ref{Viewing_angle}), we increment the probability of the corresponding ${\left( r_p, \widetilde{v} \right)}$ bin by the duration of the timestep multiplied by the relative probability of that particular viewing angle. Afterwards, we normalise the final probability distribution over ${\left( r_p, \widetilde{v} \right)}$. If a trajectory crashes or escapes, then we assign zero probability to that particular combination of model parameters.

Our approach is valid as few WBs are destroyed on an orbital timescale (Section \ref{Recently_ionized_systems}). As this is much shorter than a Hubble time, we assume the creation timescale of WBs is also much longer than an individual orbit. This leads to the ${\left( r_p, \widetilde{v} \right)}$ distribution remaining steady over many orbits.

\section{Prior distributions of binary parameters}
\label{Priors}

For the WBT, we need prior distributions for the various system parameters listed in Table \ref{Wide_binary_parameters}. The ones we consider are the semi-major axis $a$ and eccentricity $e$ (defined in Section \ref{Initial_conditions}), total system mass $M$, the angle $\theta$ between $\widehat{\bm{h}}$ and $\widehat{\bm{g}}_{ext}$ and two angles governing the direction from the WB system towards the observer, assumed to be very far from the system. To allow easy investigation of different priors without rerunning the orbital integrations, we record the resulting ${P\left(r_p, \widetilde{v}\right)}$ for the full grid of $M$, $a$ and $e$. We do not store results for different angle parameters because we assume that they all have an isotropic distribution, allowing us to marginalise over them prior to recording the results (Sections \ref{Orbital_pole} and \ref{Viewing_angle}).

\begin{table}
  \centering
		\begin{tabular}{lll}
			\hline
			Variable & Meaning & Prior range \\
			\hline
			$M$ & Total system mass & $\left( 1.2 - 2.4 \right) M_\odot$ \\ [5pt]
			$r_p$ & Sky-projected separation & $\left( 1 - 20 \right)$ kAU \\
			$a$ & Semi-major axis & $\left( 1 - 60 \right)$ kAU \\ [5pt]
			\multirow{2}{*}{e} & Orbital eccentricity (MOND) & $0 - 0.95$ \\
			 & $e$ in Newtonian models & $0 - 0.99$ \\
			$\gamma$ & See Equation \ref{P_e} & 0, 1.2 (nominal), 2 \\
			$\gamma_{_N}$ & $\gamma$ for Newtonian model & $-2$ to 2 \\
			\hline
		\end{tabular}
	\caption{Our prior ranges on wide binary orbital parameters. Although we extract probabilities for sky-projected separations $r_p$ up to 100 kAU, we assume that the WBT would be based on systems with $r_p = \left( 1 - 20 \right)$ kAU to minimise contamination by interlopers (at high $r_p$) and avoid nearly Newtonian systems (at low $r_p$). As the Newtonian versions of these simulations are much faster, we use a higher resolution and wider range in $e$. For each value of $\gamma$, we try all possible values of $\gamma_{_N}$ and take the value which minimises the detection probability (Section \ref{Statistics}). Qualitatively, this yields a Newtonian $\widetilde{v}$ distribution most similar to the MOND one.}
  \label{Wide_binary_parameters}
\end{table}

\subsection{Eccentricity}
\label{Section_P_e}

Following section 4.1 of \citet{Pittordis_2018}, we assume the WB orbital eccentricity distribution $P \left( e \right)$ has the linear form
\begin{eqnarray}
	P \left( e \right) ~=~ 1 + \gamma \left( e - \frac{1}{2} \right)
	\label{P_e}
\end{eqnarray}

The anti-symmetric factor $\left( e - \frac{1}{2} \right)$ is required to ensure the normalisation condition $\int_0^1 P \left( e \right) \, de = 1$. We assume that the constant ${\gamma = 1.2}$ for the MOND case \citep{Tokovinin_2016}. To avoid negative probabilities, ${-2 \leq \gamma \leq 2}$.

When comparing with Newtonian gravity, it is necessary to also define $\gamma_{_N}$, the corresponding value of $\gamma$ for the Newtonian model. If the WBT yielded a positive result for MOND, then astronomers would almost certainly try to fit the data with Newtonian gravity by adjusting $\gamma_{_N}$. In general, trying to match the high $\widetilde{v}$ values expected in MOND requires Newtonian models with a large $e$ as only such orbits can get $\widetilde{v}$ to significantly exceed 1. Giving a higher probability to high $e$ orbits implies a higher $\gamma_{_N}$.

As we do not a priori know $\gamma_{_N}$, we need to let it vary when estimating how easily the Newtonian and MOND $\widetilde{v}$ distributions could be distinguished using the method described in Section \ref{Statistics}. The `best-fitting' $\gamma_{_N}$ is that which makes this task the most difficult. This requires us to consider all possible values for $\gamma_{_N}$. Although $\gamma_{_N}$ can be negative, this would further reduce the probability of high $e$ orbits, likely worsening the agreement with observations of a MOND universe. Thus, we assume the optimal $\gamma_{_N}$ lies in the range $\left(0, 2 \right)$. Where it is clear that this is not the case because negative $\gamma_{_N}$ is preferred, we consider the full range of physically possible values for $\gamma_{_N}$ (Section \ref{Results}).

As the correct value of $\gamma$ is not known either, we consider the three cases of 0 (a flat distribution), 1.2 \citep{Tokovinin_2016} and 2. These were the three cases considered by \citet{Pittordis_2018}, as discussed in their section 2.1. Each time, we need to repeat our search for the best-fitting $\gamma_{_N}$. Our procedure is thus fully deterministic, avoiding uncertainties due to the use of random numbers.

\subsection{Semi-major axis}
\label{Section_P_a}

To constrain the semi-major axis distribution $P\left( a \right)$, we use the observed $P \left( r_p \right)$ distribution \citep[][section 6.2]{Andrews_2017}.\footnote{Eventually, the distribution of 3D separations $r$ will be used for this purpose, but GAIA is not expected to reach the required accuracy (Section \ref{Distance_measurement}).} Similar results were obtained by \citet{Lepine_2007}, though with a slightly smaller break radius of 4 kAU.
\begin{eqnarray}
	\label{P_r_p}
	P \left( r_p \right) dr_p ~\propto~
\left\{
	\begin{array}{ll}
		r_p^{-1} dr_p & \mbox{if } r_p \leq \text{5 kAU} \\ [5pt]
		r_p^{-1.6} dr_p & \mbox{if } r_p \geq \text{5 kAU}
	\end{array}
\right.
\end{eqnarray}

To match these results, we use a broken power law for $P\left( a \right)$ with the break at ${a = a_{_{break}}}$.
\begin{eqnarray}
	\label{P_a}
	P \left( a \right) da ~\propto~
\left\{
	\begin{array}{ll}
		a^{-\alpha} da & \mbox{if } a \leq a_{_{break}} \\
		a^{-\beta} da & \mbox{if } a \geq a_{_{break}}
	\end{array}
\right.
\end{eqnarray}

We consider $a$ in the range ${\left(1-60 \right)}$ kAU, though with reduced resolution beyond 25 kAU. The lower limit of 1 kAU is chosen because the rather gradual interpolating function we adopt \citep{Famaey_Binney_2005} implies that departures from Newtonian gravity decay rather slowly as the acceleration rises above $a_{_0}$. Moreover, tighter orbits are more common \citep{Andrews_2017}, so they might contribute something to the WBT even if they are not much different from Newtonian expectations.

Due to our imposed maximum separation of 100 kAU, orbits with $a > 60$ kAU are often terminated early and so would not contribute any statistical weight to the WBT. Such large orbits are in any case unlikely \citep{Andrews_2017}. Moreover, we only expect to perform the WBT using systems with ${r_p \leq 20}$ kAU, making it not particularly important to consider orbits for which $a$ is much larger.

As we are not a priori sure which range in $r_p$ will work best for the WBT, our algorithm is allowed to find the optimal range within the $\left( 1 - 20 \right)$ kAU range we allow (Section \ref{Statistics}). In Section \ref{Results}, we will see that the WBT does not benefit from systems with $r_p \la 3$ kAU, justifying our decision to neglect WBs with ${a < 1}$ kAU.

To determine the best fitting values of $\alpha$, $\beta$ and $a_{_{break}}$, we try a grid of models in all three parameters. For each combination, we find $P \left( a \right)$ using Equation \ref{P_a}. We then marginalise over $\widetilde{v}$ and the other model parameters to obtain a simulated $P \left( r_p \right)$. This is done over kAU-wide bins in $r_p$ over the range ${\left( 2 - 20 \right)}$ kAU, thus minimising edge effects from our lack of models with $a < 1$ kAU and our truncation of orbital separation at 100 kAU. We then normalise our simulated distribution to yield the relative frequency of WB systems in each $r_p$ bin. This is compared with the corresponding observed quantity using a $\chi^2$ statistic. We select whichever combination $\left( \alpha, \beta, a_{_{break}} \right)$ yields the lowest $\chi^2$.

\begin{table}
  \centering
		\begin{tabular}{ccccc}
			\hline
			Model & $\gamma$ & $\alpha$ & $\beta$ & $a_{_{break}}$ \\ [3pt]
			\hline
			\multirow{3}{*}{Newton} & $-2$ & 0.8 & 1.63 & 5.14 \\
			 & 0 & 0.92 & 1.66 & 5.16 \\
			 & 2 & 1 & 1.63 & 4.59 \\ [5pt]
			\multirow{3}{*}{MOND} & 0 & 0.88 & 1.96 & 7.39 \\
			 & 1.2 & 0.92 & 1.95 & 7.39 \\
			 & 2 & 0.95 & 1.94 & 7.41 \\
			\hline
		\end{tabular}
	\caption{Parameters governing our prior distribution of wide binary semi-major axes in different Newtonian and MOND models (Equation \ref{P_a}). These are chosen to best reproduce the observed distribution of sky-projected separations (Equation \ref{P_r_p}).}
  \label{Table_P_a}
\end{table}

This procedure relies on knowing the eccentricity distribution $P \left( e \right)$. For the Newtonian models, we try a range of possible distributions parameterised by $\gamma_{_N}$ (Section \ref{Section_P_e}). Thus, we need to repeat our grid search through $\left( \alpha, \beta, a_{_{break}} \right)$ for each value of $\gamma_{_N}$.

As a first approximation, we can assume that these parameters are equal to the values governing the observed $P \left( r_p \right)$. This would give ${\alpha = 1}$, ${\beta = 1.6}$ and ${a_{_{break}} = 5}$ kAU (Equation \ref{P_r_p}). Our results indicate that this estimate is reasonably accurate regardless of the adopted $\gamma$, especially for the Newtonian model (Table \ref{Table_P_a}). We are always able to match the observed ${P \left( r_p \right)}$ distribution to within a root mean square (rms) scatter of 0.3\% over the range that we fit.

Based on our results, we suggest that future work could approximate $P \left( a \right) = P \left( r_p \right)$ in order to avoid one of the most computationally intensive parts of our algorithm. This works especially well for the Newtonian model. Even if this approximation is not made, it should be possible to speed the process up by searching more efficiently through different $P \left( r_p \right)$ distributions of the form given in Equation \ref{P_r_p}. For example, a gradient descent method could be used or a multigrid approach tried that successively zooms into the region around the model with the lowest $\chi^2$.

\subsection{Total system mass}
\label{Section_M}

Due to the complexity of MOND, our orbit integrations make the simplifying assumption that all of the mass $M$ in each WB system is contained within one of its stars. The dependence of WB dynamics on the mass ratio is discussed in Section \ref{Mass_ratio_effect}, where we show that the effect is small (though not zero like in Newtonian gravity). As a result, we need a prior distribution for $M$.

To construct this prior $P \left( M \right)$, we assume the stars in each WB have independent masses \citep[][figure 2]{Belloni_2017}. This leaves us with the simpler task of obtaining the mass distribution $\widetilde{p} \left( m \right)$ for isolated stars. We assume that this follows a broken power law.
\begin{eqnarray}
	\widetilde{p} \left( m \right) dm ~\propto~
\left\{
	\begin{array}{ll}
		m^{-2.3} dm & \mbox{if } m \leq M_\odot \\
		m^{-4.7} dm & \mbox{if } m \geq M_\odot
	\end{array}
\right.
	\label{P_single_mass}
\end{eqnarray}

\begin{figure}
	\centering
		\includegraphics[width = 8.5cm] {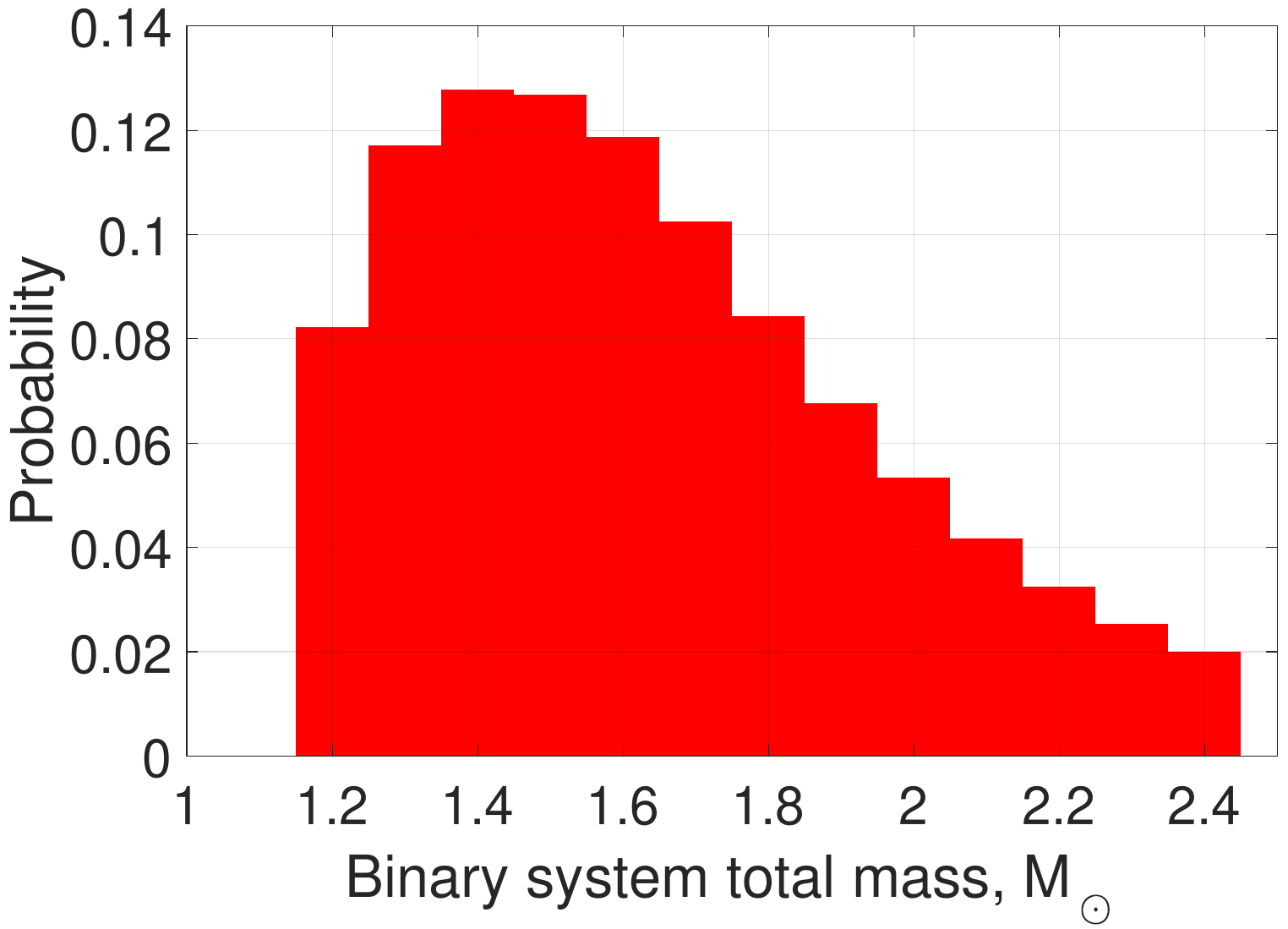}
		\caption{Our adopted prior distribution $P \left( M \right)$ for the total mass $M$ of WB systems in the Solar neighbourhood (Section \ref{Section_M}).}
	\label{Mass_prior}
\end{figure}

Here, $m$ is the mass of an individual star. We use a high-mass slope of $-4.7$ \citep[][equation 17]{Bovy_2017} and a low-mass slope of $-2.3$ \citep[][equation 2]{Kroupa_2001}. The resulting $\widetilde{p} \left( m \right)$ is used to obtain $P \left( M \right)$ by integration.
\begin{eqnarray}
	P \left( M \right) dM ~=~ \int_{m_{min}}^{M - m_{min}} \widetilde{p} \left( m \right) \widetilde{p} \left( M - m \right)\,dm
	\label{P_total_mass}
\end{eqnarray}

Following the work of \citet{Pittordis_2018}, we assume that the WBT will not use stars with $m < m_{min} = 0.55 \, M_\odot$ because of their faintness. Due to the steeply declining stellar mass function above $M_\odot$, we only consider WB systems with $M$ in the range ${\left(1.2 - 2.4 \right)M_\odot}$. The resulting $P \left( M \right)$ is shown in Figure \ref{Mass_prior}.

In Newtonian gravity, the scale invariance of the force law implies that $M$ is irrelevant for the $\left( r_p, \widetilde{v} \right)$ distribution once $a$ and $e$ are fixed. This allows our Newtonian orbit library to consider just one value for $M$. We arbitrarily set this to ${1.5 \, M_\odot}$.

The mass of each WB system has only a small effect on its expected orbital velocity. This is because MOND effects arise at smaller separations in a lower mass system, counteracting the tendency of these systems to rotate slower. Using Equation \ref{v_tilde} to estimate the circular velocity $v_c$ at the MOND radius $r_{_M}$ (Equation \ref{Deep_MOND_limit}), we see that
\begin{eqnarray}
	v_c \left( r_{_M} \right) \propto \sqrt[4]{M}
\end{eqnarray}

Consequently, systems with total mass $M = 1 \, M_\odot$ instead of ${2 \, M_\odot}$ would rotate only 16\% slower at the point where MOND effects start to become significant. This suggests that the WBT could benefit from much better statistics if it uses observations of lower mass systems. This would also allow contamination to be reduced via a tighter cut on the projected separation, as MOND effects would arise closer in (Equation \ref{Deep_MOND_limit}).

In the short term, the most serious problem with this is that lower mass stars are much less luminous \citep[e.g.][]{Mann_2015}. In the long run, this can be addressed with the use of larger telescopes and longer exposures. Using more common systems also makes it more likely that there would be a suitable background object within the same field of view whose true parallax and proper motion can be neglected, making it useful for calibration.

\subsection{Orbital plane}
\label{Orbital_pole}

Due to the presence of a preferred direction $\widehat{\bm{g}}_{_{ext}}$ induced by the EF, the behaviour of a WB system will depend somewhat on the orientation of its orbital pole $\widehat{\bm{h}}$ with respect to $\widehat{\bm{g}}_{_{ext}}$. As the WB orbital period is expected to be at most a few Myr\footnote{using Kepler's Third Law for stars similar to the Sun and a separation below 20 kAU}, we do not expect $\widehat{\bm{g}}_{_{ext}}$ to rotate significantly during a few WB orbits. Combined with our assumption that each WB system is dominated by one of its stars, this leads to an axisymmetric potential. Consequently, the only physically relevant aspect of $\widehat{\bm{h}}$ is its angle $\theta$ with $\widehat{\bm{g}}_{_{ext}}$.

We take this into account by considering a grid of possible $\theta$ whose prior distribution ${P \left( \theta \right)}$ is assigned based on the assumption that $\widehat{\bm{h}}$ is isotropically distributed.
\begin{eqnarray}
	P \left( \theta \right) d\theta ~=~ \frac{1}{2} \sin \theta\,d\theta
	\label{Orbital_pole_distribution}
\end{eqnarray}

We only consider angles $\theta \leq \frac{\mathrm{\pi}}{2}$ as larger angles are equivalent to a WB with a lower $\theta$ but with its initial velocity reversed. Because gravitational problems are time reversible, this should not affect WB characteristics like its average orbital velocity. Such properties are thus expected to be the same for $\theta \to \mathrm{\pi} - \theta$.

In the long term, the EF on each WB changes with time as it rotates around the Galaxy. However, we do not expect this to affect our results very much because the Galactic orbit is much slower than the WB orbit. As a result, the initial distribution of $\theta$ is likely preserved (Figure \ref{Centauri_adiabatic_test}), maintaining a nearly isotropic $\widehat{\bm{h}}$ distribution. This issue is discussed further in Section \ref{Secular_effects}, where we show that the distribution in $r_p$ and $\widetilde{v}$ is nearly the same whether the orbit of Proxima Cen is integrated for just 20 revolutions with a fixed EF or over 5 Gyr in a time-varying EF (Figure \ref{Centauri_control}). This is because each WB system is expected to have $\widehat{\bm{r}}$ go through a wide range of directions relative to the EF such that the gravity between its stars follows an angular average. In any case, even an EF-dominated system in the Solar neighbourhood should not have a self-gravity that depends very much on its orientation relative to the EF (using $K_0 = -0.26$ in Equation \ref{g_ratio_QUMOND} shows that the force is affected at most 9\%).


\subsection{Viewing angle}
\label{Viewing_angle}

Gaia observations are not expected to yield all six phase space co-ordinates for most WB systems it discovers. In particular, the line of sight separation between the stars would generally not be known as accurately as the other observables \citep{Pittordis_2018}, with our calculations suggesting an accuracy of $\sim 80$ kAU (Section \ref{Distance_measurement}). The radial velocity difference between the stars may also be difficult to determine at the ${\sim 0.1}$ km/s accuracy required for the WBT. In addition to accurate spectra, this also requires knowledge of the difference in convective blueshift corrections between the stars \citep{Kervella_2017}. In the short run, this makes it inevitable that what we infer about each system will depend on its orientation relative to our line of sight towards it.

To take this into account, at each timestep of our WB orbital integrations, we consider a 2D grid of possible directions $\widehat{\bm{n}}$ in which the observer lies relative to the WB system. Assuming the observer is much more distant than the WB separation $r$, we determine $r_p$ using
\begin{eqnarray}
	r_p ~\equiv ~ \left| \bm{r} - \left( \bm{r} \cdot \widehat{\bm{n}} \right) \widehat{\bm{n}} \right|
	\label{r_p_definition}
\end{eqnarray}

We use this in Equation \ref{v_tilde} to find $\widetilde{v}$, assuming masses are known regardless of the viewing angle as these should be determined from luminosities of nearly isotropic stars (Section \ref{Mass_measurement}). We then increment the appropriate $\left(r_p, \widetilde{v} \right)$ bin by the fraction of the full ${4\mathrm{\pi}}$ solid angle represented by each $\widehat{\bm{n}}$, assuming this has an isotropic distribution. This should be valid out to the ${\approx 150}$ pc distance relevant for the WBT as the MW disk scale height is larger \citep[][figure 7]{Ferguson_2017}.


\subsection{External field strength}
\label{External_field_strength}

We take the EF to point towards the Galactic centre and have a magnitude sufficient to maintain the observed Local Standard of Rest (LSR) speed of $v_{c, \odot} = 232.8$ km/s, assuming the Sun is $R_\odot = 8.2$ kpc from the Galactic centre \citep{McMillan_2017}. Gaia DR2 remains consistent with these parameters \citep{Kawata_2019}.

We use Equation \ref{g_N_ext} to find the magnitude of the Newtonian-equivalent EF $\bm{g}_{_{N,ext}}$ from $\bm{g}_{ext}$. Because $\bm{g}_{ext}$ is fixed observationally, using a different MOND interpolation function alters $\bm{g}_{_{N,ext}}$. We use the simple form of this function for reasons discussed in Section \ref{Interpolating_function}.

In principle, Equation \ref{g_N_ext} is only valid in spherical symmetry and is thus invalid near the MW disk and its resulting vertical force. However, this is expected to be rather small in the Solar neighbourhood because we are ${\approx 4}$ disk scale lengths from the Galactic Centre \citep{Bovy_2013}. We consider the accuracy of this algebraic MOND approximation in Section \ref{Galactic_disk_effect}. There, we show that the local value of $\nu_{ext}$ should be affected ${<1\%}$ by the vertical gravity due to the MW disk.

\section{The detection probability}
\label{Statistics}


To forecast the feasibility of the WBT, we need to obtain and compare the ${P \left( \widetilde{v} \right)}$ distributions expected in Newtonian and MOND gravity. We obtain these distributions by marginalising over WB parameters using the prior distributions outlined in Section \ref{Priors}. As our prior on $a$ is already chosen to get an appropriate posterior distribution for $r_p$ (Section \ref{Section_P_a}), marginalising over $r_p$ is very simple. For consistency, we use the numerically determined ${P \left( r_p \right)}$ rather than the observed distribution, though the differences are very small (rms error ${\la 0.3\%}$).

To compare the Newtonian ${P_N \left( \widetilde{v} \right)}$ with the MOND ${P_M \left( \widetilde{v} \right)}$, we use an algorithm that we make publicly available as it can be used to forecast the distinguishability of any two probability distributions.\footnote{Algorithm available at:
\href{https://uk.mathworks.com/matlabcentral/fileexchange/65465-distinguishing-two-probability-distributions-with-a-finite-number-of-data-points}{MATLAB file exchange, code 65465}} This provides a quantitative estimate of how easily we can distinguish the two theories using $N$ well-observed WB systems in different ranges of $r_p$ contained within the interval $\left(1 - 20 \right)$ kAU. In this way, we quantify how many such systems would be needed for the WBT. The actual number is likely to be somewhat larger due to observational uncertainties (Section \ref{Measurement_uncertainties}) and various systematic effects (Section \ref{Systematics}). Moreover, not all WB systems will be suitable for the WBT.

Our approach is to find the likelihood that observations drawn from ${P_M \left( \widetilde{v} \right)}$ are inconsistent with expectations based on ${P_N \left( \widetilde{v} \right)}$. Suppose we have ${N = 100}$ systems and are interested in the number $n$ of them which have ${\widetilde{v} > 1.2}$. If we expect $n = 9.7$ in Newtonian gravity but a larger number in MOND, then we begin by finding the maximum value of $n$ at the 99\% confidence level according to ${P_N \left( \widetilde{v} \right)}$. Formally, this value $n_{max}$ is the smallest integer which satisfies ${P \left( n \leq n_{max} \right) > 0.99}$. Due to the discreteness of WB systems, ${P \left( n \right)}$ follows a binomial distribution whose parameters are $\left(100, 0.097 \right)$ in this example.

This leads to the conclusion that the Newtonian model could be used to explain any observed ${n \leq n_{max} = 16}$. We then find the likelihood that $n > n_{max}$ if the observations correspond to a MOND universe. We call this likelihood the detection probability $P_{detection}$ of MOND relative to Newtonian gravity for the adopted prior distributions, $\left(r_p, \widetilde{v} \right)$ range and number of systems used.

If we use a $\widetilde{v}$ range in which ${P_M \left( \widetilde{v} \right)}$ has less probability than ${P_N \left( \widetilde{v} \right)}$, we reverse the logic outlined above. Thus, we find the 99\% confidence level \emph{lower} limit of ${P_N \left( \widetilde{v} \right)}$. We then determine the likelihood that $n$ is even smaller if the observations are drawn from ${P_M \left( \widetilde{v} \right)}$. In practice, this situation should not arise because we expect the WBT to work best by focusing on high values of $\widetilde{v}$ which are more common in MOND. Even so, our analysis is not a blind search for discrepancies with the Newtonian model but a more targeted search for discrepancies in the direction that would arise if MOND were correct. Blind analyses should also be conducted, especially if neither model describes the observations well.

When conducting our analysis, we try all possible rectangular regions in $\left( r_p, \widetilde{v} \right)$ space to see which one maximises $P_{detection}$. We expect the algorithm to use the full range of $r_p$ available to it (Table \ref{Wide_binary_parameters}), but it is not clear a priori exactly which range of $\widetilde{v}$ will work best. This is because both models predict nearly 100\% of systems within a very wide $\widetilde{v}$ range. If a very narrow range were used instead, it is quite possible that this has some probability of arising in MOND but no chance in Newtonian gravity. This is good for the WBT in the sense that a detection within the adopted $\widetilde{v}$ range would constitute very strong evidence for MOND. However, even in MOND, it may be very unlikely to observe such a system. This would lead to a low $P_{detection}$. Thus, some intermediate range of $\widetilde{v}$ is expected to be most suitable for the WBT. Our discussion so far suggests a range from the high end of the Newtonian $\widetilde{v}$ distribution to the upper limit of the MOND distribution.

Although we are a priori unsure exactly which $\widetilde{v}$ range works best, it is clear that its lower limit $\widetilde{v}_{min}$ should not be set above the maximum possible $\widetilde{v}$ in the Newtonian model. This is because raising $\widetilde{v}_{min}$ above this value does not further reduce the already zero probability of finding a Newtonian WB system with $\widetilde{v} > \widetilde{v}_{min}$. However, increasing $\widetilde{v}_{min}$ does reduce the probability of finding a system like that if gravity were governed by MOND. Thus, raising $\widetilde{v}_{min}$ above 1.42 can only ever reduce $P_{detection}$. For this reason, we restrict the algorithm to only consider $\widetilde{v}_{min} \leq 1.42$.

The upper limit on $\widetilde{v}$ is not restricted apart from the basic requirements to exceed $\widetilde{v}_{min}$ and to lie below the maximum of 1.68 which arises in our MOND models.\footnote{We raise this cap to 3.2 when considering MOND without the EFE (Section \ref{MOND_no_EFE}).} In theory, selecting a larger value will not affect $P_{detection}$. However, in the real world, this would lead to additional sources of contamination that could hamper the WBT (Section \ref{Systematics}).

If the data give any hint of a MOND signal, this will be highly controversial and immediately raise many observational and theoretical questions. On the theory side, astronomers would inevitably try a different $\gamma_{_N}$, thus changing the eccentricity distribution for the Newtonian model. In particular, higher values of $\gamma_{_N}$ would increase the weight given to highly eccentric orbits, making it more likely that $\widetilde{v}$ significantly exceeds 1. In future, it may be possible to predict the Newtonian eccentricity distribution. As this is not currently possible, we use the most conservative case where the value of $\gamma_{_N}$ is that which makes the WBT as difficult as possible. We find this by trying a grid of possible values for $\gamma_{_N}$, each time recording $P_{detection}$. Whichever $\gamma_{_N}$ yields the lowest $P_{detection}$ then sets $P_{detection}$ for that particular value of $N$. In this way, we quantify how well the WBT can be expected to work for different values of $N$ and different model assumptions, both for Newtonian gravity and for MOND.

\section{Results}
\label{Results}

\begin{figure}
	\centering
		\includegraphics[width = 8.5cm] {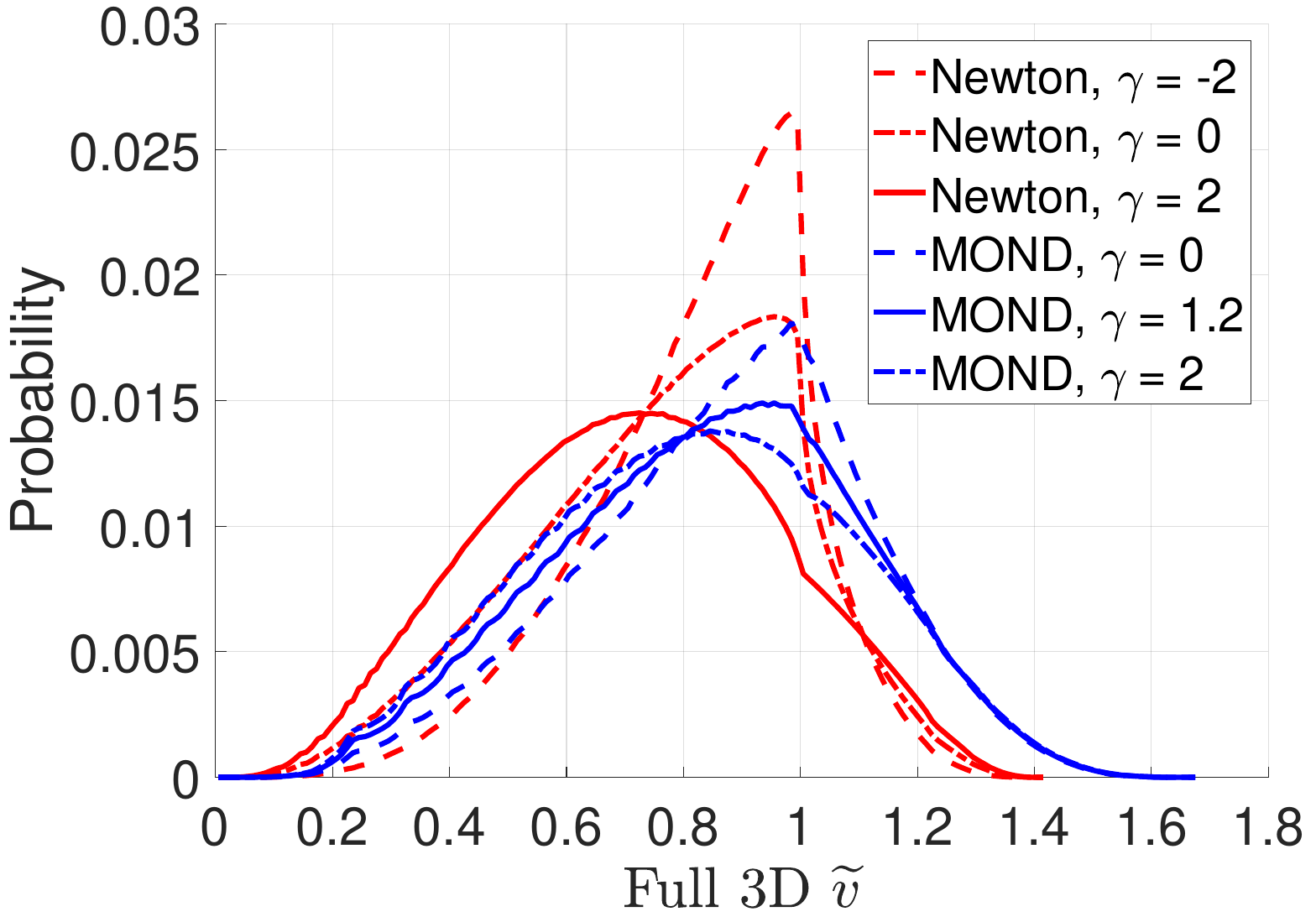}
		\includegraphics[width = 8.5cm] {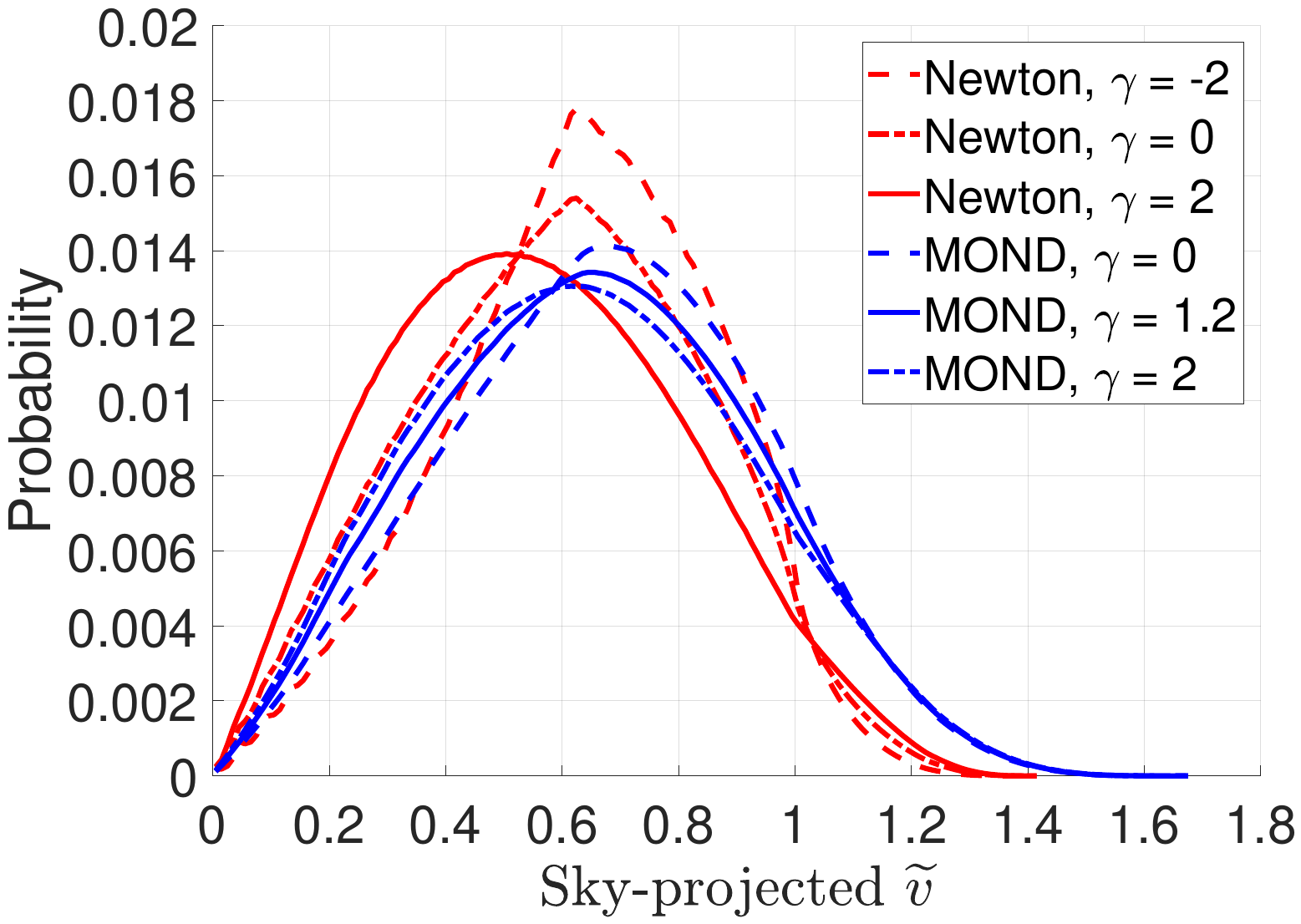}
		\caption{Distributions of the scaled WB relative velocity $\widetilde{v}$ (Equation \ref{v_tilde}) in Newtonian and MOND gravity, using the prior distributions from Section \ref{Priors}. We only consider WB systems with sky-projected separation $r_p = \left( 1 - 20 \right)$ kAU. Different eccentricity distributions are shown by varying $\gamma$ (Equation \ref{P_e}). Notice how $\gamma$ and $\gamma_{_N}$ do not much affect the results for $\widetilde{v} \ga 1.1$. \emph{Top}: Using the full 3D relative velocity in the definition of $\widetilde{v}$ (Equation \ref{v_tilde}). \emph{Bottom}: Using the sky-projected velocity only.}
	\label{v_tilde_comparison_gamma}
\end{figure}

We begin by showing the $\widetilde{v}$ distributions $P \left( \widetilde{v} \right)$ for the Newtonian and MOND models under different assumptions about $\gamma$ (Figure \ref{v_tilde_comparison_gamma}). The distributions are rather insensitive to $\gamma$ (Equation \ref{P_e}) in the region $\widetilde{v} \ga 1.1$, a result also evident from figure 2 of \citet{Pittordis_2018}. Clearly, a much larger fraction of WB systems have such a high $\widetilde{v}$ in MOND than in any plausible Newtonian model.

It may initially seem surprising that $\gamma_{_N}$ does not much affect the Newtonian $\widetilde{v}$ distribution for $\widetilde{v} \ga 1.1$. After all, such high values of $\widetilde{v}$ are impossible for nearly circular orbits but quite possible for elliptical orbits. However, a highly elliptical orbit spends the majority of its time near apocentre, where $\widetilde{v}$ is very low. Consequently, such an orbit will not contribute much probability to the region $\widetilde{v} > 1.1$. This is why Newtonian models with any value of $\gamma_{_N}$ can never perfectly mimic a modified gravity theory with a higher circular speed (Equation \ref{a_definition}).

In Section \ref{Boost_to_Newton}, we used Equation \ref{eta_EFE} to estimate how much MOND would typically enhance the gravity binding a WB if the EF were dominant. This is a reasonable approximation towards the upper limit of the WB separations we consider. Therefore, we expect that
\begin{eqnarray}
	\widetilde{v} ~\leq~ \sqrt{2 \nu_{_{ext}} \left(1 + \frac{K_0}{3} \right)}
	\label{v_tilde_max}
\end{eqnarray}

In the Solar neighbourhood, this suggests that the MOND $\widetilde{v}$ distribution extends up to 1.68. This is indeed the upper limit of our much more rigorously determined MOND $\widetilde{v}$ distributions (Figure \ref{v_tilde_comparison_gamma}).


So far, we assumed that the WBT requires the full 3D relative velocity $\bm{v}$ of each system. However, Gaia is expected to release proper motions for a large number of stars before there is time to follow them up and take accurate radial velocity measurements. Moreover, these can be hampered by uncertainties in convective blueshift corrections \citep[][section 2.2]{Kervella_2017}. Thus, we redo our analysis using only the sky-projected velocity, which we find in a similar way to $r_p$ (Equation \ref{r_p_definition}). The resulting $\widetilde{v}$ distributions are shown in the bottom panel of Figure \ref{v_tilde_comparison_gamma}.

\begin{figure}
	\centering
		\includegraphics[width = 8.5cm] {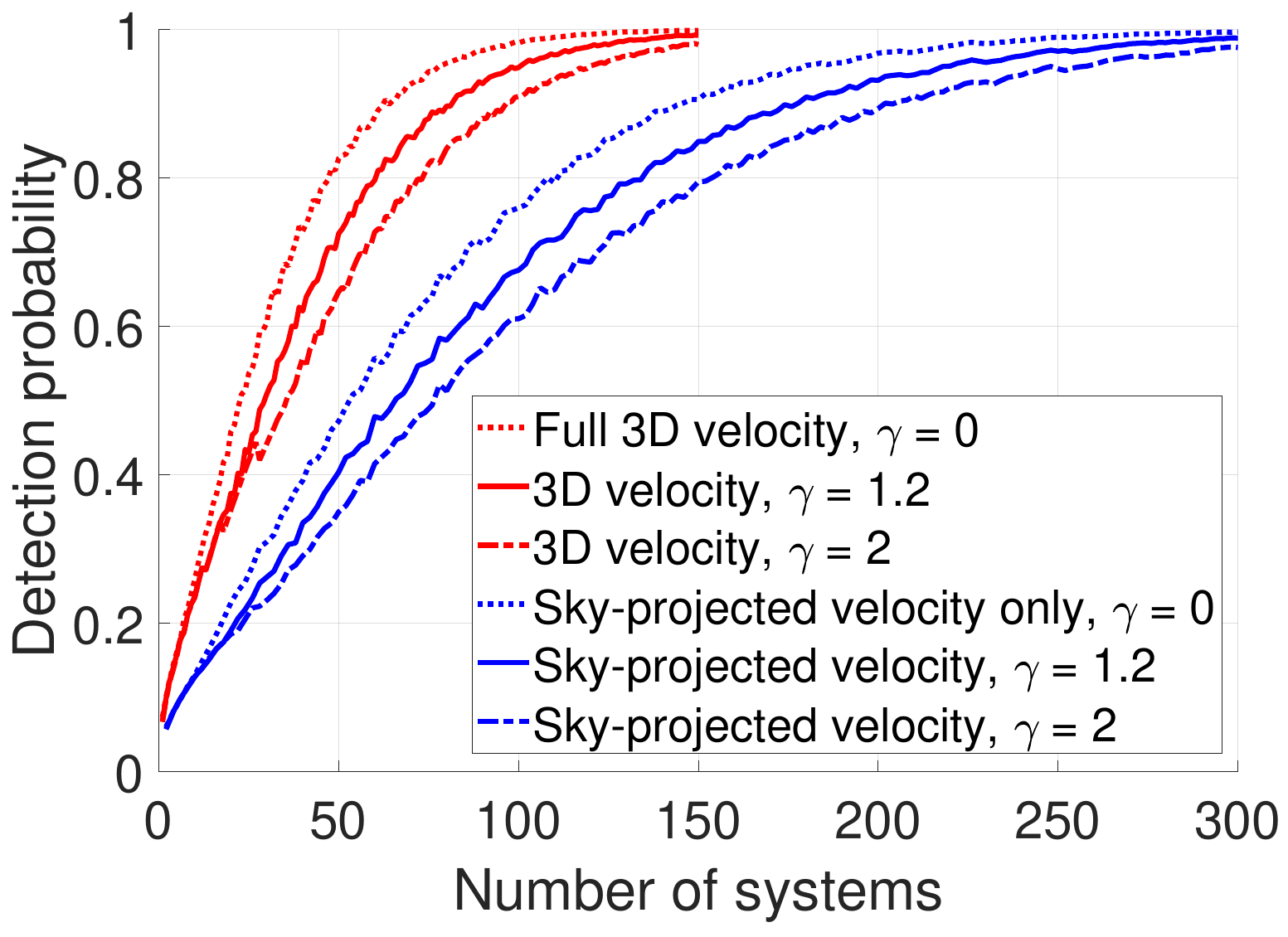}
		\caption{The probability $P_{detection}$ of detecting a significant departure from Newtonian expectations if the underlying $\widetilde{v}$ distribution follows the MOND model. The method used is explained in Section \ref{Statistics}. For each value of $\gamma$, we choose the value of $\gamma_{_N}$ which minimises $P_{detection}$. This `best-fitting' $\gamma_{_N}$ is ${\approx 0.5}$ below the value of $\gamma$ adopted for the MOND model. Thus, we consider all possible $\gamma_{_N}$ for the models where ${\gamma = 0}$.}
	\label{P_detection_gamma}
\end{figure}

Having obtained Newtonian and MOND $\widetilde{v}$ distributions, we compare them using the method explained in Section \ref{Statistics}. This allows us to quantify the detectability of MOND effects for different numbers of WB systems. Our results show that the WBT should be feasible with a few hundred well-observed systems (Figure \ref{P_detection_gamma}).

The jagged nature of the $P_{detection}$ curves arise from discreteness effects. To understand this, suppose that 99.1\% of the time, binomial statistics tells us that there will be ${\leq15}$ systems in the chosen range of parameters $\left( r_p, \widetilde{v} \right)$ for ${N = 100}$ systems. Thus, $P_{detection}$ is based on observing $\geq$16/100 systems in that parameter range under the MOND model. If $N$ is raised slightly, then the chance of getting $\leq$15 systems like that might drop to 98.9\% in the Newtonian model. As this is below 99\%, we would be forced to consider that Newtonian gravity can explain the existence of $\leq$16 systems in the selected parameter range at the adopted 99\% confidence level. This causes a sudden drop in $P_{detection}$ because this is now based on observing ${\geq 17}$ systems in the chosen parameter range rather than the previous ${\geq 16}$. Consequently, although $P_{detection}$ generally increases with $N$, it occasionally decreases because it is impossible to exactly maintain a fixed confidence level with a finite number of data points.

When using the full 3D relative velocity and an intermediate value for $\gamma$ of 1.2, our calculations show that the WBT is best done by considering systems with $\widetilde{v} > 1.05 \pm 0.02$, with the scatter arising from discreteness effects. As expected, the analysis prefers not to impose an upper limit on $\widetilde{v}$, thus considering systems with $\widetilde{v}$ all the way up to the maximum value of 1.68 which arises in our MOND simulations. The probability of finding a WB system in this range is ${\approx 16 \pm 2\%}$ for the MOND model but only ${\approx 4.5 \pm 1\%}$ for the Newtonian model.

If using sky-projected velocities only, it becomes best to consider the range $\widetilde{v} > 0.97 \pm 0.02$ because the sky-projected velocity is generally smaller than the full velocity. The probability of finding a WB system in this range is ${\approx 9 \pm 1\%}$ for the MOND model but only ${\approx 3 \pm 1\%}$ for the Newtonian model.

\begin{figure}
	\centering
		\includegraphics[width = 8.5cm] {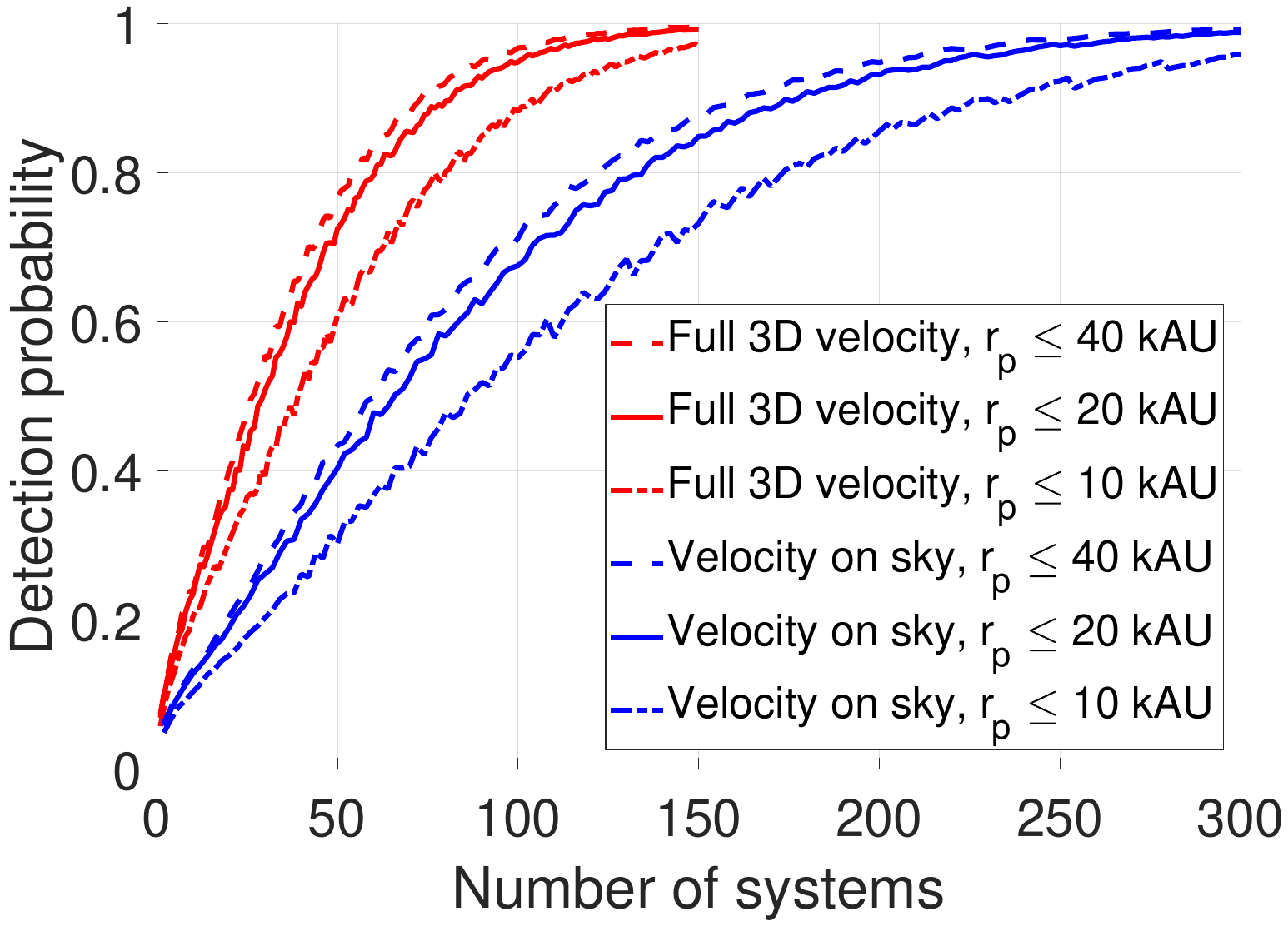}
		\caption{Similar to Figure \ref{P_detection_gamma} for the case $\gamma = 1.2$, but with different upper limits on $r_p$. Probabilities are normalised to the number of systems with $r_p$ between 1 kAU and the maximum considered for each curve, avoiding changes to $P_{detection}$ due to a different number of WB systems in each range of $r_p$ (Figure \ref{r_p_distribution}). Once the outer limit on $r_p \ga 10$ kAU, raising it further has only a small effect on our analysis because the MOND boost to gravity is limited by the Galactic external field (Figure \ref{Pittordis_forces}).}
	\label{P_detection_40_kAU}
\end{figure}

As well as giving guidance on what $\widetilde{v}$ range is best for the WBT, our algorithm also provides information about the best $r_p$ range. At the lower limit, the algorithm generally prefers to use 3 kAU even though it could have extended this down to 1 kAU. The fact that it does not means that the WBT is worsened by including such systems, presumably because they are very nearly Newtonian.

At the upper limit, the algorithm behaves as expected by preferring to use systems with $r_p$ up to the maximum of 20 kAU that we allow. This shows that the WBT would benefit from including even more widely separated WBs if they could be accurately identified and their slower relative velocities accurately measured. The work of \citet{Andrews_2018} suggests that this may be feasible (they went up to 40 kAU).

To investigate how much this would help the WBT, we repeat our analysis with the upper limit on $r_p$ raised to 40 kAU but fix $\gamma$ at our nominal value of 1.2. To avoid the increased number of WB systems automatically improving $P_{detection}$, we normalise our probability distribution to the number of WB systems whose $r_p = \left(1 - 40 \right)$ kAU.

The increased range in $r_p$ has only a small effect on $P_{detection}$ (Figure \ref{P_detection_40_kAU}). In fact, our algorithm sometimes prefers not to use this extra information, instead restricting itself to $r_p < 37$ kAU to exploit discreteness effects. Clearly, there is only a marginal benefit to doing the WBT with systems that have $r_p \leq 40$ kAU instead of 20 kAU.

This is also evident from Figure \ref{Pittordis_forces}, where we see that the MOND boost to gravity does not increase much for systems more widely separated than their MOND radius (Equation \ref{Deep_MOND_limit}). As this is only 11 kAU for the heaviest systems we consider (${2.4 \, M_\odot}$), it is not very helpful to consider systems which are much more widely separated.

\begin{figure}
	\centering
		\includegraphics[width = 8.5cm] {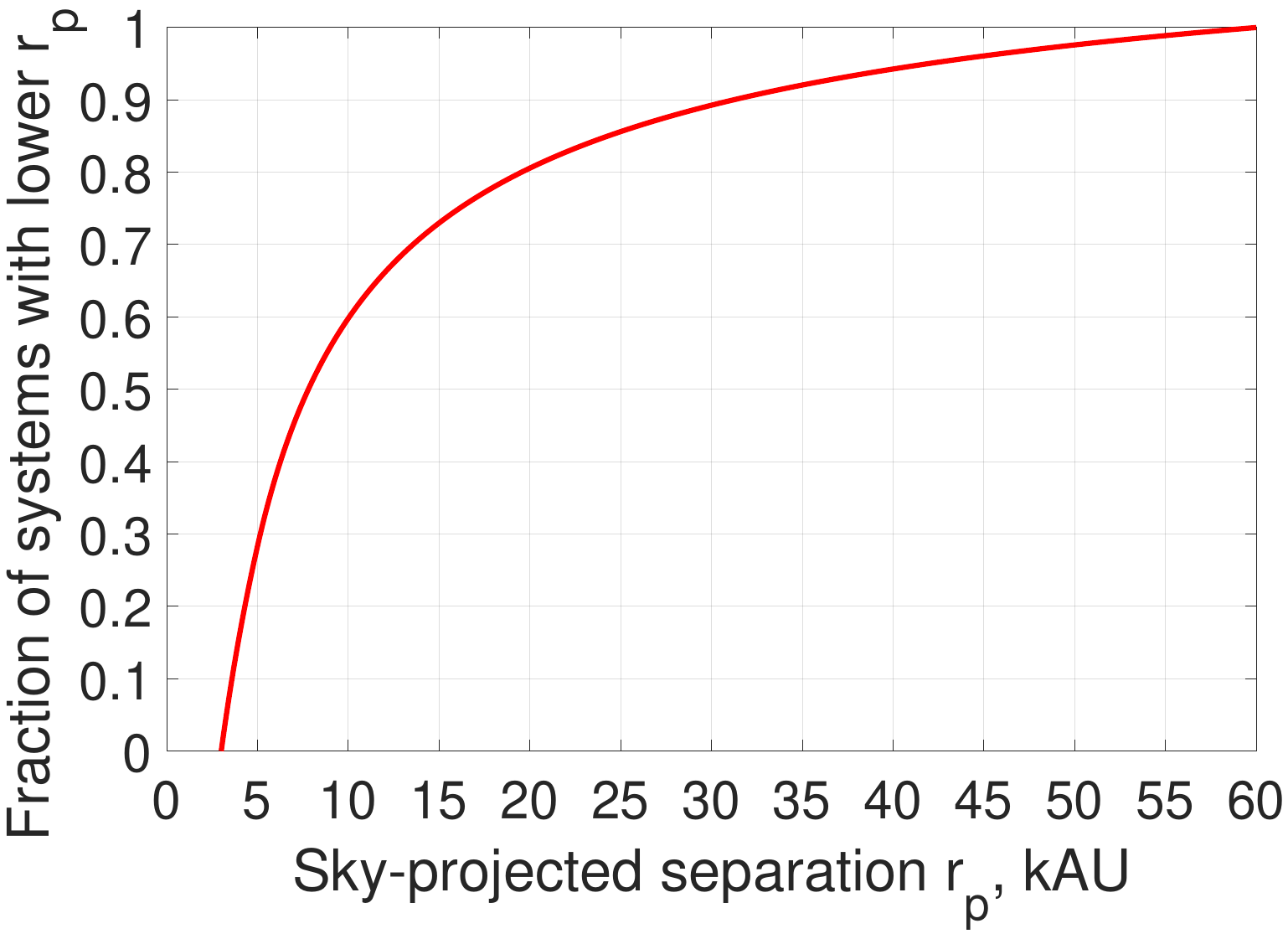}
		\caption{The cumulative probability distribution of $r_p$ for WB systems with $r_p = \left(3 - 60 \right)$ kAU, according to the empirically determined Equation \ref{P_r_p}.}
	\label{r_p_distribution}
\end{figure}

Although the MOND effect is not much enhanced by going out to 40 kAU instead of 20 kAU, this would increase the number of systems available for the WBT. The way in which this occurs can be quantified based on existing observations of the WB $r_p$ distribution (Section \ref{Section_P_a}). By integrating Equation \ref{P_r_p} and considering only systems where $r_p = \left(3 - 60 \right)$ kAU, we get the cumulative distribution of $r_p$ shown in Figure \ref{r_p_distribution}. This shows that the number of WB systems is unlikely to increase much as a result of increasing the upper limit on $r_p$ from 20 kAU to 40 kAU.

However, allowing systems with larger $r_p$ increases the chance of contamination and requires more accurate proper motions due to a slower WB orbit. Given these challenges, it might be preferable to reduce the limit on $r_p$. Thus, Figure \ref{P_detection_40_kAU} also shows how the WBT would be affected if using only systems where $r_p \leq 10$ kAU. In this case, the WBT is hampered by the smaller MOND effects at low separations. Moreover, Figure \ref{r_p_distribution} also shows that there would be a discernible reduction in the number of available WB systems.

We therefore recommend the use of an intermediate upper limit to $r_p$ of ${\approx 20}$ kAU. Without more information on how observations get more difficult at larger $r_p$, it is impossible to provide any more quantitative guidance regarding the best $r_p$ range for the WBT.


\section{Measurement uncertainties}
\label{Measurement_uncertainties}

In this section, we consider whether the basic parameters of each WB system could be constrained accurately enough for the WBT. Because $v_c \appropto \frac{1}{\sqrt{r}}$ but $g \appropto \frac{1}{r^2}$, the WB orbital acceleration $\appropto {v_c}^4$. Consequently, a ten-fold improvement in the accuracy of velocity measurements allows us to probe systems whose orbital accelerations are ${10^4 \times}$ smaller. This makes the WBT very well placed to benefit from accurate proper motions.

In the short term, these will come from the Gaia mission \citep{Perryman_2001}. Its performance has become much clearer due to the recent Gaia DR2 \citep{GAIA_2018}. However, some of the inputs to the WBT will be based on data collected in other ways and on our theoretical understanding of stars (Section \ref{Mass_measurement}).

\subsection{Relative velocity}

\begin{figure}
	\centering
		\includegraphics[width = 8.5cm] {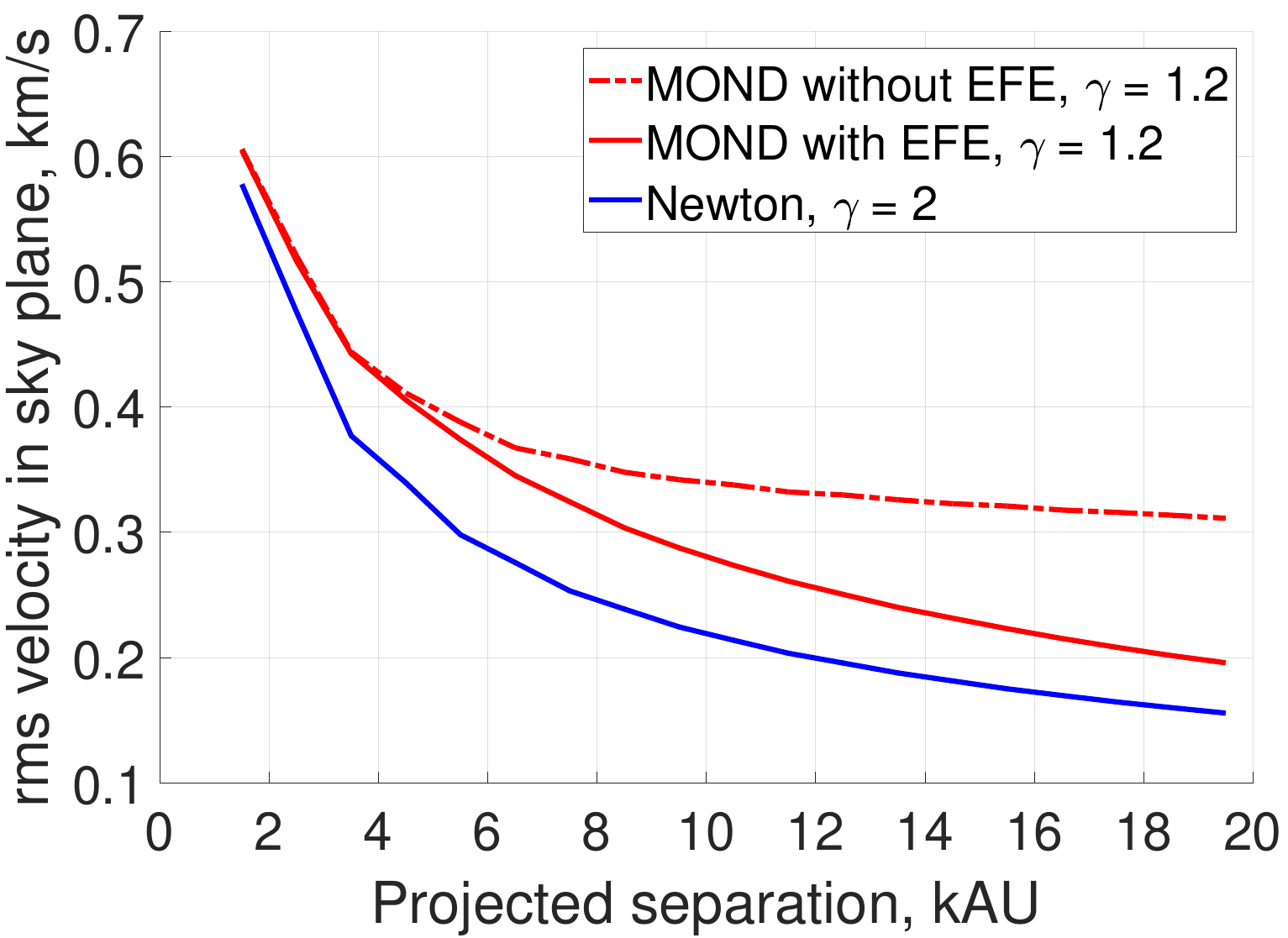}
		\caption{The rms sky-projected relative velocity between the components of a WB system with total mass ${M = 1.5 \, M_\odot}$, shown as a function of its projected separation. The legends indicate the adopted values of $\gamma$ (Equation \ref{P_e}). Some speculative versions of MOND lack an external field effect, as shown with a dot-dashed line and discussed in Section \ref{MOND_no_EFE}.}
	\label{v_rms_no_EFE}
\end{figure}

To get a feel for the actual relative velocities of WB systems relevant for the WBT, we use Figure \ref{v_rms_no_EFE} to show the rms sky-projected relative velocity for systems with ${M = 1.5 \, M_\odot}$. For completeness, we also show results for MOND without an EFE (Section \ref{MOND_no_EFE}). This is based on applying Equation \ref{g_near_field} at all radii, with $\nu$ calculated assuming no EF.

Once accurate radial velocity measurements become available, we can measure the rms 3D relative velocity between WB stars. Due to isotropy, this is expected to be ${\sqrt{\frac{3}{2}} \times}$ larger than just its sky-projected component.

\subsubsection{Proper motion}

For a WB system with $M = 1.5 \, M_\odot$ and separation ${r = 20}$ kAU, Newtonian gravity predicts a circular velocity of 260 m/s (Equation \ref{v_tilde}). If MOND is correct, we expect the circular velocity to typically be ${\approx 20\%}$ higher (Section \ref{Results}). Thus, a velocity accuracy of 10 m/s should be enough for a ${5 \sigma}$ detection of MOND effects with the WBT. This accuracy needs to be achieved for a sufficiently large number of systems and thus out to a large enough heliocentric distance. According to \citet[][section 5.1]{Pittordis_2018}, the required distance is ${\approx 100}$ pc. At this distance, a velocity of 10 m/s corresponds to $21 \, \mu$as/yr.

Fortunately, Gaia DR2 has already achieved an accuracy of $70 \, \mu$as/yr for stars of 15\textsuperscript{th} magnitude based on $T = 22$ months of observations \citep[][table 3]{GAIA_2018}. This is the expected brightness of stars relevant to the WBT \citep[][section 5.1]{Pittordis_2018}. The accuracy of proper motions improves as $T^{1.5}$ because the actual signal (shift in sky position) grows linearly with $T$ whereas the collection of more data reduces the position error as $\frac{1}{\sqrt{T}}$. Thus, a future release of Gaia data collected over $T = 5$ years should achieve a proper motion accuracy better than $20 \, \mu$as/yr. This will enable the WBT.

\subsubsection{Radial velocity}

Although radial velocities are not strictly essential for the WBT (Figure \ref{P_detection_gamma}), they could contribute meaningfully if their accuracy is comparable to that achieved with proper motions. The main difficulty would likely be in determining how much the gravitational redshift and convective blueshift corrections differ between the stars in a WB. After careful study of the 11\textsuperscript{th} magnitude star Proxima Cen, \citet{Kervella_2017} considered such effects and measured its radial velocity to within 32 m/s.

These corrections could be known much more accurately using a larger sample of binary stars whose separation is small enough that their dynamics would definitely be Newtonian \citep{Pittordis_2018}. This is based on the idea that the stars in each binary statistically have the same radial velocity. The same is also true of the stars within an individual binary system, if their radial velocities are averaged over a full orbital period. In practice, this technique can be implemented as long as spectroscopic observations are taken over a sufficiently large fraction of its orbit.

At present, it is not clear whether these methods will allow the WBT to benefit significantly from radial velocity information. Much depends on observations of close binaries, the sample of which is expected to be greatly enlarged in the Gaia era \citep{Zwitter_2003}.

\subsection{Distance}
\label{Distance_measurement}

At a distance of 100 pc, a WB system will by definition have a parallax of 10 milliarcseconds (mas). Gaia DR2 has allowed the determination of parallaxes to within $40 \, \mu$as \citep[][table 3]{GAIA_2018}. As this is $250\times$ smaller than the parallax of a WB at 100 pc, its distance can be measured to an accuracy of 0.4\%. This represents a line of sight distance uncertainty of 0.4 pc or 82.5 kAU. Although this is much larger than the WB separations relevant to the WBT, it is small enough to greatly reduce the chance of a falsely detected WB. Also requiring a common proper motion and (if known) a similar radial velocity makes it extremely unlikely that any pair of stars will be misidentified as a WB. This is probably why the WB sample of \citet{Andrews_2018} has a very low contamination rate of ${\approx 6\%}$.

As well as false positive detections of WBs, another concern is false negatives that make the WBT unnecessarily difficult. This could arise if a system has a faint red background galaxy which appears point-like to Gaia but is too faint for an accurate parallax measurement. Thus, the system might appear like a triple star configuration unsuitable for the WBT. Fortunately, such a situation appears simple to rectify with a modest amount of follow-up observations.

\subsection{Mass}
\label{Mass_measurement}

Based on the parallax and apparent magnitude of each star in a WB system, we will have accurate knowledge of its absolute magnitude. This can be converted into a stellar mass using empirical mass-luminosity relations. Such relations are often based on eclipsing binaries with a separation small enough that modified gravity would not affect the system \citep[e.g.][]{Spada_2013}. Using this and other techniques, it is already possible to constrain the mass of a star to within 6\% using just its luminosity \citep{Mann_2015, Eker_2015}. Gaia observations are expected to tighten this considerably.

A fractional uncertainty of 6\% in stellar mass translates to a 3\% uncertainty in $\widetilde{v}$ (Equation \ref{v_tilde}), much smaller than the ${\approx 20\%}$ boost to $\widetilde{v}$ expected in MOND (Section \ref{Results}). Thus, the total mass of each WB system should be known accurately enough for the WBT.

\section{Theoretical uncertainties}
\label{Theoretical_uncertainties}

Our results are sensitive to the way in which MOND is formulated and its interpolating function. This is because WBs have internal accelerations $\sim a_{_0}$ and the EF is also of this order. As a result, forecasting the WBT requires careful numerical simulations. Despite this, we showed in Section \ref{Results} that the extent of the $\widetilde{v}$ distributions resulting from our simulations (Figure \ref{v_tilde_comparison_gamma}) can be captured rather accurately by Equation \ref{v_tilde_max}. This allows us to quickly get a reasonable idea of how the WBT would be affected by different interpolating functions (Section \ref{Interpolating_function}) and MOND formulations (Section \ref{MOND_formulation}).

\subsection{The MOND interpolating function}
\label{Interpolating_function}

There are already tight constraints on the MOND interpolating function, in particular from rotation curves of our Galaxy and others. Although the standard $\nu$ function \citep[][section 6]{Kent_1987} was the first to be actively studied, there is much evidence to argue against its rather sharp transition between the Newtonian and modified regimes. For example, 21 cm neutral hydrogen observations of nearby galaxy rotation curves clearly prefer the simple $\nu$ function used in this contribution \citep{Gentile_2011}. It also fits the Galactic rotation curve well while the standard function does not, for a wide range of assumptions about the MW baryonic mass distribution \citep{Iocco_Bertone_2015}.

In recent years, astronomers have exploited reduced variability in stellar mass to light ratios at near-infrared wavelengths \citep{Bell_de_Jong_2001, Norris_2016} by using Spitzer data. In particular, the Spitzer Photometry and Accurate Rotation Curve dataset \citep{SPARC} has given a much better idea of the relation between the kinematically inferred acceleration $\bm{g}$ in galaxies and the value $\bm{g}_{_N}$ predicted by Newtonian gravity based solely on the actually observed baryons. There is very little intrinsic scatter in the relation between $\bm{g}$ and $\bm{g}_{_N}$, with \citet{Li_2018} finding that it is almost certainly below 13\%.

\begin{table}
  \centering
		\begin{tabular}{llcc}
			\hline
			Interpolating & \multirow{2}{*}{$\nu_{ext}$} & \multirow{2}{*}{$\eta$ in QUMOND} & \multirow{2}{*}{$\eta$ in AQUAL}\\
			function & & \\
			\hline
			Standard & 1.1462 & 1.0726 & 1.0661 \\
			Simple & 1.5602 & 1.4228 & 1.4056 \\
			MLS & 1.5081 & 1.3692 & 1.3508 \\
			\hline
		\end{tabular}
	\caption{The expected MOND boost to gravity in the Solar neighbourhood for the standard \citep[][section 6]{Kent_1987} and simple \citep{Famaey_Binney_2005} interpolating functions as well as the one labelled `MLS', an empirical fit to high-accuracy rotation curves and photometry of 153 spiral galaxies \citep[][equation 4]{McGaugh_Lelli_2016}. The different functions require different values for $g_{_{N,ext}}$ in order to match the EF strength $g_{ext}$ implied by the position and velocity of the Local Standard of Rest \citep{McMillan_2017}. The last two columns show angular averages $\eta$ for QUMOND (Equation \ref{eta_EFE}) and AQUAL (Equation \ref{eta_EFE_AQUAL}), both based on applying Equation \ref{eta} to a system dominated by the EF \citep{Banik_2015}.}
  \label{Interpolating_function_effect}
\end{table}

\citet{McGaugh_Lelli_2016} fit this relation using an interpolating function that we call MLS (see their equation 4). Numerically, this is rather similar to the simple interpolating function, especially in the Solar neighbourhood. However, the MLS function has $\nu \to 1$ much more rapidly at high accelerations, thereby better satisfying Solar System constraints on departures from Newtonian gravity \citep{Hees_2016}. This function also fits MW kinematics rather well \citep{McGaugh_2016_MW}.

Recently, data on elliptical galaxies has started to have a bearing on the appropriate MOND interpolating function. Using data on nearly 4000 ellipticals covering accelerations of ${\left(1 - 30 \right) a_{_0}}$, \citet{Chae_2019} found that the simple function was strongly preferred over the standard one. This is particularly evident in their figure 4, which shows that the standard $\nu$ function significantly under-predicts $g$ across the entire dataset.


In Section \ref{Results}, we showed that Equation \ref{v_tilde_max} accurately captures the maximum value of $\widetilde{v}$ that arises in more rigorous MOND simulations of WB systems. It is important to bear in mind that this requires knowledge of the unknown $g_{_{N,ext}}$ which enters the governing Equation \ref{QUMOND_equation}. However, kinematic observations of the MW constrain the Solar neighbourhood value of $g_{ext}$. These quantities are related via Equation \ref{g_N_ext}. For some MOND interpolating functions like the simple and standard ones, this can be analytically inverted to determine $g_{_{N,ext}}$ from $g_{ext}$. However, this is not possible with the MLS function. We therefore use the Newton-Raphson root-finding algorithm to invert Equation \ref{g_N_ext} and so determine the MLS value of $g_{_{N,ext}}$ in the Solar neighbourhood. This lets us determine the local values of $\nu_{ext}$ and $K_0$ in the MLS function, allowing a comparison with the other $\nu$ functions (Table \ref{Interpolating_function_effect}).

Our results show that the standard function would be hard to rule out with the WBT or other tests in the Solar neighbourhood. However, it already faces severe challenges further afield and could eventually be ruled out with further improvements, especially in the Gaia era. This shows that the WBT must work in conjunction with the more traditional and more accurate rotation curve-based tests of MOND. These will continue to provide a vital anchor for novel tests of MOND like the WBT, which are likely to prove less accurate in the short term but may be more direct.

\subsection{The MOND formulation}
\label{MOND_formulation}

So far, we have focused on a particular formulation of MOND called QUMOND \citep{QUMOND}. However, MOND can also be formulated using an aquadratic Lagrangian \citep[AQUAL,][]{Bekenstein_Milgrom_1984}. In spherical symmetry, AQUAL and QUMOND take opposite approaches to relate $g$ and $g_{_N}$.
\begin{eqnarray}
	\label{AQUAL_spherical_symmetry}
	x \mu \left( x \right) &=& y~\text{ (AQUAL) } \nonumber \\
	y \nu \left( y \right) &=& x ~\text{ (QUMOND) } \nonumber \\
	x &\equiv& \frac{g}{a_{_0}} ~\text{ and } y \equiv \frac{g_{_N}}{a_{_0}}
\end{eqnarray}

Following equation 19 of \citet{Banik_2015} and defining $\mu_{_{ext}} \equiv \mu \left( \frac{g_{_{ext}}}{a_{_0}}\right)$, the AQUAL analogue of Equation \ref{g_ratio_QUMOND} is
\begin{eqnarray}
	\label{g_ratio_AQUAL}
	g_r ~&=&~ \frac{g_{_{N,r}}}{\mu_{ext} \sqrt{1 + L_0 \sin^2 \theta}} \\
	L_0 &\equiv & \frac{\partial Ln \, \mu_{_{ext}}}{\partial Ln \, g_{_{ext}}}
\end{eqnarray}

This implies that the AQUAL boost to Newtonian gravity is maximised when $\theta = 0$ or $\pi$, similarly to QUMOND.
\begin{eqnarray}
	\frac{g_r}{g_{_{N,r}}} ~\leq~ \frac{1}{\mu_{ext}}
	\label{eta_max_EFE_AQUAL}
\end{eqnarray}

Although AQUAL and QUMOND formally use different interpolating functions $\mu \left( x \right)$ and $\nu \left( y \right)$, we can say that the two formulations use the `same' interpolating function when they have the same relation between $\bm{g}$ and $\bm{g}_{_N}$ in spherical symmetry. This arises if ${\mu \nu = 1}$ for corresponding values of $g$ and $g_{_N}$. In this case, both theories give the same result for the maximum boost to Newtonian gravity in the EF-dominated regime (compare Equations \ref{eta_max_EFE} and \ref{eta_max_EFE_AQUAL}).

Using the EF-dominated solution for AQUAL (Equation \ref{g_ratio_AQUAL}) in our definition for the angle-averaged boost to Newtonian gravity (Equation \ref{eta}), we get that\footnote{This is based on solving Equation \ref{eta} by substituting $u = \cos \theta$ to simplify the integral and then letting $u = \sqrt{\frac{1 + L_0}{L_0}} \sin \phi$.}
\begin{eqnarray}
	\eta ~=~ \frac{1}{\mu_{ext}\sqrt{L_0}} \tan^{-1}\sqrt{L_0}
	\label{eta_EFE_AQUAL}
\end{eqnarray}

To relate the AQUAL $L_0$ with the QUMOND $K_0$, we differentiate Equation \ref{AQUAL_spherical_symmetry} with respect to $x$.
\begin{eqnarray}
	x \frac{\partial \mu}{\partial x} + \mu ~&=&~ \frac{\partial y}{\partial x} \\
	 ~&=&~ \frac{1}{\nu + \underbrace{y\frac{\partial \nu}{\partial y}}_{\nu K_0}}
\end{eqnarray}

Because $L_0 \equiv \frac{x}{\mu} \frac{\partial \mu}{\partial x}$ and $K_0 \equiv \frac{y}{\nu} \frac{\partial \nu}{\partial y}$, we obtain that
\begin{eqnarray}
	\overbrace{\mu \nu}^1 \left( 1 + L_0 \right) \left( 1 + K_0 \right) ~&=&~ 1
	\label{K_0_L_0_relation}
\end{eqnarray}

If we think of $L_0$ in terms of an angle $\omega$ by setting ${L_0 \equiv \tan^2 \omega}$, then ${K_0 = -\sin^2 \omega}$. For interpolating functions that avoid very rapid transitions, $\omega$ lies between its values in the Newtonian limit (0) and the deep-MOND limit ($\frac{\pi}{4}$).

We use Equations \ref{eta_EFE_AQUAL} and \ref{K_0_L_0_relation} to obtain the results shown in the last column of Table \ref{Interpolating_function_effect}. The different MOND formulations yield results which are numerically very similar, a fact which is also evident from figure 1 of \citet{Banik_2015}. Thus, the WBT would likely work very similarly in either case. In this contribution, we use the computationally simpler QUMOND.

To understand why QUMOND and AQUAL give such similar results, we Taylor expand Equation \ref{eta_EFE_AQUAL} and use Equation \ref{K_0_L_0_relation} to write the result in terms of $K_0$.
\begin{eqnarray}
	\eta ~=~ \frac{1}{\mu_{ext}} \left( 1 + \frac{K_0}{3} - \frac{2 {K_0}^2}{15} \ldots \right)
	\label{eta_EFE_AQUAL_Taylor}
\end{eqnarray}

At leading order, this is the same as the QUMOND Equation \ref{eta_EFE}. Therefore, we expect AQUAL and QUMOND to yield very similar predictions for the WBT (within ${\approx 5\%}$) once the interpolating function is independently fixed, most likely by extragalactic observations (Section \ref{Interpolating_function}).

\subsection{The wide binary mass ratio}
\label{Mass_ratio_effect}

Our analysis has focused exclusively on the situation where one of the stars in a WB dominates the mass of the system, making the gravitational field axisymmetric. More generally, our results apply as long as the relative acceleration can be found by taking the difference between the accelerations that each star causes on a test particle placed at the location of the other star. This superposition principle applies to Newtonian gravity. A necessary condition for it to hold is that the gravity due to a point mass be linear in the mass. This is not true at low accelerations in MOND as $g \propto \sqrt{M}$ (Equation \ref{Deep_MOND_limit}). Thus, we expect that the gravity $\bm{g}$ binding a WB system depends on the fraction $q_{_1}$ of its total mass $M$ in its least massive star.

Neglecting the EF and assuming ${g \ll a_{_0}}$, the two-body force in QUMOND has previously been derived analytically \citep[][equation 53]{QUMOND}. The result is the same in AQUAL \citep{Zhao_2010}. The force is weakened by at most 21.9\% if $q_{_1} = \frac{1}{2}$. The effect arises essentially because the gravity from one star creates an `external' field on the other, thereby reducing the phantom dark matter density in its vicinity. In the isolated deep-MOND limit, this is quite a significant effect because there is no EF otherwise. Without any EF, the phantom dark matter halo would extend for ever with a divergent enclosed mass. \footnote{This still leads to a finite force on star $A$ due to star $B$ because, roughly speaking, we only need to consider phantom dark matter around star $B$ that is closer to it than star $A$.}

For the WBT, we expect a considerably smaller effect given that the Solar neighbourhood is never very far into the deep-MOND regime due to the Galactic EF (Table \ref{Interpolating_function_effect}). To estimate the effect of a non-zero $q_{_1}$, suppose we have a WB system whose stars are equally massive and have a relative $g_{_N}$ of $\frac{1}{2}g_{_{N,ext}}$. Due to the EF, the phantom dark matter halo around each star is effectively truncated at a radius smaller than the separation $r$ between the stars.

In this case, each star exerts a Newtonian gravity of $\frac{1}{4}g_{_{N,ext}}$ on the other star. If the stars were aligned with $\widehat{\bm{g}}_{ext}$, then the total EF on each star would be $\left(1 \pm \frac{1}{4} \right) g_{_{N,ext}}$ depending on which star was on the side facing the Galactic Centre. Assuming that $\nu$ is linear in $g_{_{N}}$, we see that the reduction in phantom dark matter around one star would be compensated by an increase around the other star. This is not true once we allow $\nu$ to be non-linear in $g_{_{N}}$, suggesting that a non-zero value of $q_{_1}$ only affects $\bm{g}$ at second order.

A much more common geometrical configuration is one in which the stars are aligned orthogonally to $\widehat{\bm{g}}_{ext}$. In this case, the Newtonian gravity of one star on the other adds in quadrature to the EF. As a result, the total EF on each star is increased by $\approx \frac{1}{2\times4^2} = \frac{1}{32}$. Given that $K_0 \approx -\frac{1}{4}$ in the Solar neighbourhood, the phantom dark matter around each star would be reduced by $\approx \frac{1}{128}$. This rather small figure arises because the two sources of the EF add in quadrature, causing $q_{_1}$ to only become relevant at second order.

For a more widely separated WB, the mass ratio has an even smaller effect. Doubling the separation $r$ reduces $g_{_N}$ by a factor of 4, thereby reducing ${16\times}$ the effect of each star on the total EF felt by the other star and thus on the mass of its phantom dark matter halo. In the limit that the accelerations are everywhere dominated by the EF, the superposition principle applies once again so that $q_{_1}$ has no effect \citep[][equation 24]{Banik_2015}.

\begin{figure}
	\centering
		\includegraphics[width = 8.5cm] {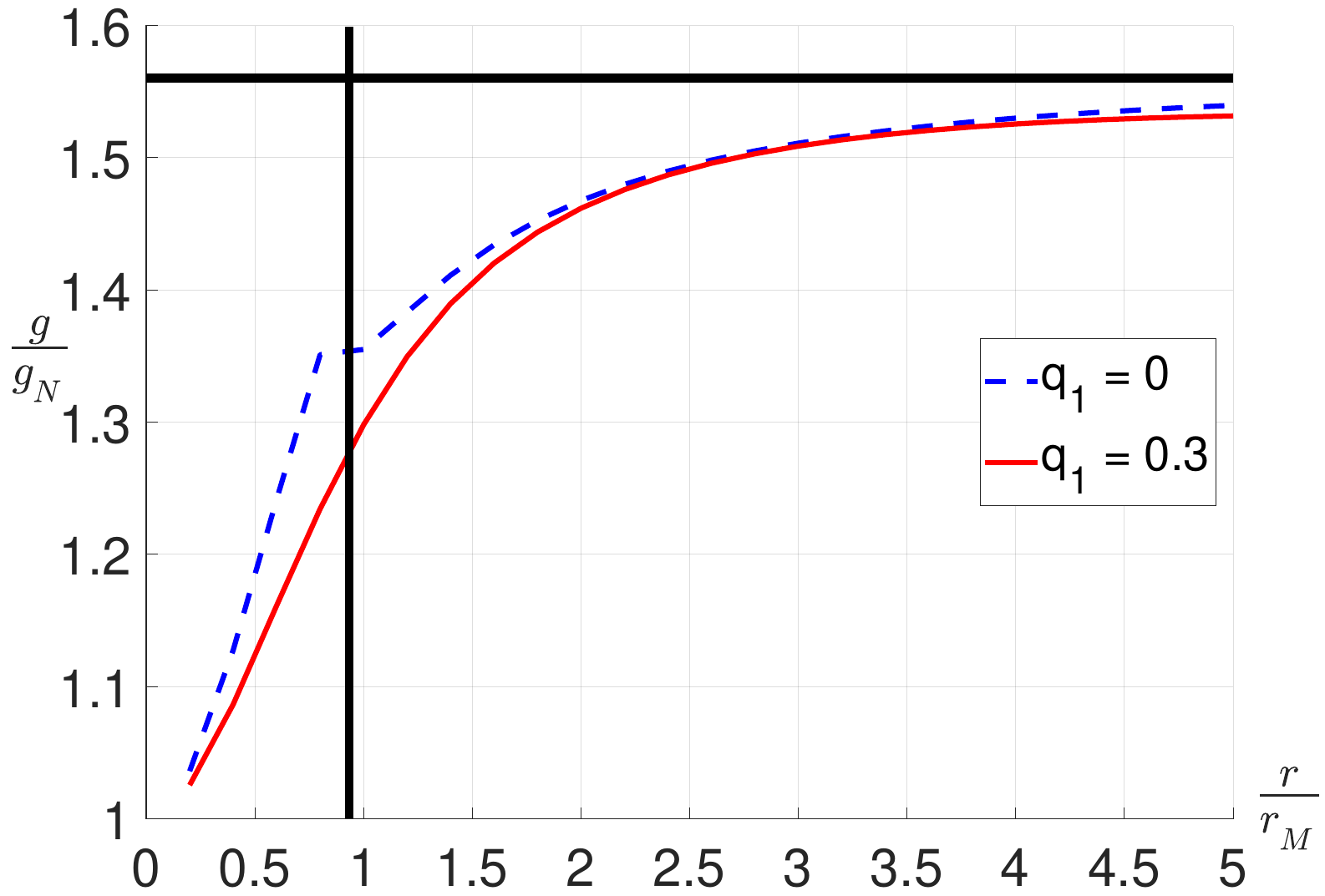}
		\caption{The boost to Newtonian gravity for WB systems aligned with $\widehat{\bm{g}}_{ext}$ as a function of their separation $r$ relative to their MOND radius $r_{_M}$ (Equation \ref{Deep_MOND_limit}). We show an average over the two possible orientations of the WB for the cases $q_{_1} = 0$ (dashed blue) and $q_{_1} = 0.3$ (solid red), assuming the simple MOND interpolating function. The black vertical line shows the separation beyond which the Galactic EF dominates. For much wider systems, the boost to Newtonian gravity should asymptotically reach $\nu_{ext}$ (black horizontal line).}
	\label{Mass_ratio_effect_figure}
\end{figure}

One could argue that $\bm{g}$ does depend on $q_{_1}$ if the separation is smaller, making the EF sub-dominant. However, given that the local Galactic EF is itself stronger than $a_{_0}$, this is only possible if the internal acceleration is above $a_{_0}$. In the Newtonian regime, we know that $q_{_1}$ does not affect the internal gravity in a WB. Thus, in the Solar neighbourhood, there is no WB in a regime where the total acceleration is everywhere below $a_{_0}$ and arises mostly internally (rather than due to the Galactic EF).

To check our argument that $q_{_1}$ does not much affect $\bm{g}$, we use our ring library procedure to find the mutual acceleration between two comparable masses if they are embedded in an aligned EF. Specifically, we assume $q_{_1} = 0.3$ and that $\widehat{\bm{r}} = \pm \widehat{\bm{g}}_{ext}$, taking advantage of an algorithm used in \citet{Banik_Ryan_2018}. Because WB systems rotate, we average $g$ in these two orientations. The factor by which this averaged $g$ exceeds the Newtonian result $g_{_N}$ is shown in Figure \ref{Mass_ratio_effect_figure}, where we also show the result for our previously discussed force library in which $q_{_1} = 0$ (Section \ref{Governing_equations}). We get rather similar relative accelerations for the different values of $q_{_1}$. In both models, the barycentre of the masses accelerates by a very small amount, consistent with numerical uncertainties.

These results are valid only for an EF aligned with the WB. Our preceding discussion suggests that the effect of a finite $q_{_1}$ would also be very small in the more common scenario where the WB separation is orthogonal to the EF. Consequently, the WBT should not be much affected by details regarding how the two-body force is weakened if the stars in a WB have comparable masses. The resulting uncertainties are almost certainly smaller than those arising from other effects, especially the astrophysical systematics discussed in Section \ref{Systematics}.


\subsection{MOND without an external field effect}
\label{MOND_no_EFE}

The EFE is a somewhat controversial aspect of MOND that arises because the theory depends non-linearly on the total acceleration \citep[][Section 2g]{Milgrom_1986}. It states that the internal dynamics of a system are affected by the gravitational acceleration of its centre of mass, even in the absence of tidal effects. The EFE may be understood by considering a dwarf galaxy with low internal accelerations ($\ll a_{_0}$) freely falling in the strong acceleration ($\gg a_{_0}$) environment of a distant massive galaxy such that there are no tidal effects. The overall acceleration at any point in the dwarf is rather high due to the dominant EF of the massive galaxy. Thus, the dwarf would obey Newtonian dynamics and forces in its vicinity would follow the usual inverse square law rather than Equation \ref{Deep_MOND_limit}. However, without the massive galaxy, the internal dynamics of the dwarf would be very non-Newtonian. Mathematically, the EFE arises because the Newtonian gravity $\bm{g}_{_N}$ entering Equation \ref{QUMOND_equation} consists of contributions from the rest of the dwarf \emph{and} from its massive neighbour, regardless of whether that raises any tides across the dwarf.

Using the principle of continuity, the rotation curve of a galaxy must be slightly affected even if the EF on it is much weaker than its internal gravity. Applying this idea, \citet{Haghi_2016} analyzed whether MOND achieves a better fit to the rotation curves of a sample of 18 disk galaxies once the EFE is considered. Their work relied on a plausible analytic estimate of how the EFE would weaken the internal gravity of these galaxies. In most of the cases considered, non-zero values of the EF were preferred due to the outer parts of the rotation curves declining faster than expected for isolated galaxies. Moreover, the preferred EF strengths were roughly consistent with the expected gravity from other known galaxies in the vicinity of the 18 they considered (see their figure 7).

Perhaps the clearest demonstration of the EFE is in the velocity dispersion of the MW satellite Crater 2, which was predicted to be only ${2.1^{+0.6}_{-0.3}}$ km/s in MOND \citep[][Section 3]{McGaugh_2016_Crater}. The rather low value is partly due to the EFE of the nearby MW, without which the prediction would have been ${\approx 4}$ km/s. This is in tension with the observed value of ${2.7 \pm 0.3}$ km/s \citep{Caldwell_2017}. Thus, the internal dynamics of Crater 2 are not consistent with a naive application of the RAR but are consistent with a more rigorous treatment of MOND that includes the EFE.

\begin{figure}
	\centering
		\includegraphics[width = 8.5cm] {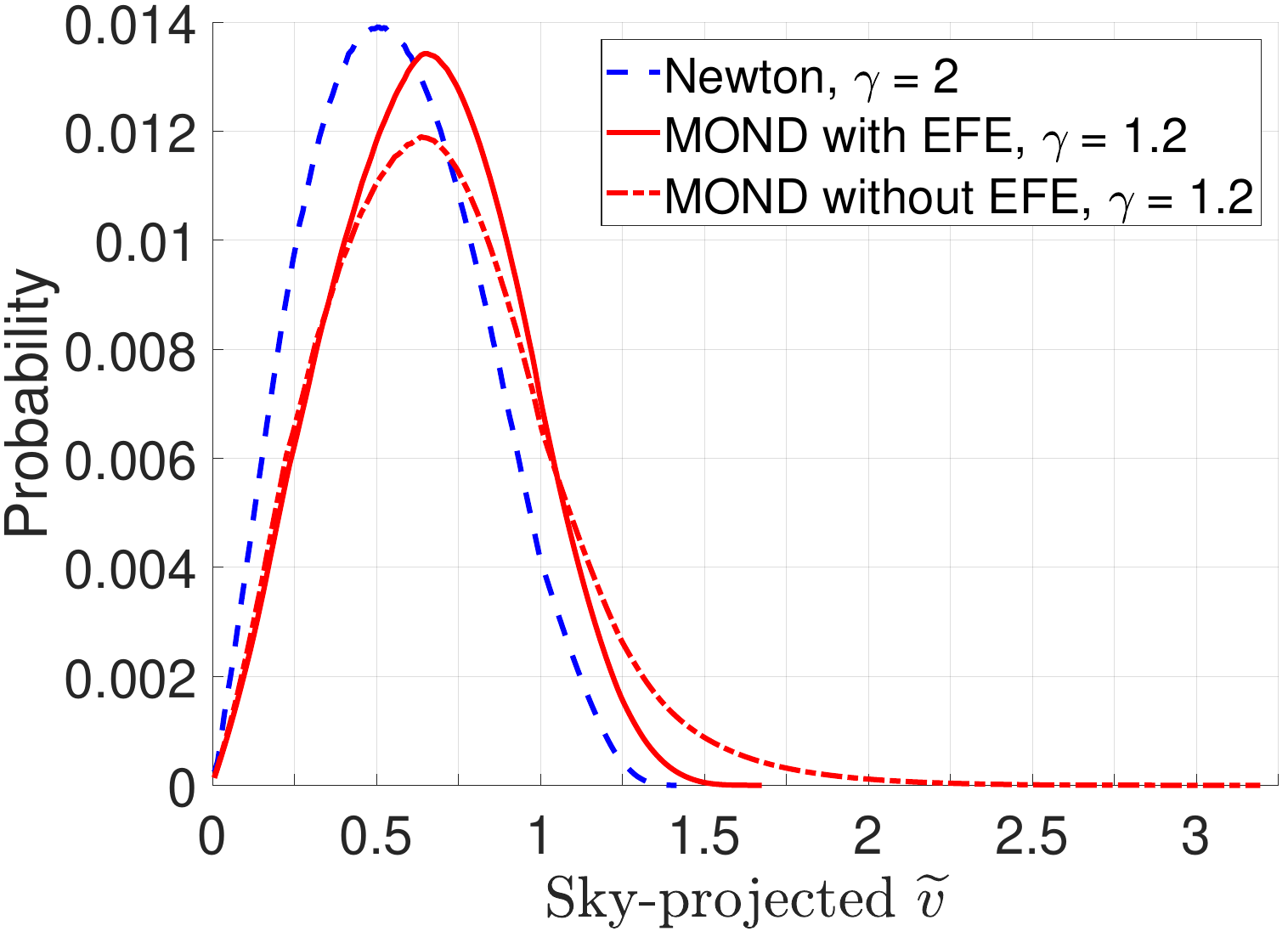}
		\caption{Distributions of the scaled WB relative velocity $\widetilde{v}$ (Equation \ref{v_tilde}) for the indicated models. The results shown here use only the sky-projected velocity. Much higher values of $\widetilde{v}$ are possible in MOND without the EFE.}
	\label{v_tilde_comparison_no_EFE}
\end{figure}

A similar case is the ultra-faint dwarf galaxy NGC1052-DF2 \citep{Van_der_Marel_2018}. The isolated MOND prediction for its internal velocity dispersion is close to 20 km/s. This is in some tension with the observed 90\% confidence level upper limit of 14.6 km/s using the `likelihood' method \citep{Van_der_Marel_2018_RNAAS}. Other methods generally give similar values \citep{Martin_2018}. However, the massive elliptical galaxy NGC 1052 is at a projected distance of only 80 kpc from this object. Including the resulting EFE reduces its MOND-predicted velocity dispersion to 13 km/s \citep{Famaey_2018} based on the most recent stellar mass estimate for NGC 1052 \citep[][table 2]{Bellstedt_2018}.

An important consequence of the EFE is that the force from a point mass eventually returns to inverse square decay with distance once the EF dominates (Equations \ref{g_ratio_QUMOND} and \ref{g_ratio_AQUAL}). This leads to a finite escape velocity from any bounded mass distribution as there is always some EF. The work of \citet{Banik_2018_escape} showed that MOND with the EFE could accurately explain the amplitude and radial gradient of the MW escape velocity curve recently measured by \citet{Williams_2017}.

Despite these successes, some formulations of MOND do not have an EFE. This is particularly true for modified inertia theories, where the different frequencies of motion due to the internal and external fields could allow them to superpose much like in Newtonian gravity \citep{Milgrom_2011}. Observationally, there is some evidence for MOND without an EFE from the fact that the velocity dispersion profiles in the outer parts of Galactic globular clusters generally reach some asymptotic value ${\sigma_\infty > 0}$ \citep{Hernandez_2013}. Interestingly, their figure 7 shows that ${\sigma_\infty \appropto M^\frac{1}{4}}$ for globular clusters with baryonic mass $M$, as would be expected from Equation \ref{Deep_MOND_limit}. However, that equation only applies to \emph{isolated} systems. Thus, the flattened dispersion profiles of globular clusters suggest that their internal dynamics are not much affected by the EF from the rest of the Galaxy, despite the EF often being quite significant compared to the internal acceleration \citep{Durazo_2017}. Although this may support versions of MOND without the EFE, flattened dispersion profiles also arise naturally in Newtonian gravity due to the effect of Galactic tides \citep{Kupper_2010, Claydon_2017}. It is unclear whether this is the correct explanation given that the radii at which the dispersion profile flattens out is often much smaller than the tidal radius of the cluster in question \citep[][figure 5]{Hernandez_2013}.

\begin{figure}
	\centering
		\includegraphics[width = 8.5cm] {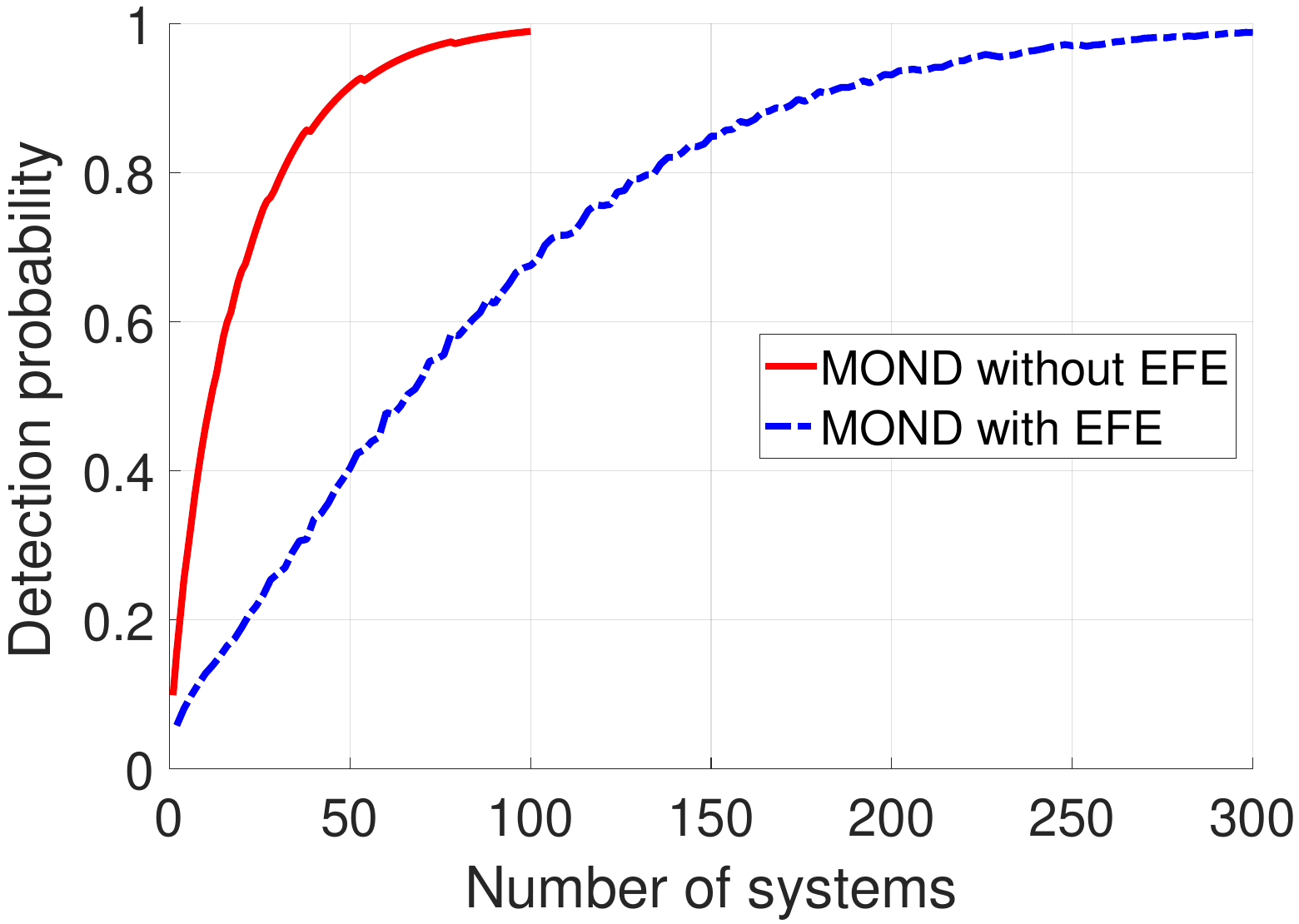}
		\caption{The detection probability of MOND with and without the EFE for different numbers of WB systems, based on the sky-projected $\widetilde{v}$ distributions shown in Figure \ref{v_tilde_comparison_no_EFE}.}
	\label{P_detection_no_EFE}
\end{figure}


To allow for the possibility that WB systems are governed by a version of MOND that lacks the EFE, we run our simulations without this effect. This makes the equation of motion much simpler as each system can be advanced using Equation \ref{g_near_field}, avoiding the need for numerical force calculations. The resulting $\widetilde{v}$ distribution is shown in Figure \ref{v_tilde_comparison_no_EFE}.

Although the lack of an EFE allows much higher orbital velocities, this only occurs for systems with large $r_p$. Such systems are much less common than systems with low $r_p$ (Equation \ref{P_r_p}). This reduces the difference in the $\widetilde{v}$ distribution compared to the case of MOND with the EFE.

Even so, the difference is quite noticeable. As a result, it is much easier to distinguish MOND from Newtonian gravity if there is no EFE (Figure \ref{P_detection_no_EFE}). Our calculations suggest that only $\sim 100$ well-observed systems would be required, even if only their sky-projected velocities were available. Thus, it should not be necessary to obtain their radial velocities. Our results in Section \ref{Results} suggest that doing so might halve the required number of systems.

Because $\widetilde{v}$ reaches much larger values without the EFE, it is possible to increase the $\widetilde{v}_{min}$ used for the WBT, reducing the fraction of systems with higher $\widetilde{v}$ in Newtonian gravity. Thus, our analysis indicates that the WBT is best done by focusing on the proportion of systems with $\widetilde{v} \geq 1.30$. The $\widetilde{v}$ distribution in MOND extends up to 3.2, though with very little probability beyond 3. Therefore, we recommend focusing on systems with $\widetilde{v} = \left( 1.30 - 3.04\right)$. This contains ${4.5 \%}$ of the WB systems in MOND if it is not weakened by the EFE. However, our Newtonian models predict that ${< 10^{-4}}$ of the systems have $\widetilde{v}$ in this range.

Our investigation suggests that the WBT without the EFE should focus on systems with $r_p > 4.5$ kAU. Although our calculations assume $r_p < 20$ kAU, it would be beneficial to increase the upper limit if observations allow. This is especially true given how MOND without the EFE deviates significantly from Newtonian gravity at even larger separations (Figure \ref{v_rms_no_EFE}).

\section{Astrophysical systematics}
\label{Systematics}

The WBT will be somewhat complicated by various astrophysical systematic effects \citep[][section 5.2]{Pittordis_2018}. In this section, we consider recently ionized WBs (Section \ref{Recently_ionized_systems}) and undetected close companions to one of the stars in a WB (Section \ref{Hierarchical_systems}). While the former turns out to not be much of an issue, the latter will require some care to properly account for.

\subsection{Recently ionized wide binaries}
\label{Recently_ionized_systems}

\subsubsection{Encounters with stars}
\label{Stellar_encounters}

WBs are only weakly bound, allowing a small perturbation to significantly affect an individual system. However, we are not concerned with a perturbation that merely alters the WB orbit as the revised orbit can just as well be used to test gravity. The major concern is if the system became unbound. Such an ionized WB would contribute to the tail of the $\widetilde{v}$ distribution, potentially skewing the WBT.

Fortunately, such a system will by definition disperse. Assuming the velocity of the system `at $\infty$' is comparable to the WB circular velocity of ${v_c \sim 0.3}$ km/s, the WB separation $r$ would rise to 63 kAU after only 1 Myr. As 3D position measurements should attain roughly this level of precision in the Gaia era (Section \ref{Distance_measurement}), only very recently ionized WBs could be problematic for the WBT. In particular, we need to consider WBs ionized within about an orbital time of the present epoch.

We now estimate what proportion $\alpha_*$ of WBs might have been ionized recently by a passing field star. If ionizing a WB requires the flyby to have an impact parameter ${\leq b}$, we get that
\begin{eqnarray}
	\alpha_* ~=~ \frac{2\mathrm{\pi} b^2 v t}{\underbrace{{d_*}^3}_{\text{Volume per star}}}
	\label{alpha_star_preliminary}
\end{eqnarray}

Here, the typical relative velocity between field stars is $v$ and their typical separation is $d_*$, causing them to have a number density of ${d_*}^{-3}$. For a WB separation of $r$ and mass $M$, the Newtonian orbital timescale
\begin{eqnarray}
	t ~=~ \sqrt{\frac{r^3}{GM}}
\end{eqnarray}

This is only slightly altered in MOND, an effect we do not consider here as we are only interested in obtaining $\alpha_*$ approximately. To do so, we need to estimate the threshold impact parameter $b$. For the impulse on one of the WB stars to be comparable to the WB orbital velocity, we require
\begin{eqnarray}
	\overbrace{\frac{GM}{bv}}^{\text{Impulse}} ~&=&~ \overbrace{\sqrt{\frac{GM}{r}}}^{\text{Newtonian }v_c} \\
	b ~&=&~ \frac{\sqrt{GMr}}{v}
	\label{b_min}
\end{eqnarray}

We estimate the impulse using the impulse approximation, which requires the perturber to move so fast that its own trajectory is unaffected by the encounter. This is valid because we expect the impulse to be ${\ssim v_c = 0.3}$ km/s, much below typical interstellar velocities which are very likely above ${15}$ km/s \citep{Figueras_2018}.\footnote{This also means there would not be much gravitational focusing of perturber trajectories.} We assume the WB has a system mass $M$ similar to that of the perturber, which passes very close to one of the stars while leaving the other star almost unaffected. This is easily justified if we consider the realistic parameters $M = M_\odot$, $r = 7$ kAU and $v = 20$ km/s. In this case, $b = 170$ AU, much below the WB separations of interest for the WBT.

Using these assumptions, we combine Equations \ref{alpha_star_preliminary} and \ref{b_min} to estimate that
\begin{eqnarray}
	\alpha_* ~=~ 2\mathrm{\pi} \left( \frac{r}{d_*} \right)^3 \frac{v_c}{v}
	\label{alpha_star}
\end{eqnarray}

For the parameters in Table \ref{Wide_binary_parameters_rough}, this gives $\alpha_* \approx 1.2 \times 10^{-4}$. If we allow an impulse $5\times$ weaker to ionize the WB, this allows $b$ to be $5 \times$ larger, thus raising $\alpha_*$ by a factor of $5^2$. Doubling the time required for the ionized WB to become sufficiently dispersed has a proportionate effect on $\alpha_*$. A lower estimate for $v$ also raises $\alpha_*$. However, the local stellar velocity dispersion in each direction exceeds 15 km/s \citep{Figueras_2018}, making it likely that the 3D encounter velocity $v$ is faster. Even with the assumption that $v$ is only 15 km/s, it is clear that $\alpha_* < 0.01$ at a very high degree of confidence.

Our calculations assume a stellar number density close to 1/pc$^3$ and a typical perturber mass of $M_\odot$, yielding a local stellar mass density of $1 \, M_\odot$/pc$^3$. In reality, the observed value is only ${\left(0.040 \pm 0.002\right)} \, M_\odot$/pc$^3$ \citep{Bovy_2017}, reducing the impact of stellar encounters on the WBT.

If the typical mass of each star is increased by some factor $k$, then their number density must decrease by this factor to maintain a fixed mass density. However, larger mass perturbers can achieve the same impulse despite a more distant encounter. In particular, Equation \ref{b_min} implies that $b \propto k$, so the `collision cross-section' rises as $k^2$. As a result, $\alpha_*$ would be increased by a factor of $k$. Given that most stars are less massive than the Sun \citep[e.g.][equation 2]{Kroupa_2001}, this strongly suggests that only a very small proportion of WBs were ionized so recently that the system has not yet dispersed. Consequently, we do not expect such systems to seriously hamper the WBT.

\begin{table}
  \centering
		\begin{tabular}{lll}
			\hline
			Parameter & Meaning & Value\\
			\hline
			$M$ & Wide binary system mass & $M_\odot$ \\
			$r$ & Wide binary orbital separation & 20 kAU \\
			$v_c$ & Wide binary orbital velocity & 0.3 km/s \\
			$d_*$ & Typical interstellar separation & 200 kAU \\
			$v$ & Typical interstellar velocity & 15 km/s \\
			$f_{_{MC}}$ & Molecular cloud mass fraction & 0.1 \\
			\hline
		\end{tabular}
	\caption{Parameters used to estimate how much the proposed wide binary test of gravity might be affected by recently ionized wide binaries. Our estimated $v_c$ exceeds the Newtonian value to account for possible MOND effects. We use a low estimate for $v$ because this allows more distant encounters to ionize a WB, thereby increasing our estimated $\alpha_*$ (Equation \ref{alpha_star}).}
  \label{Wide_binary_parameters_rough}
\end{table}

\subsubsection{Encounters with molecular clouds}

As well as stars, WBs can also be perturbed by molecular clouds (MCs). In this case, the higher perturber mass $M_p$ implies a much lower number density. Given also that MCs have a finite size, WB-MC encounters are not in the regime where only one of the WB stars is significantly affected. Instead, the encounter would be sufficiently distant that both stars in the WB would be almost equally affected. However, even a small difference could unbind the WB. In this regime, the impulse on the WB as a whole is ${\ssim \frac{GM_p}{bv}}$, making the tidal effect on the WB internal dynamics ${\ssim \frac{GM_pr}{b^2v}}$. Requiring this to be ${\ssim v_c}$ and assuming this is roughly given by the Newtonian value $\sqrt{\frac{GM}{r}}$, we get that
\begin{eqnarray}
	b ~=~ r \sqrt{\frac{M_p}{M} \frac{v_c}{v}}
	\label{Threshold_b_MC}
\end{eqnarray}

If we take $v_c = 0.01 v$ and assume that $M_p$ is at least ${1000 \, M_\odot}$, it is clear that $b \gg r$, justifying our distant tide approximation. Proceeding in a similar manner to Section \ref{Stellar_encounters} and using $d_*$ for the typical interstellar separation, we get that
\begin{eqnarray}
	\alpha_{_{MC}} ~=~ 2\mathrm{\pi} \left( \frac{r}{d_*} \right)^3 f_{_{MC}}
\end{eqnarray}

Fortunately, the uncertain mass of the perturbing MC cancels in the equations because $b^2 \propto M_p$ (Equation \ref{Threshold_b_MC}). Even so, we still need to know the fractional contribution $f_{_{MC}}$ of MCs to the mass distribution in the Solar neighbourhood. Assuming ${f_{_{MC}} = 0.1}$ and that $\frac{r}{d_*}$ has a similar value, we get that $\alpha_{_{MC}} \approx 10^{-4}$. Allowing encounters that cause a relative impulse $5 \times$ weaker raises $\alpha_{_{MC}}$ by the same factor. Thus, it is very likely that $\alpha_{_{MC}} < 0.01$, similar to our result for $\alpha_*$ (Section \ref{Stellar_encounters}).

MCs can also be detected, allowing regions near them to be avoided for the WBT. In the long run, this can be achieved simply by avoiding regions too close to the Galactic disk. However, the molecular gas has a scale height of ${\approx 100}$ pc \citep{Nakanishi_2006}, so this will only be possible when WBs can be observed accurately out to ${\approx 500}$ pc. This is too far in the Gaia era, so it is currently most sensible to simply avoid WBs within a few pc of known MCs.

The effect of MCs on the WBT is further reduced by the fact that these tend to be concentrated in spiral arms, where the density of gas is higher. The inter-arm location of the Sun \citep[][figure 2]{Vallee_2014} suggests that their number density should be low in the region relevant for the WBT. The low frequency of MCs in the Solar neighbourhood is also hinted at by the 140 pc distance to the Taurus MC, our nearest large MC \citep{Gudel_2007}. To find out if a WB might have been ionized by this MC, one could check the 3D positions and velocities of both systems. Integrating their trajectories backwards for a few Myr should reveal how likely it is that the WB approached the MC closely enough to have been ionized by it.

This only needs to be done for WBs sufficiently close to the Taurus MC because ionized WBs further away had more time to disperse. Assuming that $\frac{v_c}{v} = 0.01$, any recently ionized WB would have separated by $> 100$ kAU if its constituent stars now lie ${> 50}$ pc from the location where the WB was ionized. Even a 50 pc sphere around the Taurus MC represents only $\left( \frac{50}{150} \right)^3 = \frac{1}{27}$ of the survey volume used for the WBT, assuming this extends out to a heliocentric distance of 150 pc. Thus, we expect that only a very small fraction of WB systems might need to be rejected from the WBT due to possible recent ionization by a nearby MC.


\subsection{Hierarchical systems}
\label{Hierarchical_systems}

A significant fraction of stars have a binary companion \citep[e.g.][]{Reid_2006}, making it likely that at least some WB systems will contain a third bound star. This can be detected if it is sufficiently massive, though one must be careful not to reject a genuine WB due to e.g. a background red galaxy with insufficiently precise astrometry.

One way for a massive object to avoid detection is for it to be very dark e.g. a black hole. However, stellar mass black holes have a fairly high minimum mass of ${\approx 4 \, M_\odot}$, presumably due to how they form \citep{Farr_2011}. Such a massive object would create a large change in $\widetilde{v}$ that would not easily get mistaken for a MOND effect. Moreover, a WB system would likely be unstable if it contained three massive objects at comparable separations.

Of concern are the more subtle effects of stars with masses ${\la 0.1 \, M_\odot}$ particularly close to one of the stars in a WB system. Such dwarf stars could conceivably broaden the $\widetilde{v}$ distribution. Though this is also the signature of MOND, it can only get ${\widetilde{v} \la 1.7}$. Thus, no system should be discovered with a much larger value if $\widetilde{v}$ arises from WB orbital motion.\footnote{\citet{Pittordis_2018} consider various other modified gravity theories and how MOND would work without the EFE. Their figure 14 shows that $\widetilde{v}$ could not exceed 3 under a very wide range of assumptions. We reach a similar conclusion for MOND without the EFE (Section \ref{MOND_no_EFE}). This is a more conservative upper limit on $\widetilde{v}$ in uncontaminated systems.} However, a third object in the system can very easily cause $\widetilde{v}$ to exceed 1.7 as this only requires orbital motions of ${\ga 0.5}$ km/s. If such systems are found, then it is extremely likely that they have been `contaminated' in some way. This contamination could be modelled and the model extrapolated to $\widetilde{v} < 1.7$, giving an idea of how many systems in the critical $\widetilde{v}$ range $\left(0.9 - 1.7 \right)$ would be expected due to contamination alone. A robust claim of modified gravity must involve the actual number of such systems being larger in a statistically significant way.

Inferring contamination rates like this would be somewhat model-dependent, so we investigate if it may be feasible to obtain more conclusive evidence of a third low-mass star. For this purpose, we consider a WB system with separation $r$ and total mass $M$, of which a fraction $\alpha$ is in the contaminated star. The dynamics of this star are perturbed by a companion with a low mass ${\beta \alpha M \ll M}$ some distance $d$ away. We will see that $d \ll r_{_M}$, making the close binary orbit nearly Newtonian with circular velocity
\begin{eqnarray}
	v_{_{close}} ~=~ \sqrt{\frac{GM\alpha}{d}}
\end{eqnarray}

Of this velocity, only a fraction ${\beta \ll 1}$ arises due to motion of the contaminated star, with the rest due to motion of its undetected companion. For the velocity of the contaminated star to be significantly perturbed, we therefore require $\beta v_{_{close}}$ to exceed the WB orbital velocity, which we can approximate using Newtonian gravity (Equation \ref{v_tilde}). This occurs if $d$ is sufficiently small.
\begin{eqnarray}
	\frac{d}{r} ~\leq~ \beta^2 \alpha
	\label{Required_distance_ratio}
\end{eqnarray}

In the plausible scenario of a ${0.1 \, M_\odot}$ star contaminating one of the two Sun-like stars in a WB, we get ${\beta = 0.1}$ and ${\alpha = 0.5}$, allowing us to safely assume that $d \ll r$. As a result, the orbital period of the close binary is much shorter than for the WB. Both orbital motions of the contaminated star have a similar velocity, so its acceleration $g_{_{close}}$ for the close binary orbit must be much larger than $g_{_{wide}}$ for the lower frequency WB orbit. The ratio of these accelerations is
\begin{eqnarray}
	\frac{g_{_{close}}}{g_{_{wide}}} ~&=&~ \frac{GM \beta \alpha}{d^2} \div \frac{GM\left(1 - \alpha \right)}{r^2} \\
	&\geq&~ \frac{1}{\beta^3 \alpha \left(1 - \alpha \right)}
	\label{g_close_wide_ratio}
\end{eqnarray}

Using ${\alpha = 0.5}$ and ${\beta = 0.1}$ as before, this ratio is 4000 because the close binary orbital motion is only problematic for the WBT when $\frac{d}{r} \leq 0.005$ (Equation \ref{Required_distance_ratio}). An even higher $\frac{g_{_{close}}}{g_{_{wide}}}$ would be obtained if $\alpha \neq \frac{1}{2}$. As the WBT requires $g_{_{wide}} \sim a_{_0}$, we conservatively assume $g_{_{close}} = 1000 \, a_{_0}$. Even this is not large enough for the close binary to complete an orbit within a typical observing program $-$ assuming $r = 10$ kAU gives $d = 50$ AU, leading to an orbital period of 350 years. Consequently, $g_{_{close}}$ would hardly change during the course of observations spanning a decade or less, making the contaminated star's orbital motion a simple parabola.

The secular acceleration in radial velocity would only be 3.8 m/s each year if the full $1000 \, a_{_0}$ happened to be along the line of sight. The latest high-accuracy radial velocity planet searches do reach roughly this level of precision \citep[e.g.][]{Trifonov_2018}. Even so, this entails a lot of detailed follow-up observations, partly because the radial velocity only changes linearly with time.

A better method might be to look at the sky position, which would change quadratically with time. The necessary high-precision astrometric observations could be provided by Gaia itself. As an example, if the initial position and proper motion of the contaminated star were known exactly and it lies 100 pc away, an acceleration within the sky plane of $1000 \, a_{_0}$ over 5 years would cause it to appear $100 \, \mu$as off from where it would be if it was unaccelerated. In reality, the initial conditions would not be known exactly, so astronomers would try to fit the positions with a linear trend in accordance with Occam's Razor. This would reduce the maximum deviation from the unaccelerated model by a factor of 6, but ultimately a parabola can't be fit by a straight line. Given that DR2 of Gaia shows that it can already measure the positions of some 15\textsuperscript{th} magnitude stars to within $20 \, \mu$as, it seems likely that observers could remove a significant fraction of stars contaminated by close binary companions.

$g_{_{close}}$ would fall below the Gaia detection threshold if $d$ were much larger. Although this would reduce the resulting reflex velocity of the contaminated star, the undetected star would still affect the total mass of the system. For the typical parameters just discussed, this would be a rather small effect as the system mass would be ${2.1 \, M_\odot}$, 5\% more than expected based on the two visible Solar mass stars. This raises the Newtonian expectation for the WB orbital velocity by 2.5\%, reducing $\widetilde{v}$ by essentially the same proportion (Equation \ref{v_tilde}). The MOND effect is very likely much larger $-$ our calculations show that $\widetilde{v}$ can exceed the Newtonian limit of $\sqrt{2}$ by ${\approx 20\%}$ (Section \ref{Results}) for plausible assumptions about the MOND interpolating function (Section \ref{Interpolating_function}).

A WB system could also contain planetary mass objects. If $\beta$ was reduced ten-fold to 0.01, then the perturber would be a ${\approx 10}$ Jupiter mass object. In this case, Equation \ref{Required_distance_ratio} shows that the orbital separation must be only 0.5 AU for the perturber to significantly affect $\widetilde{v}$. At this distance, the orbital period would be short enough that the radial velocity would change significantly over a few months. Detecting this would not require us to know the absolute radial velocity of the star, thereby avoiding uncertainty from the convective blueshift correction. Any time variations in this correction would still be relevant, but these are expected to be rather small \citep[e.g.][]{Kurster_2003}. The WBs of interest for the WBT have orbital velocities of ${\sim 0.2}$ km/s (Figure \ref{v_rms_no_EFE}), much larger than the stellar reflex motion used to discover one of the first known exoplanets, 51 Pegasi b \citep{Mayor_1995}. As technology has improved greatly in the subsequent generation, it should be straightforward to discover systems in such a configuration. Rather than removing them from the sample, it may be preferable to average over several orbital periods to deduce the velocity of the star-planet system's barycentre.

Secular accelerations decrease for companions at larger $d$. At these distances, the companion must have a higher mass to have the same effect on $\widetilde{v}$ (Equation \ref{Required_distance_ratio}). Thus, some initially undetected companions might be easier to discover directly using deeper exposures of the system. Even objects as faint as white dwarfs are routinely discovered nowadays, with the discovery rate accelerating in the Gaia era \citep{Kilic_2018}. Of particular relevance is the recent discovery of a 5 Gyr old ${0.034 \, M_\odot}$ object $0.535 \arcsec$ from a ${1.05 \, M_\odot}$ star that lies 42 pc from the Sun \citep{Cheetham_2018}. Their magnitude 18.5 detection in the near-infrared $J$-band suggests that, once some follow-up observations are taken of high-$\widetilde{v}$ systems, only companions even fainter than this could evade direct detection. Such an object would generally be less massive, reducing its effect on $\widetilde{v}$ unless it is much closer in (${\la 50}$ AU for a ${0.1 \, M_\odot}$ companion). This would cause a larger orbital acceleration (lower $\beta$ in Equation \ref{g_close_wide_ratio}), likely allowing for indirect detection in a manner analogous to exoplanets.

Our estimates suggest that a companion marginally too faint for direct detection yet close enough to affect the WBT could be detected via secular astrometric acceleration with ${\approx 5}$ years of Gaia observations.\footnote{We assume a future Gaia data release will provide sky positions at different epochs rather than just quantities derived from them. Alternatively, the data release could include the fitted secular acceleration, which would be useful for studies of distant companions more generally.} This may be preferable to searches for radial velocity trends as the latter only measures motion along one dimension and requires additional follow-up. Therefore, the combination of direct detection and secular acceleration (in proper motion or radial velocity) should be able to definitively confirm or rule out the undetected companion hypothesis for any high-$\widetilde{v}$ systems that may be discovered.

\subsubsection{The wide binary orthogonality test}

\begin{figure}
	\centering
		\includegraphics[width = 8.5cm] {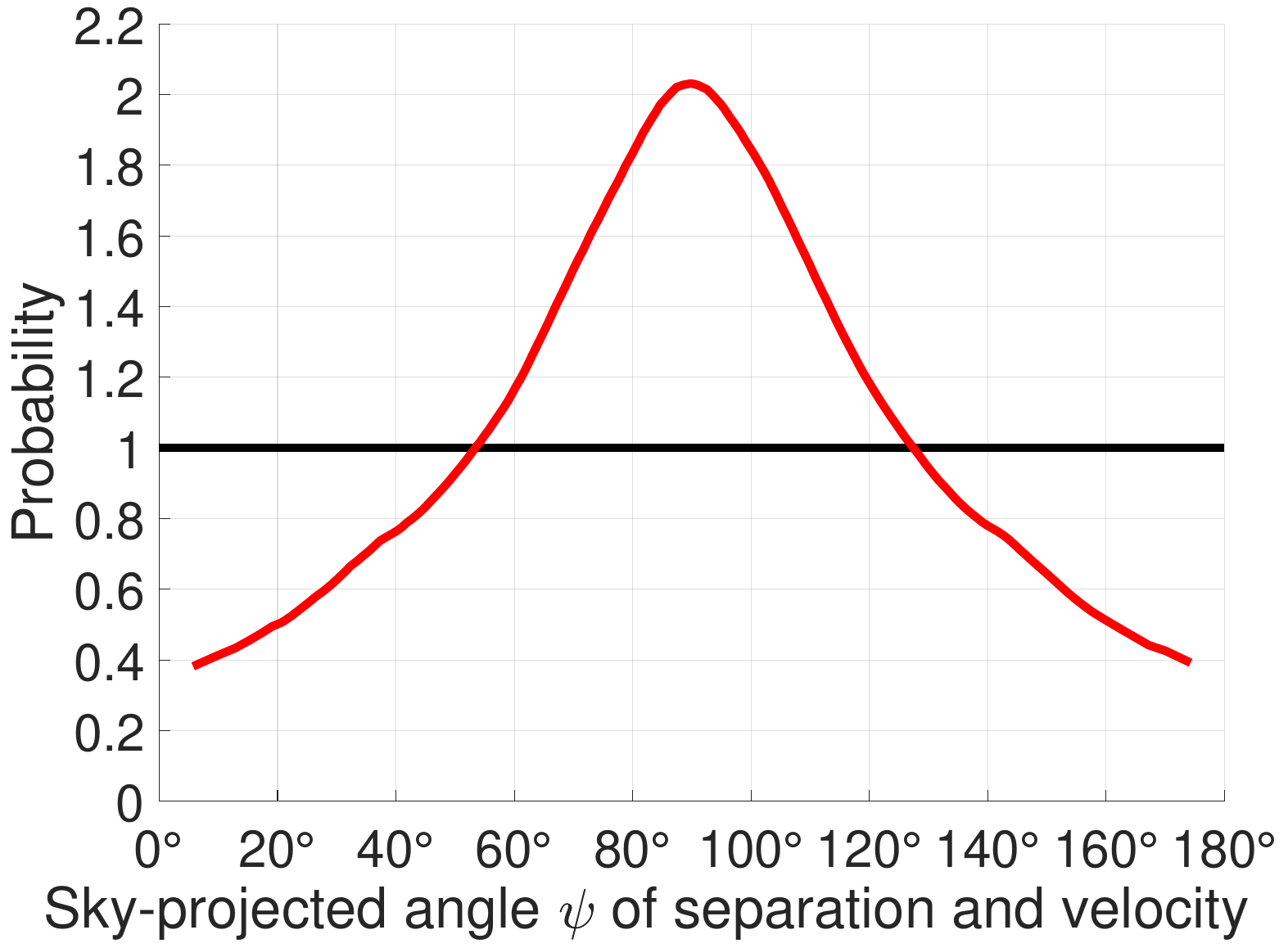}
		\caption{Distribution of the angle $\psi$ between the sky-projected separation and relative velocity of WB systems governed by MOND (red curve). Our result is based on systems with $\widetilde{v} > 0.97$ and $r_p = \left(3 - 20 \right)$ kAU, the optimal range for the WBT (Section \ref{Results}). Different orbital eccentricities are weighted according to ${\gamma = 1.2}$ (Equation \ref{P_e}). If the relative velocity does not arise from the WB orbit but due to e.g. an undetected close companion, we expect $\psi$ to have a uniform distribution (black line).}
	\label{Psi_distribution_MOND_20_kAU}
\end{figure}

If many high-$\widetilde{v}$ systems are found but these systems are contaminated by an additional low-mass object (to evade direct detection), then this object would have to be quite close to one of the stars forming the WB. Thus, the extra object could be located in any direction with respect to the star whose velocity it is contaminating. This would lead to an isotropic relative velocity between WB stars with high $\widetilde{v}$ as $\widetilde{v}$ would not be measuring the WB orbital motion.

However, this is not true if the high measured $\widetilde{v}$ is a genuine consequence of the WB orbit. The systems with the highest $\widetilde{v}$ would be those observed close to pericentre, with an orbital plane aligned closely with the sky plane. Consequently, we expect these systems to have ${\psi \approx \frac{\pi}{2}}$, where $\psi \equiv \cos^{-1} \left( \widehat{\bm{r}} \cdot \widehat{\bm{v}} \right)$ is the angle between the sky-projected separation and relative velocity of the stars in a WB.\footnote{${\psi \approx \frac{\pi}{2}}$ near apocentre as well, when $\widetilde{v}$ would be very low.} This is unexpected if the high-$\widetilde{v}$ systems arise due to contamination. Consequently, the distribution of $\psi$ could serve as an important check on any future claim of detecting modified gravity using the WBT.

To see how this might work, we determine the $\psi$ distribution using our MOND orbit library, assuming only systems with ${r_p \geq 3}$ kAU will be used for the WBT and restricting ourselves to systems with sky-projected ${\widetilde{v} > 0.97}$. This is because the important thing for the WBT is verifying the orbital nature of the relative velocity in the controversial high-$\widetilde{v}$ systems. Our results in Section \ref{Results} suggest that the optimal $\widetilde{v}$ range for the WBT is ${0.97 - 1.68}$ in the more conservative scenario where only sky-projected information is available. After applying these restrictions on $r_p$ and $\widetilde{v}$, we obtain the result shown in Figure \ref{Psi_distribution_MOND_20_kAU}.

\begin{figure}
	\centering
		\includegraphics[width = 8.5cm] {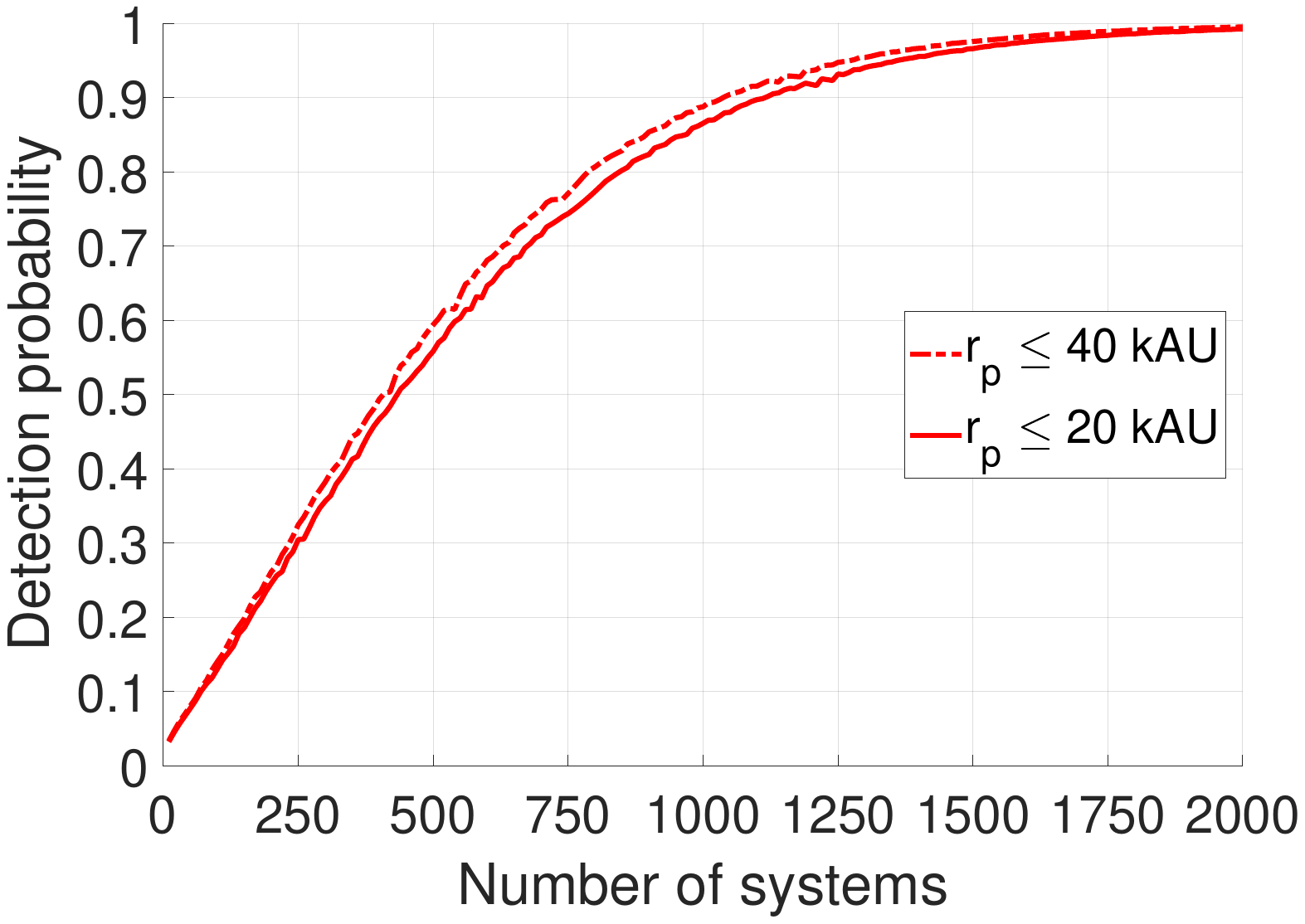}
		\caption{The probability of detecting the difference in $\psi$ distributions shown in Figure \ref{Psi_distribution_MOND_20_kAU}. Our results are shown against the number of systems with $r_p = \left(1 - 20 \right)$ kAU (solid red) or $r_p = \left(1 - 40 \right)$ kAU (dashed red). In both cases, we only use systems with $r_p > 3$ kAU in the comparison to avoid nearly Newtonian systems. Importantly, we also restrict to systems with $\widetilde{v} > 0.97$ based on our results in Section \ref{Results}. Because only a small fraction of WB systems have such a high $\widetilde{v}$ (even in MOND), it is rather difficult to detect the seemingly obvious anisotropy in their $\psi$ distribution. Our results show that the best way to do so is to focus on the fraction of systems with $\psi$ between ${64 \pm 2 ^\circ}$ and ${116 \pm 2 ^\circ}$.}
	\label{P_detection_psi}
\end{figure}

To quantify how many WB systems would be needed to detect this anisotropic distribution in $\psi$, we use the method described in Section \ref{Statistics}. Our results are shown in Figure \ref{P_detection_psi} for two different upper limits on $r_p$. The probability distributions we compare have been normalised to the number of systems with $r_p$ between 1 kAU and the indicated upper limit, though we assume only systems with ${r_p > 3}$ kAU are used in the comparison to avoid nearly Newtonian systems (Section \ref{Results}). Importantly, we do not restrict the range in sky-projected $\widetilde{v}$ when normalising. This captures the fact that only a rather small fraction of WB systems are expected to have ${\widetilde{v} > 0.97}$. As a result, accurate observations of ${\approx 1000}$ systems would be required to detect the anisotropy in $\psi$ \emph{for the high-velocity systems}. This rather high number arises from the fact that binning WB systems in both $\widetilde{v}$ and $\psi$ requires more systems than binning in $\widetilde{v}$ alone, which is sufficient for the WBT.

Our results in Figure \ref{P_detection_psi} assume that $\psi$ would be distributed isotropically if the relative velocity between the stars in a WB mostly arose from contamination by a close undetected companion to one of the stars. This is because we expect only a weak tidal influence from the distant star in the WB. If we suppose it is ${200 \times}$ further away, then its gravity on the close binary is $\approx 200^2 \times$ weaker than the gravity binding the close binary. However, the close binary is expected to be in the Newtonian regime (Equation \ref{g_close_wide_ratio}) such that a uniform external gravitational field merely moves it without affecting its internal dynamics. Thus, only tides raised by the distant star can be relevant.

This tide is another factor of $\frac{d}{r}$ weaker, making it $8\times 10^6$ times weaker than the gravity binding the close binary. It therefore needs to complete about that many orbits to `notice' the distant star. As the orbital period would be ${\approx 350}$ years at a 50 AU separation, this would take 2.8 Gyr. However, the WB orbit is much shorter than this. Even the Galactic orbit is much shorter. We will see later that the WB orbital pole precesses to maintain its angle with the Galactic EF (Figure \ref{Centauri_adiabatic_test}). It also precesses on shorter timescales (Figure \ref{Centauri_single_orbit}). Thus, the WB orbital motion and precession do not give the close binary enough time to significantly adjust to the presence of its distant companion.

\section{Proxima Centauri as a wide binary}
\label{Proxima_Centauri}

For the WBT to work, it is important that WB systems are fairly common. Fortunately, this seems very likely if we consider our nearest external star, Proxima Cen. This orbits the close binary $\alpha$ Cen A and B at a current distance of 13 kAU \citep{Kervella_2017}. As first pointed out by \citet{Beech_2009}, this puts the Proxima Cen orbit well within the regime where MOND would have a significant effect.

\subsection{Short-term evolution}
\label{Proxima_Cen_short_term}

Due to its proximity, even this single system could allow a direct test of MOND with the proposed Theia mission \citep{Theia_2017}. To see how this might work, we use the method described in Section \ref{Governing_equations} to find the orbital acceleration of Proxima Cen. We treat it as a test particle orbiting the much more massive $\alpha$ Cen A and B, which we consider as a single point mass of ${2.043 \, M_\odot}$ at their barycentre given that they are in a tight orbit separated by only ${\approx 18}$ AU \citep{Kervella_2016}. This is $< \frac{1}{700}$ of their distance to Proxima Cen, so the force on it should be affected by $\approx \frac{1}{700^2}$ due to the finite separation of $\alpha$ Cen A and B. This would only significantly affect the WB orbit of Proxima Cen after $\approx 5 \times 10^5$ orbits. However, its long orbital period means that it has only completed $\approx 1.3 \times 10^4$ orbits in the past 5 Gyr, the estimated age of the system \citep{Bazot_2016}.

Based on the mass-luminosity relation of \citet{Mann_2015}, Proxima Cen has a mass of only ${0.122 \, M_\odot}$, justifying our assumption that it can be treated as a test particle. To account for the fact that the MOND gravity between two bodies is weakened if their total mass is split more equally (Section \ref{Mass_ratio_effect}), we neglect the mass of Proxima Cen altogether rather than add it to the mass of $\alpha$ Cen AB. However, when we require the Galactocentric velocity of the system as a whole (Section \ref{Full_Galactic_orbit}), we take a weighted mean of the values for Proxima Cen and the $\alpha$ Cen AB barycentre, thus using the barycentre of all three stars.

\begin{table}
  \centering
		\begin{tabular}{lll}
			\hline
			Parameter & $\alpha$ Cen A+B & Proxima Cen\\ \hline
			Right ascension & $14^h39^m40.2068\arcsec$ & $14^h29^m47.7474\arcsec$ \\
			Declination & $-60^\circ 50\arcmin 13.673\arcsec$ & $-62^\circ 40\arcmin 52.868\arcsec$\\
			Parallax & 747.17 mas & 768.77 mas \\ [5pt]
			$\mu_{\alpha,*}$ & -3619.9 mas/yr & -3773.8 mas/yr \\
			$\mu_\delta$ & 693.8 mas/yr &  770.5 mas/yr\\
			Heliocentric& \multirow{2}{*}{-22.332 km/s} & \multirow{2}{*}{-22.204 km/s} \\
			 radial velocity & & \\ [5pt]
			Mass & ${2.0429 \, M_\odot}$ & ${0.1221 \, M_\odot}$ \\ \hline
		\end{tabular}
	\caption{Observed parameters of the barycentre of the $\alpha$ Cen A+B close binary \citep[][table 1]{Kervella_2016} and their companion Proxima Cen \citep[][table 2]{Kervella_2017}. $\mu_{\alpha,*}$ is the time derivative of the right ascension multiplied by the cosine of the declination, making it a true (East-West) angular velocity.}
  \label{Proxima_Cen_parameters}
\end{table}

We determine the gravitational field of $\alpha$ Cen AB using the method described in Section \ref{Governing_equations}, assuming it feels the same $\bm{g}_{ext}$ as the Sun because both are in much the same part of the Galaxy. We then integrate the binary orbit forwards, starting with the initial conditions given in Table \ref{Proxima_Cen_parameters}. We also find the Proxima Cen trajectory in Newtonian gravity. In both cases, any near-term observations would span a negligibly short fraction of the ${\approx 400}$ kyr orbital period, allowing us to assume a parabolic trajectory during the observing campaign. The acceleration would be ${0.60 \, a_{_0}}$ in Newtonian gravity but ${0.87 \, a_{_0}}$ in MOND (though the MOND force points ${3.1^\circ}$ away from the radial direction).

\begin{figure}
	\centering
		\includegraphics[width = 8.5cm] {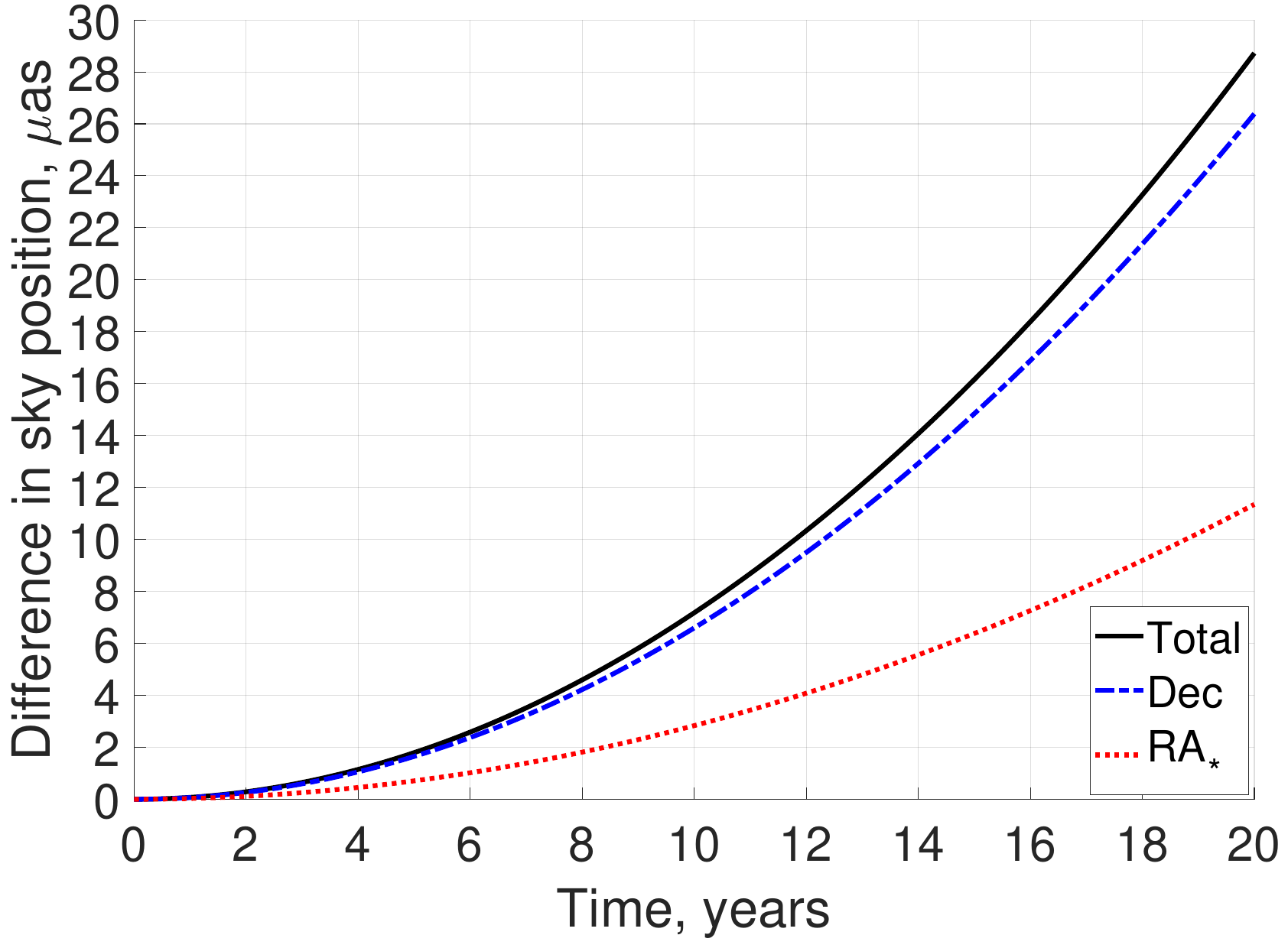}
		\caption{Difference in the sky position of Proxima Cen depending on whether Newtonian gravity or MOND governs its orbit about $\alpha$ Cen A and B. The same initial conditions are used for both trajectories (Table \ref{Proxima_Cen_parameters}). $RA_*$ is the difference in right ascension multiplied by the cosine of the declination. The total angular difference grows quadratically with time and is $7.18 \, \mu$as after a decade. Astronomers might try to fit the data by varying the initial conditions, in which case the angular differences would be ${\approx \frac{1}{6}}$ that shown here.}
	\label{Theia_test}
\end{figure}

To see if this might be detectable, we show the angular difference between the trajectories on our sky (Figure \ref{Theia_test}). By the time Theia is flown, we assume the LSR parameters would be known very accurately, allowing a reliable adjustment for the Galactic acceleration of Proxima Cen. This also has a vertical component towards the MW disk, but this is expected to be very small because Proxima Cen lies very close to the disk mid-plane \citep{Ferguson_2017}. These uncertainties could be reduced with accurate astrometric observations of both Proxima Cen and $\alpha$ Cen AB.

Unless the initial conditions were known exactly, the difference in sky position would actually be $\frac{1}{6}$ that shown because astronomers would try to fit the data using different initial conditions.\footnote{The exact ratio will depend on spacecraft performance and other factors. We assume the fit to data is designed to minimise its $\chi^2$ with respect to observations taken at regular intervals with equal accuracy. In this case, the best linear fit to the parabola $y = t^2$ over the range ${t = 0-1}$ is given by $y = t - \frac{1}{6}$.} Even so, a parabola can only be fit with a straight line for so long. Thus, if Theia is flown and achieves $\mu$as astrometric precision over a few years, it should be able to directly measure how much Proxima Cen accelerates towards $\alpha$ Cen AB. This would yield a much-needed strong yet direct constraint on gravity at low accelerations.

In principle, the radial velocity $v_r$ of Proxima Cen could also be used to distinguish these theories. However, a constant acceleration causes $v_r$ to change linearly with time, whereas the position would respond quadratically. Thus, $v_r$ would only differ by 0.5 cm/s between the models after a decade of observations. This would be very challenging to detect, making it a much less plausible test of MOND than using precise astrometry of Proxima Cen.

One possible complication with such tests is that an undetected exoplanet could also cause an extra acceleration. However, as perceived at Proxima Cen, the exoplanet is quite likely to be in a different direction than $\alpha$ Cen. If the acceleration of Proxima Cen is directed to within a few degrees of $\alpha$ Cen but is stronger than expected in Newtonian gravity, then this would be strong evidence against it. Moreover, a short period exoplanet would show up in multi-epoch observations. This would not be the case for a sufficiently long period, but in this case the greater distance implies the exoplanet must be more massive and so more likely to be detected in other ways. This is especially true given our proximity to the system enlarging the angles involved, thus making it easier to efficiently suppress the light from a star already at the faint end of the main sequence. If an anomalous acceleration was detected, then intensive observations could be taken in its direction from Proxima Cen.


\subsection{Long-term evolution and orbital stability}
\label{Secular_effects}

\begin{figure}
	\centering
		\includegraphics[width = 8.5cm] {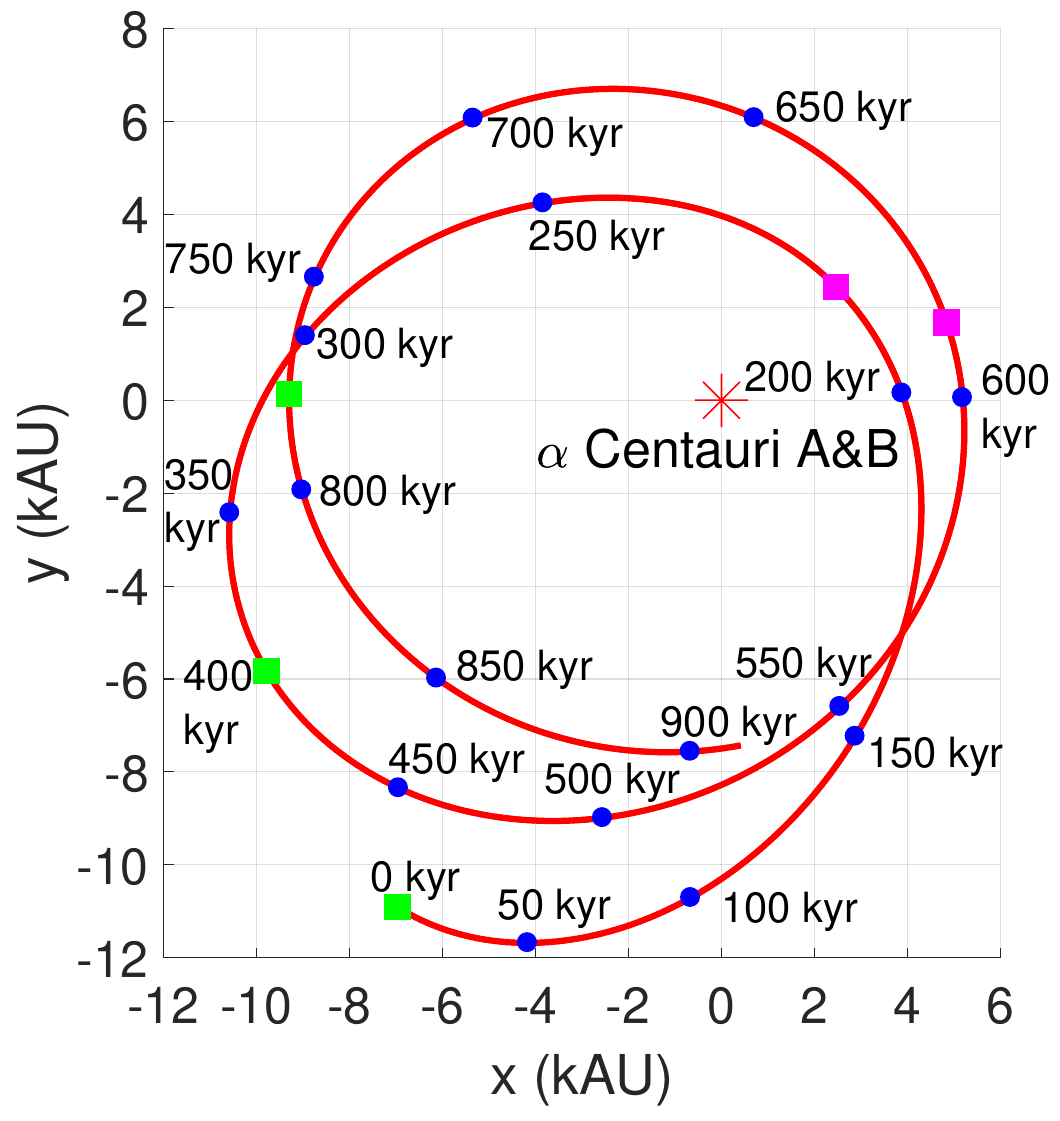}
		\caption{Part of the Proxima Cen orbit relative to $\alpha$ Cen, treating its A and B components as a single point mass at their barycentre (red * at the origin). The $x$ and $y$ axes used here are orthogonal to each other and to the initial orbital angular momentum $\bm{h}$. We show the orbit over the next 909 kyr, starting with the initial conditions in Table \ref{Proxima_Cen_parameters}. By the end of the simulation, $\widehat{\bm{h}}$ precesses ${8.0^\circ}$ from its initial orientation (not shown). We indicate the position of each pericentre (pink squares) and apocentre (green squares). The radial period is $\approx$ 210 kyr, though the azimuthal period is slightly longer because the system rotates ${< 360^\circ}$ between successive pericentres/apocentres.}
	\label{Centauri_single_orbit}
\end{figure}

In this section, we consider the evolution of the $\alpha$ Cen system over a longer period. We begin by generalising the calculations of Section \ref{Proxima_Cen_short_term} to cover a little over one full revolution, a period still short enough that the Galactic orbit of the system is expected to have a negligible effect on the EF it experiences. In Figure \ref{Centauri_single_orbit}, we show the trajectory within the plane orthogonal to the initial $\bm{h}$ of the system. The orbit is not closed, with a small amount of precession evident.

Over much longer periods, the system rotates around the Galaxy, changing the direction towards the Galactic Centre and thus $\widehat{\bm{g}}_{ext}$. The magnitude of the EF might also change, something we consider in Section \ref{Full_Galactic_orbit}. Gravity from the Galactic disk would have some effect, but this is not expected to be significant in the Solar neighbourhood (Section \ref{Galactic_disk_effect}).

To explore these issues, we integrate the WB orbit of Proxima Cen backwards for 5 Gyr, allowing $\widehat{\bm{g}}_{ext}$ to rotate at the angular rate $-\frac{v_{c, \odot}}{R_\odot}$ appropriate for the LSR. We make the approximation that $\bm{g}_{ext}$ rotates at a constant rate and does not change in magnitude, as appropriate for a purely circular Galactic orbit. A more rigorously calculated Galactic orbit is considered in Section \ref{Full_Galactic_orbit}.

\begin{figure}
	\centering
		\includegraphics[width = 8.5cm] {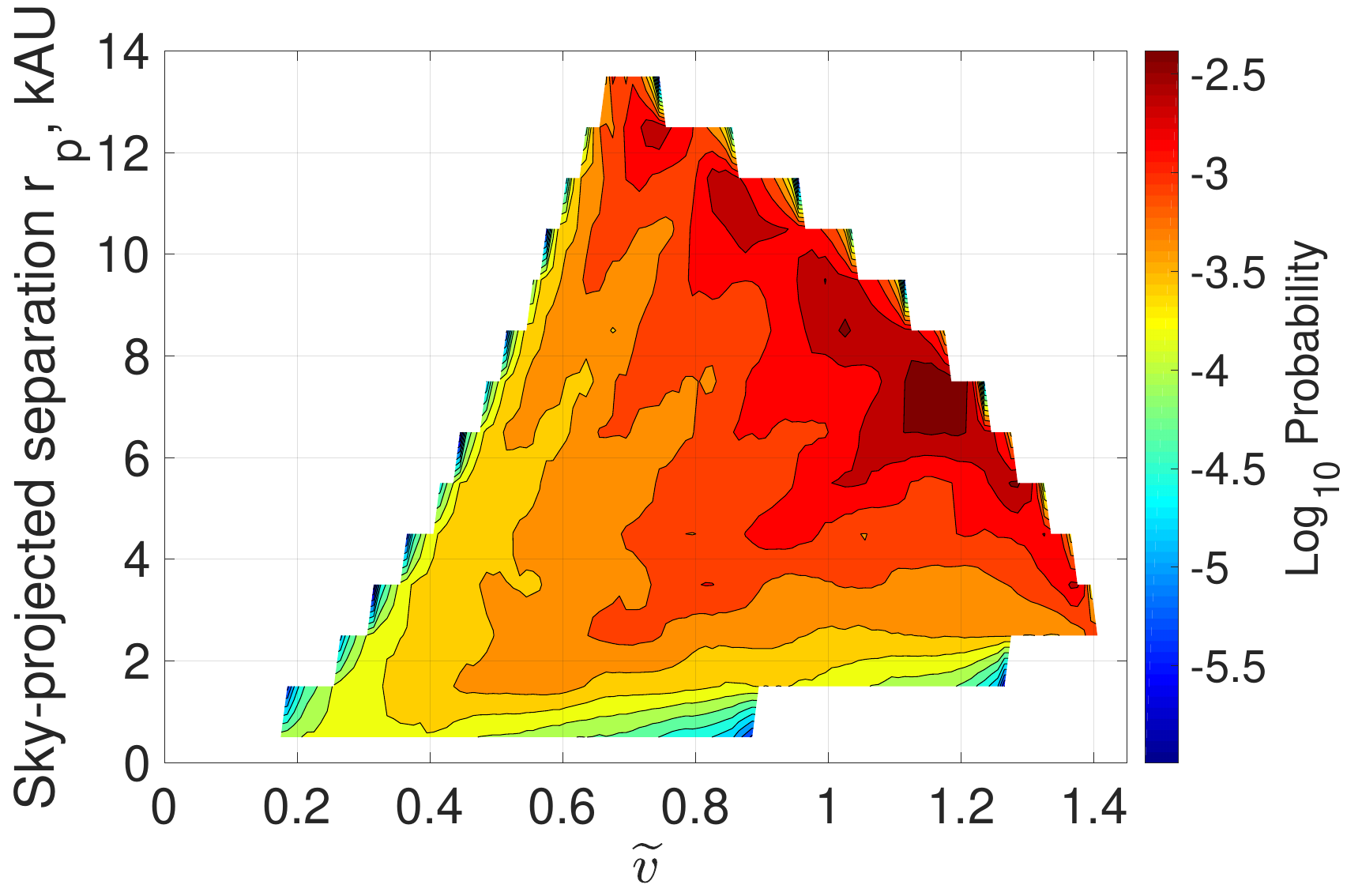}
		\includegraphics[width = 8.5cm] {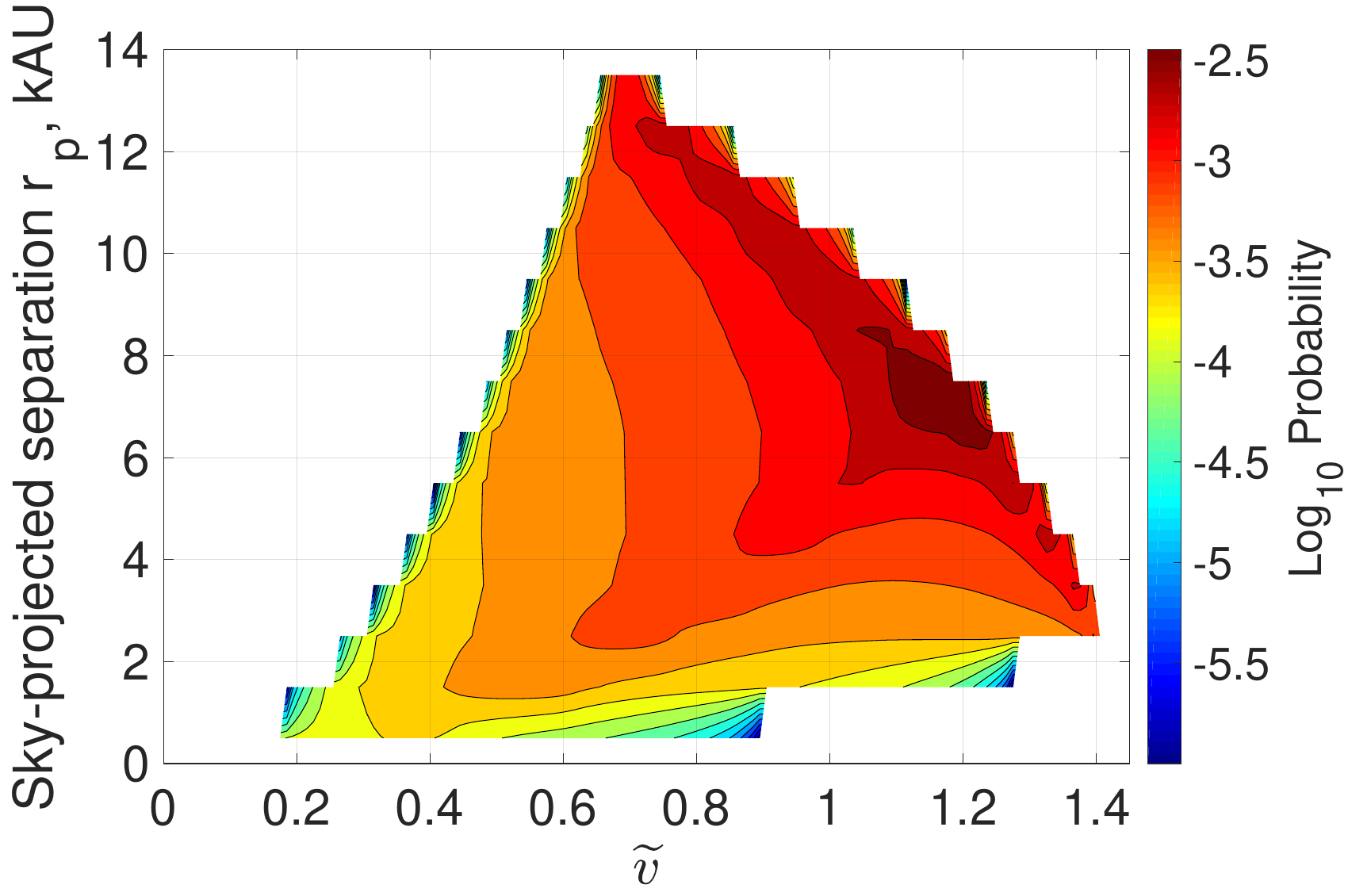}
		\caption{The probability distribution for a randomly located observer to see different combinations of $r_p$ and $\widetilde{v}$ if they observe the motion of Proxima Cen relative to $\alpha$ Cen. \emph{Top}: Based on integrating over 20 revolutions. \emph{Bottom}: Integrating backwards for 5 Gyr, assuming the whole system is on a circular orbit about the Galactic Centre at the speed of the LSR. The different simulation durations give rather similar results, suggesting that 20 orbits is long enough to accurately estimate the statistical properties of the system.}
	\label{Centauri_control}
\end{figure}

Our calculations show that the $\alpha$ Cen system ought to be stable over its lifetime. This allows us to address whether just 20 revolutions is enough to accurately determine the probability distribution of observing different $\left( r_p, \widetilde{v} \right)$ combinations. Figure \ref{Centauri_control} suggests that this should be sufficient as a similar distribution is obtained when integrating backwards over 5 Gyr instead. Moreover, when discussing the WBT, we consider a wide range of WB orbital parameters (Table \ref{Wide_binary_parameters}) rather than just the values currently appropriate for the $\alpha$ Cen system. Even if 20 revolutions is too short to consider how the shape and size of an individual orbit changes, this should still get accounted for statistically when considering a sufficiently broad range of initial conditions. Our results also indicate that the pericentre and apocentre of the $\alpha$ Cen system typically cycle around a closed loop in only ${\approx 8}$ radial periods (blue curve in Figure \ref{Centauri_pericentre_apocentre_comparison}).

Although the $\alpha$ Cen system is stable in our model, this may not be the case for WB systems with different parameters. To investigate this, we explore rotated versions of the system that preserve $r$, $v$ and $\bm{r} \cdot \bm{v}$. The tangential velocity is rotated to explore a range of $\widehat{\bm{h}}$ along the great circle orthogonal to the present separation $\bm{r}$. To investigate other orbital poles, we first rotate both $\bm{r}$ and $\bm{v}$ about the axis $\bm{r} \times \bm{h}$ before rotating the tangential velocity. This lets us explore the full range of $\widehat{\bm{h}}$.

\begin{figure}
	\centering
		\includegraphics[width = 8.5cm] {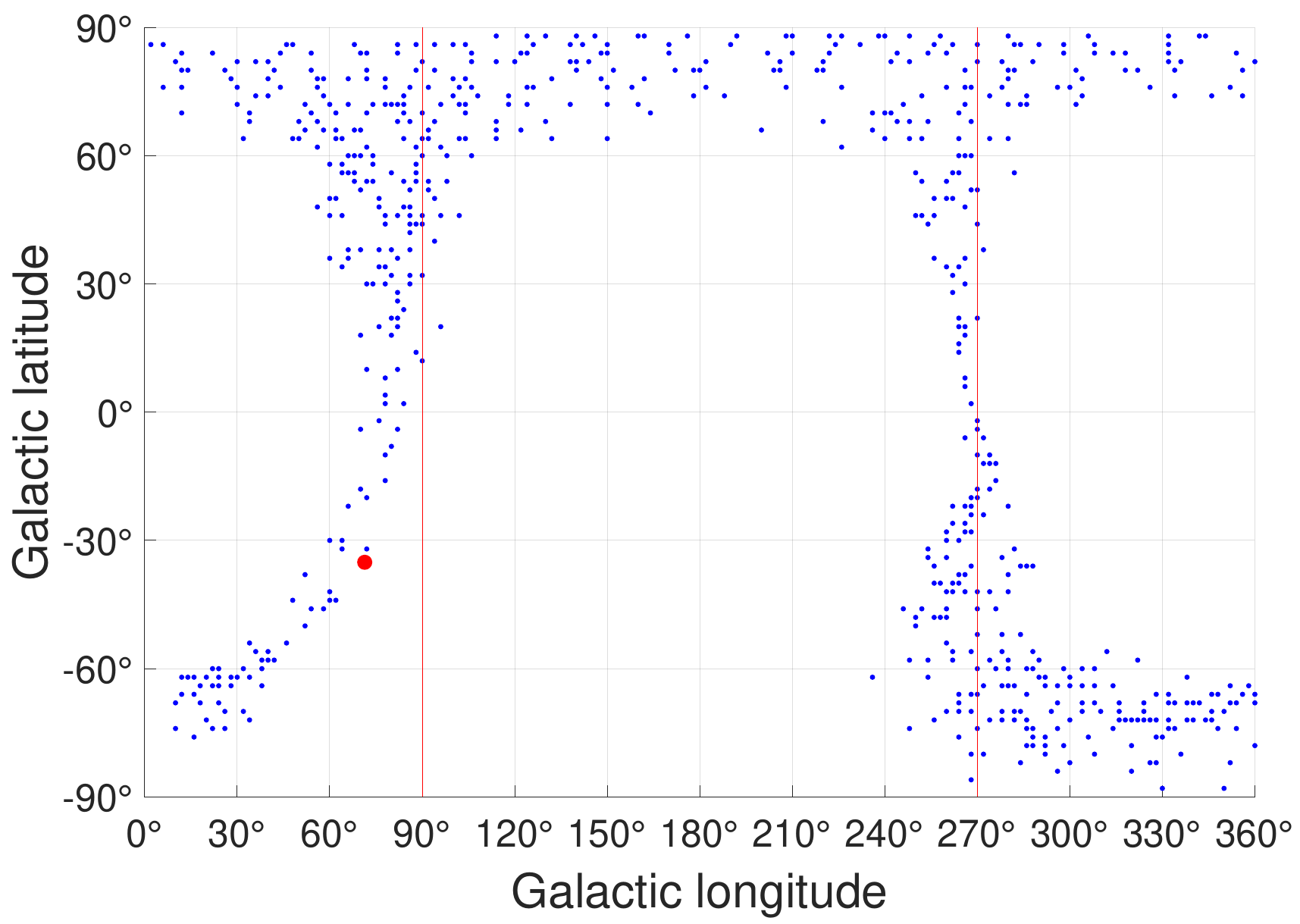}
		\caption{Blue dots show orbital poles for which a rotated version of the $\alpha$ Cen WB orbit would have undergone a close passage ($<50$ AU) when integrated backwards for 5 Gyr, assuming a purely circular Galactic orbit. The observed orbital pole (large red dot) yields a stable orbit that lacks such a close encounter. We also show red vertical gridlines at Galactic longitudes of ${90^\circ}$ and ${270^\circ}$ as these correspond to orbits for which $\bm{h}$ is initially orthogonal to the EF direction. These orbits are expected to be most vulnerable to a collision. As no system ever reaches a separation exceeding 19 kAU, we infer that instability rarely arises in the sense of a system becoming unbound by gaining energy from the time-dependent potential.}
	\label{Centauri_pole_stability}
\end{figure}

We use Figure \ref{Centauri_pole_stability} to show the values of $\widehat{\bm{h}}$ which cause the orbit to reach $r < 50$ AU at some point in the last 5 Gyr. Although some orbits `crash' in this sense, we find no orbital poles for which $r$ rises to very large values such that the system could be considered unbound. In fact, $r$ never exceeds 19 kAU.

If full 6D phase space information becomes available for more WB systems, then we could find whether $\widehat{\bm{h}}$ always falls in the region allowed by MOND. Discovering systems where this is not the case would challenge the theory. Equally, as not all directions are allowed by it, finding no such systems would lend it some credence. This is especially true as Newtonian gravity does not readily provide an explanation for why some WB orbital poles should be preferred over others. The main reason is that Newtonian gravity lacks an EFE, so only the Galactic tide can affect the internal dynamics of a WB. Because both the WB and Galactic orbits have accelerations ${\sim a_{_0}}$, the Galactic tide across a WB is smaller than its internal acceleration by $\sim \frac{r}{R_\odot} \approx 10^{-5}$. Thus, ${\approx 10^5}$ WB orbits would be needed for Galactic tides to have a significant effect. As a single orbit takes ${\approx 1}$ Myr at 10 kAU separation, there is probably insufficient time since the Big Bang for Galactic tides to become important. Even so, some anisotropic tidal effects may arise from the flattened distribution of MCs and of the Galactic disk in general.

\begin{figure}
	\centering
		\includegraphics[width = 8.5cm] {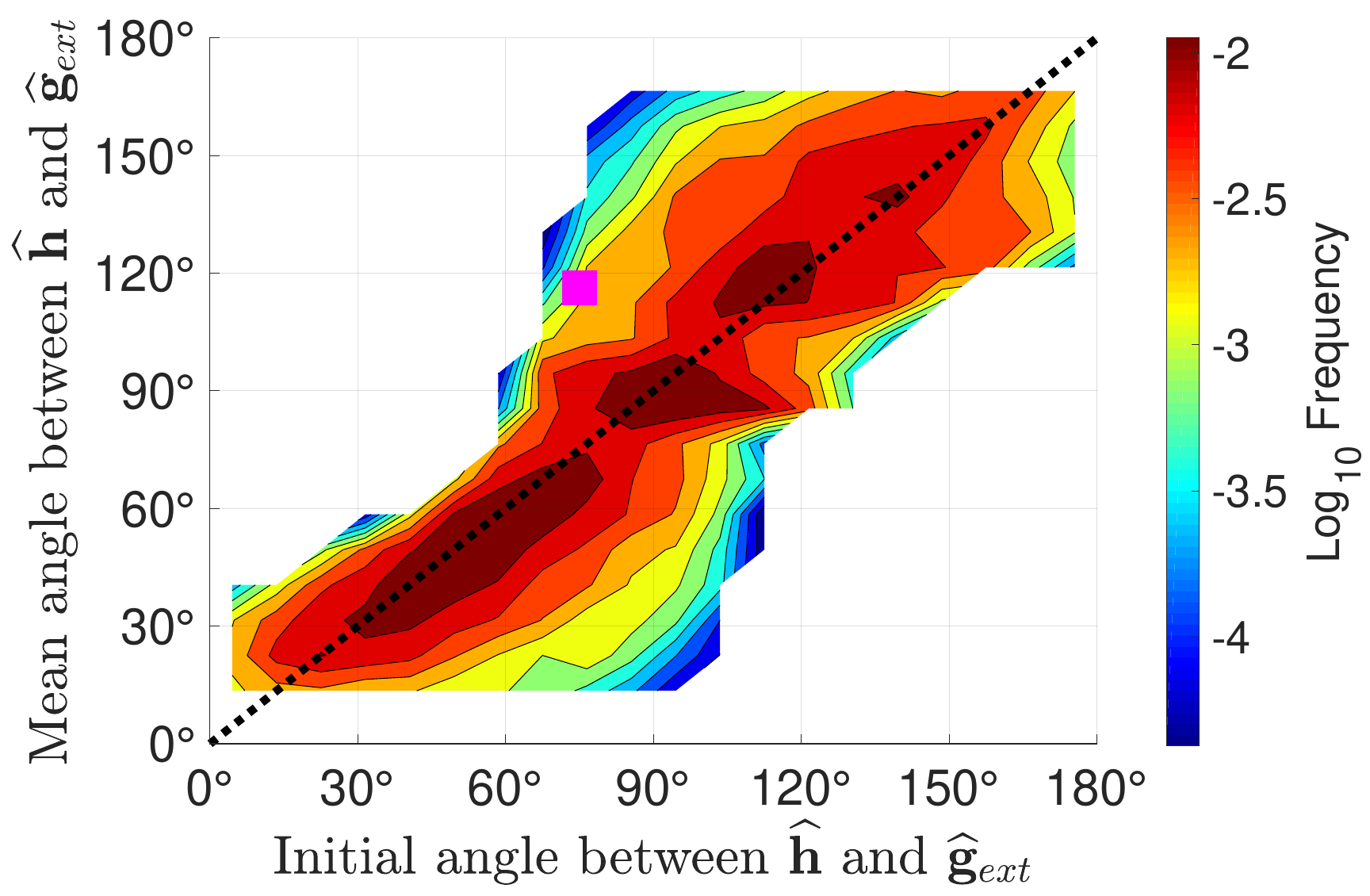}
		\includegraphics[width = 8.5cm] {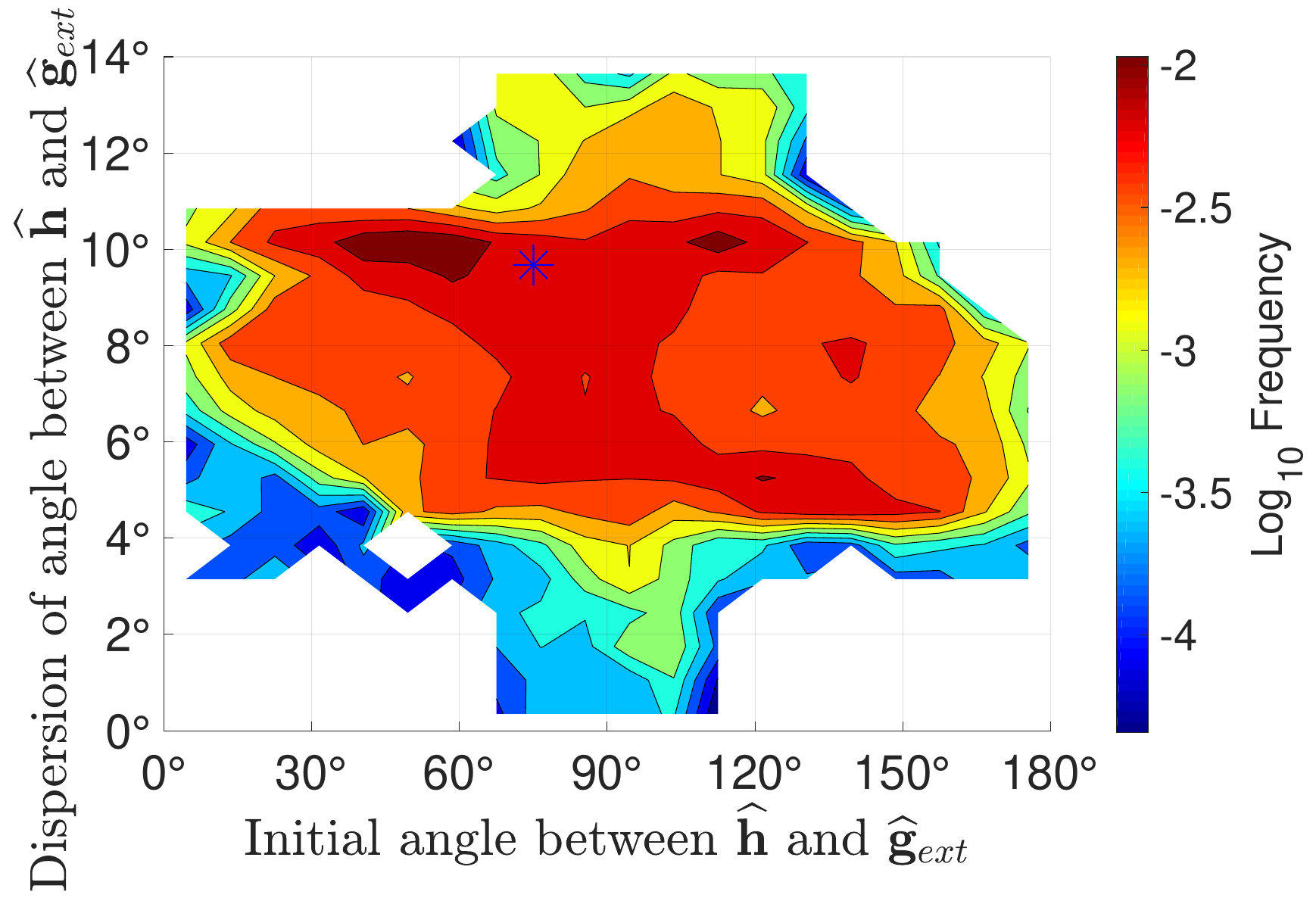}
		\caption{\emph{Top}: Frequency distribution of the mean angle between $\widehat{\bm{h}}$ and $\widehat{\bm{g}}_{ext}$ as a function of its initial value for rotated versions of the $\alpha$ Cen system integrated backwards 5 Gyr. Different initial spin vectors $\widehat{\bm{h}}$ are weighted to have an isotropic distribution. The line of equality is shown as a black dotted line. We use a pink square to show the result for the system whose $\widehat{\bm{h}}$ most closely matches observations of the $\alpha$ Cen system (Table \ref{Proxima_Cen_parameters}). For each system, the mean is taken by considering the angle at each pericentre and apocentre as it would be too intensive to consider the value at every timestep. \emph{Bottom}: The standard deviation of these angles is shown for each system. The blue * represents the system whose initial conditions most closely resemble $\alpha$ Cen.}
	\label{Centauri_adiabatic_test}
\end{figure}

Because the Galactic orbit of $\alpha$ Cen is much slower than its WB orbit, we expect approximate conservation of the component of $\bm{h}$ in the $\widehat{\bm{g}}_{ext}$ direction. We might also expect the magnitude of $\bm{h}$ to be roughly conserved, as often happens in Galactic dynamics. Thus, we show the mean and dispersion in $\widehat{\bm{h}} \cdot \widehat{\bm{g}}_{ext}$ for rotated versions of the $\alpha$ Cen system (Figure \ref{Centauri_adiabatic_test}). This reveals that $\widehat{\bm{h}} \cdot \widehat{\bm{g}}_{ext}$ is indeed approximately conserved despite each system rotating around the Galaxy many times. As a result, systems with high $\left| \widehat{\bm{h}} \cdot \widehat{\bm{g}}_{ext} \right|$ should remain stable over the long term whereas systems with low $\left| \widehat{\bm{h}} \cdot \widehat{\bm{g}}_{ext} \right|$ would be more vulnerable to crashing.

It is interesting to consider whether such $\widehat{\bm{h}}$-dependent effects could be detected using WB systems without full 6D phase space information. One possibility is to focus on ${\bm{h}}_{_{LOS}}$, the line of sight (LOS) component of ${\bm{h}}$. This should be known fairly accurately as it requires only sky-projected positions and velocities. For a WB observed towards the Galactic anti-centre, ${\bm{h}}_{_{LOS}}$ should be approximately conserved as the LOS is aligned with $\widehat{\bm{g}}_{ext}$, the direction about which the gravitational field should be nearly axisymmetric. Thus, if a WB observed in this direction has a large ${\bm{h}}_{_{LOS}}$, it is unlikely to ever crash. This is much more plausible if it has a very low ${\bm{h}}_{_{LOS}}$, perhaps leading to a paucity of systems with low ${\bm{h}}_{_{LOS}}$ in the directions towards or away from the Galactic Centre.

For WBs in the orthogonal direction on our sky but still within the Galactic disk, the LOS is nearly orthogonal to $\widehat{\bm{g}}_{ext}$. Thus, the ${\bm{h}}_{_{LOS}}$ of a WB system would not be so strongly correlated with its long-term orbital stability. This should lead to the frequency distribution of ${\bm{h}}_{_{LOS}}$ varying with sky position. Such an effect would arise even though we expect the intrinsic physical characteristics of nearby (within ${\ssim 100}$ pc) WB systems to not depend on viewing direction as the disk scale length is much longer \citep{Bovy_2013}.

\begin{figure}
	\centering
		\includegraphics[width = 8.5cm] {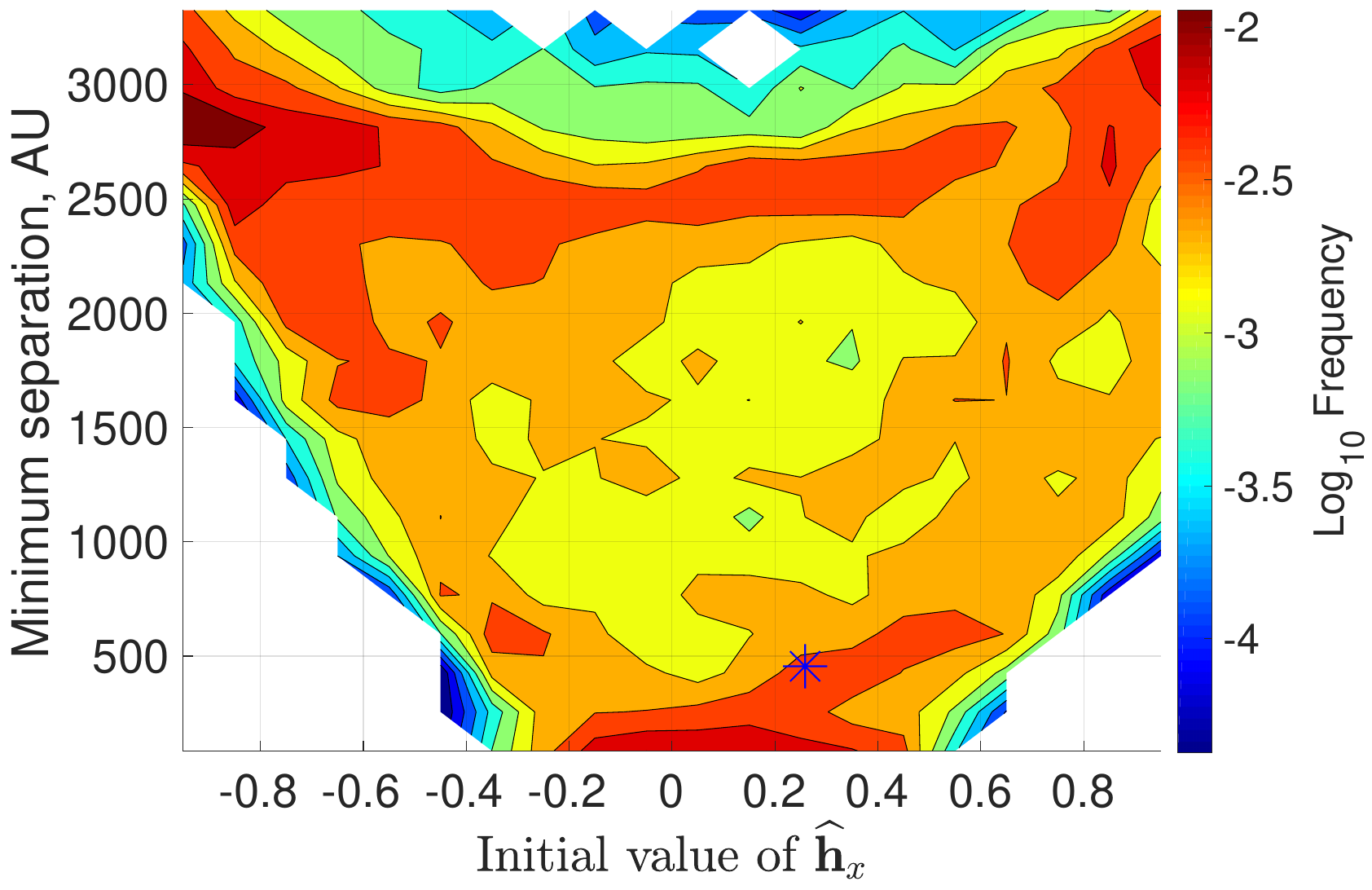}
		\caption{The minimum orbital separation of the $\alpha$ Cen system during the last 5 Gyr for rotated versions of it integrated backwards. The blue * shows the system with $\widehat{\bm{h}}$ most closely matching the observed system. The overdensity at very low separations arises from orbits which were terminated once they came within 50 AU. This only occurs when $\widehat{\bm{h}}$ has a sufficiently small component along the present EF direction $\widehat{\bm{x}}$, a consequence of approximate conservation of ${\bm{h} \cdot \widehat{\bm{g}}_{ext}}$ due to the potential being instantaneously axisymmetric about the very slowly varying EF direction (Figure \ref{Centauri_adiabatic_test}). For systems whose $\widehat{\bm{h}}_x$ is currently high, this provides an angular momentum barrier that precludes very small separations.}
	\label{Centauri_r_min}
\end{figure}

Of course, there are numerous other complications that could arise with this test. The main problem is that $\bm{h}_{_{LOS}}$ is also correlated with the total $h$ of the system, very low values for which necessarily render a system liable to a destructive close approach. However, this is true regardless of the observing direction. The benefit of this test is that, when observing along $\pm \widehat{\bm{g}}_{ext}$, a system with low $\bm{h}_{_{LOS}}$ is vulnerable to a crash regardless of how large $h$ is. This is because other components of $\bm{h}$ are not conserved, making it quite possible that they reached very low values at some time in the history of the system.

Though we do not consider this in much detail, we use Figure \ref{Centauri_r_min} to show the frequency distribution of different closest approach distances $r_{min}$ as a function of $\widehat{\bm{h}} \cdot \widehat{\bm{g}}_{ext} \equiv \widehat{\bm{h}}_x$ for the same grid of models as are shown in Figures \ref{Centauri_pole_stability} and \ref{Centauri_adiabatic_test}. There is a strong correlation between $r_{min}$ and the initial value of $\widehat{\bm{h}}_x$ despite each system having undergone ${\ga 10^4}$ orbits and initially having the same $h$. This is evident from the fact that systems which reach very low $r_{min}$ (blue dots in Figure \ref{Centauri_pole_stability}) initially had low values for $\widehat{\bm{h}}_x$ (near $y$-axis in Figure \ref{Centauri_r_min}), a consequence of the binary orbit adiabatically adjusting to the much slower Galactic orbit.

\subsection{Towards a full Galactic orbit}
\label{Full_Galactic_orbit}

To consider the $\alpha$ Cen Galactic orbit in more detail, we need to find its Galactic velocity by adding the heliocentric velocity of the system to the Galactocentric velocity of the Sun, $\bm{v_\odot}$. In the usual Galactic Cartesian co-ordinates (Section \ref{Initial_conditions}), this Solar velocity is
\begin{eqnarray}
 \bm v_\odot = \begin{bmatrix}
 U_\odot \\
 V_\odot + v_{c, \odot} \\
 W_\odot
 \end{bmatrix}
 \label{Solar_velocity}
\end{eqnarray}

Here, $\left( U_\odot, V_\odot, W\odot \right)$ is the velocity of the Sun with respect to the LSR, which has a speed of $v_{c, \odot}$. Our adopted values for these parameters are given in Table \ref{Solar_velocity_table}.

\begin{table}
  \centering
		\begin{tabular}{lll}
			\hline
			Parameter & Meaning & Value\\
			\hline
			$R_\odot$ & Galactocentric radius of Sun & 8.2 kpc \\
			$v_{c, \odot}$ & Local Standard of Rest velocity & 232.8 km/s \\ [5pt]
			$U_\odot$ & See Equation \ref{Solar_velocity} & 14.1 km/s \\
			$V_\odot$ & See Equation \ref{Solar_velocity} & 14.6 km/s \\
			$W_\odot$ & See Equation \ref{Solar_velocity} & 6.9 km/s \\
			\hline
		\end{tabular}
	\caption{Parameters governing the position and velocity of the Sun with respect to the Galaxy. The position and velocity of the Local Standard of Rest are obtained from \citet{McMillan_2017}, which yields similar results to more recent estimates that use Gaia data \citep{Kawata_2019}. The non-circular velocity of the Sun is obtained from \citet{Francis_2014}.}
  \label{Solar_velocity_table}
\end{table}

The $\alpha$ Cen system is assumed to orbit the Galactic centre in a fixed plane, though the motion within this plane need not be circular. Although vertical forces due to the disk must matter somewhat, we neglect these as the Sun is ${\approx 4}$ disk scale lengths from the Galactic Centre \citep{Bovy_2013}. We discuss the accuracy of our spherical potential approximation in Section \ref{Galactic_disk_effect}.

The velocity of $\alpha$ Cen out of the MW disk means that its orbit is tilted with respect to the Galactic plane. We find that the tilt angle is ${4.61^\circ}$, causing the $\alpha$ Cen orbit to rise ${\approx 700}$ pc out of the MW. Given that the Sun is only ${\approx 15}$ pc from the MW disk plane \citep[][figure 6]{Ferguson_2017} and that $\alpha$ Cen is only 1.3 pc from the Sun \citep{Kervella_2016}, we assume that $\alpha$ Cen currently lies within the line of nodes between its Galactic orbital plane and the MW disk.

To estimate how the Galactocentric distance $R$ of $\alpha$ Cen varies with time, we rotate its velocity into the Galactic disk plane. We then integrate its orbit using the potential of the nominal MOND Galactic model from \citet{Banik_2018_escape}. This shows that the mean of the perigalacticon and apogalacticon distances is $R_0 = 8.94$ kpc, with oscillations in $R$ of amplitude $dR = 0.84$ kpc. The fact that ${dR \ll R_0}$ means that the orbit is nearly circular, allowing us to make an epicyclic approximation.
\begin{eqnarray}
	R ~\approx~ R_0 + dR \, \sin \left( \frac{2\mathrm{\pi} t}{P_R} + \phi_0 \right)
	\label{Epicyclic_approximation}
\end{eqnarray}

\begin{figure}
	\centering
		\includegraphics[width = 8.5cm] {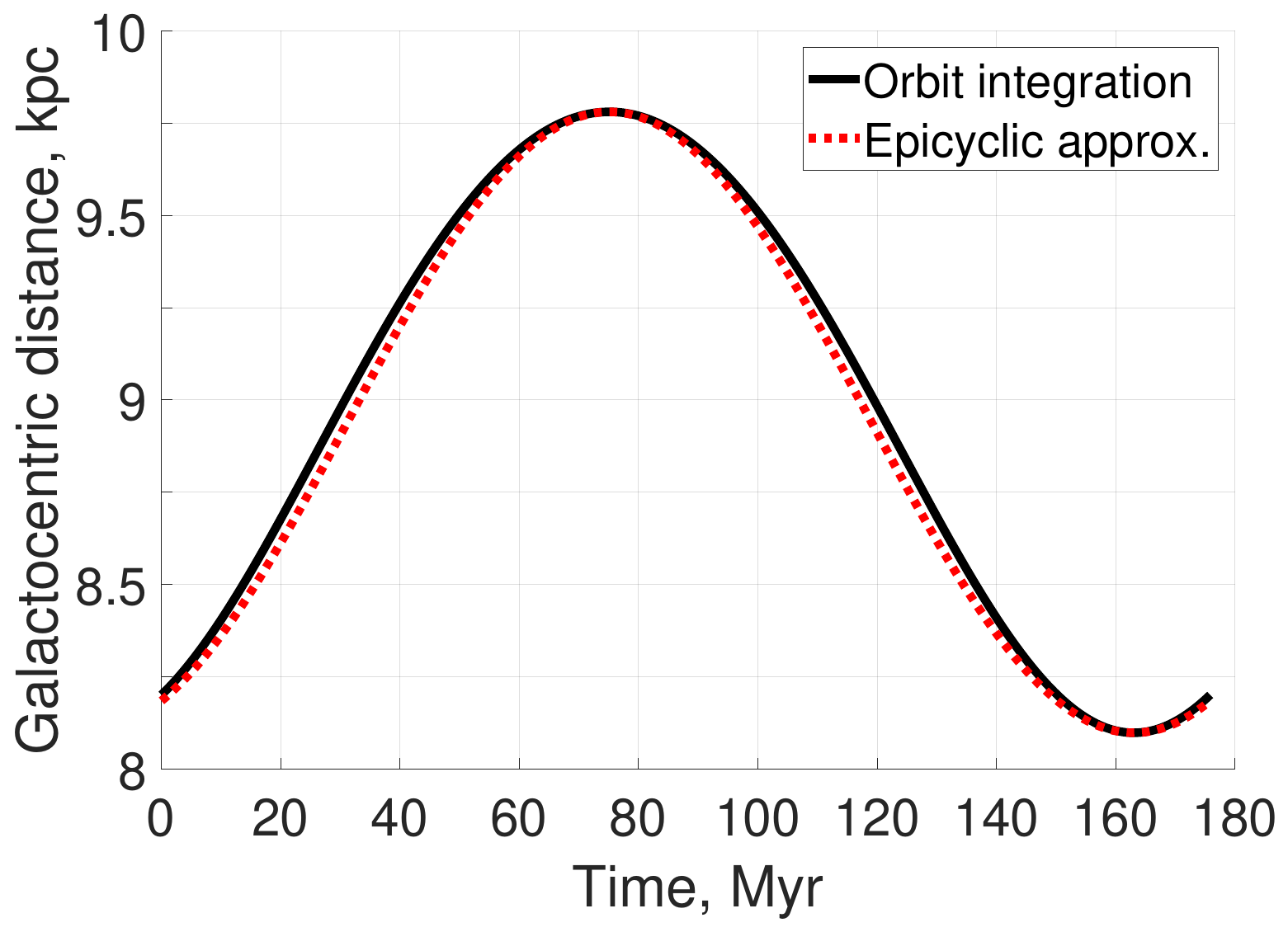}
		\caption{The Galactocentric radius $R$ of the $\alpha$ Cen system over one radial period using orbit integration (solid black line) and in our epicyclic approximation (Equation \ref{Epicyclic_approximation}), shown with a dotted red line.}
	\label{Epicyclic_approximation_Centauri}
\end{figure}

To find the radial period $P_R$, we numerically advance the orbit until $R$ and $\dot{R}$ get back to their initial value, with $\dot{R}$ denoting a time derivative. The presently observed values of $R$ and $\dot{R}$ fix the current radial phase $\phi_0$ of the $\alpha$ Cen Galactic orbit. We demonstrate the accuracy of the epicyclic approximation in Figure \ref{Epicyclic_approximation_Centauri}.

As $R$ varies, we also vary the tangential speed of $\alpha$ Cen around the Galaxy to maintain angular momentum conservation. The Galactic latitude of $\alpha$ Cen is always rather small, allowing us to make the approximation that it varies sinusoidally with the Galactic longitude. In a spherical potential, the maximum Galactic latitude of $\alpha$ Cen is directly determined by its present position and velocity because the orbit is confined to a plane.

\begin{figure}
	\centering
		\includegraphics[width = 8.5cm] {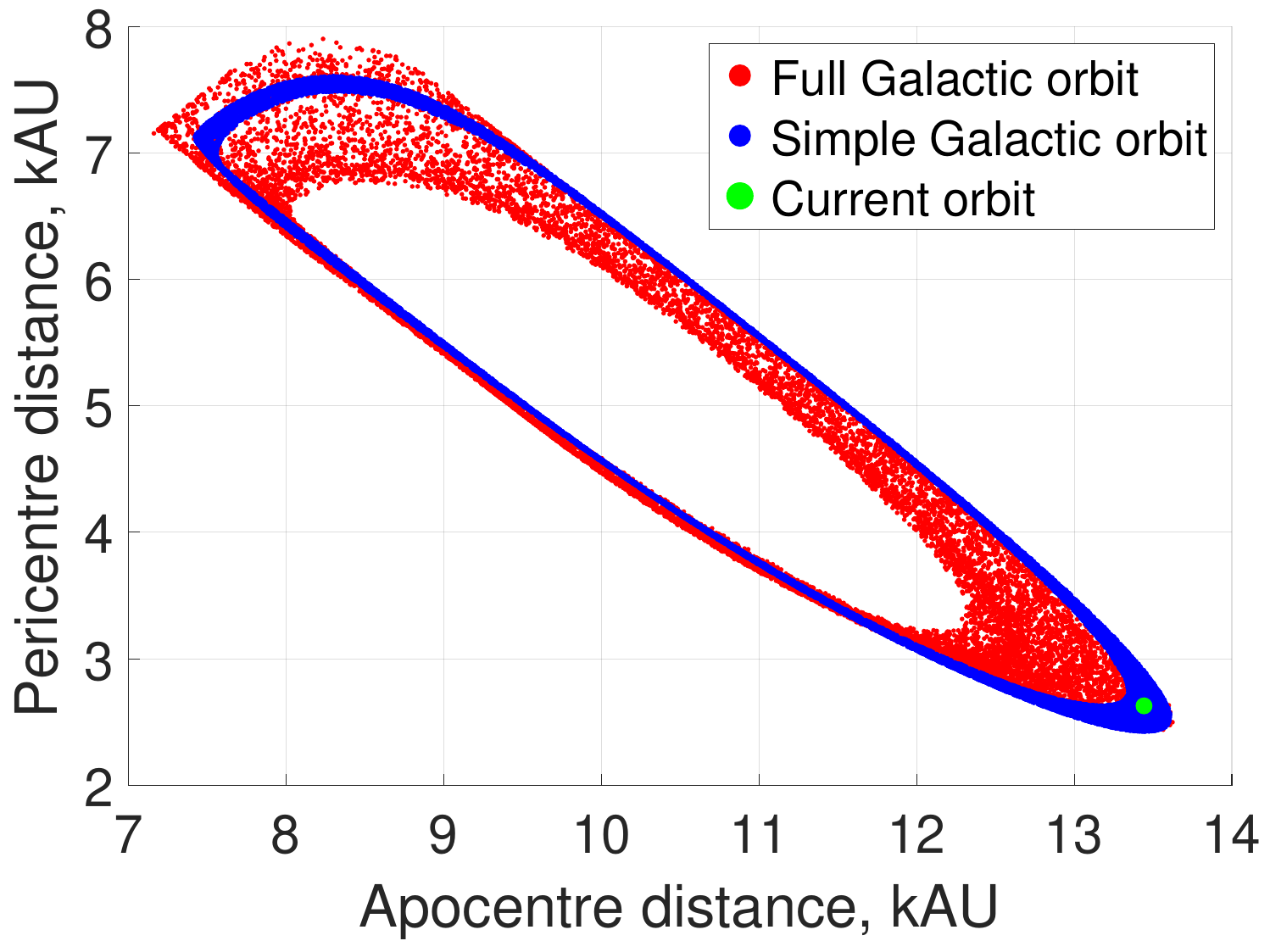}
		\caption{The separation of $\alpha$ and Proxima Cen at each pericentre against their distance at the successive apocentre, with the current orbit shown as a large green dot. The blue dots show results for a simple model where the whole system rotates around the Galaxy for 5 Gyr on a purely circular orbit at the LSR speed. The red dots are based on a more realistic Galactic orbit (Section \ref{Full_Galactic_orbit}). The points fall on a well-defined curve which is thicker for the latter case. In both cases, the system cycles around this curve in ${\approx 8}$ radial periods.}
		\label{Centauri_pericentre_apocentre_comparison}
\end{figure}

An important objective of considering the Galactic orbit of $\alpha$ Cen in more detail is to include the time variation of the EF strength. This is done self-consistently with our numerical integration of the $\alpha$ Cen Galactic orbit. As a result, the relation between the orbital separations at each pericentre and the successive apocentre becomes less tight than for a purely circular Galactic orbit (Figure \ref{Centauri_pericentre_apocentre_comparison}). However, the result remains broadly similar. In both cases, significant evolution of the orbit is apparent due to the non-spherical nature of the potential. Although its time dependence must also have some effect, the $\alpha$ Cen system takes only ${\approx 8}$ radial periods to cycle around the curve apparent in Figure \ref{Centauri_pericentre_apocentre_comparison}. This is too short for the Galactic orbit of the system to have much effect.

Less extreme effects would be expected in the Solar System, where the stronger $g$ makes for more nearly Newtonian behaviour. Even so, MOND combined with the Galactic EF could conceivably explain objects like Sedna with perihelion well outside the orbit of Neptune, the most distant Solar System planet \citep{Brown_2004}. Significant MOND effects are only possible if Sedna had a fairly large apocentre, most likely due to interaction with a large planet. Although such an interaction is possible in Newtonian gravity too, this would cause Sedna to return to the heliocentric distance where it strongly interacted with the planet. This requirement is relaxed in MOND as its Keplerian orbital parameters are no longer fixed, perhaps providing a way to explain Sedna's orbit. In fact, its perihelion would take only ${\approx 10}$ Myr to double from Neptune's orbital radius to its observed perihelion \citep{Pauco_2016}. However, we note that there are alternative Newtonian explanations for Sedna such as capture from the planetesimal disk of a passing star \citep{Jilkova_2015}.

As well as increasing the perihelion distance, MOND effects can reduce it, explaining why some orbits crash in Figure \ref{Centauri_pole_stability}. This could provide an additional mechanism by which Oort Cloud comets get perturbed onto orbits that reach the inner Solar System. It is unclear how this would affect observations as there are also conventional mechanisms for generating such perturbations.

\subsubsection{The effect of the Galactic disk}
\label{Galactic_disk_effect}

So far, we have not considered the fact that the MW is a disk rather than a point mass. This is because the Sun lies several disk scale lengths out. To test how the extended nature of the MW might affect our results, we improve our algebraic MOND approximation in Equation \ref{g_N_ext} by including the vertical force. Specifically, we assume that
\begin{eqnarray}
	g_{_{N,r}} \, \nu \left( \frac{\sqrt{{g_{_{N,r}}}^2 + {g_{_{N,z}}}^2}}{a_{_0}} \right) ~=~ g_r ~=~ \frac{{v_{c,\odot}}^2}{R_\odot}
	\label{Algebraic_MOND_disk}
\end{eqnarray}

This is known to work rather well in disk galaxies, at least once the vertical Newtonian gravity $g_{_{N,z}}$ is taken into account when determining $\nu$ so as to more rigorously implement the algebraic MOND approximation \citep{Jones_2018}. For a thin disk, $g_{_{N,z}}$ may be estimated as
\begin{eqnarray}
	g_{_{N,z}} ~=~ -2\mathrm{\pi} G \Sigma \sign{ \left(z \right)}
	\label{g_N_z}
\end{eqnarray}

We obtain the local Galactic surface density $\Sigma$ using MW parameters from table 1 of \citet{Banik_2018_escape}. Using our previous estimate of $g_{_{N,r}}$, we see that $g_{_{N,z}} \approx \frac{1}{3}g_{_{N,r}}$. This means that including the vertical component of $\bm{g}_{_N}$ raises its magnitude by $\approx \frac{1}{2\times3^2}$ fractionally. As $K_0 \approx -0.26$, we expect that this reduces $\nu$ by only ${\approx 1.4\%}$. The lower value of $\nu$ makes $\nu g_{_{N,r}}$ fall below $g_r$, even though these are equal when neglecting $g_{_{N,z}}$ (Equation \ref{g_N_ext}). As a result, $g_{_{N,r}}$ would have to increase ${\approx 1.4 \%}$, thereby reducing $\nu$ by ${\approx 0.35\%}$.

To check this, we solve Equation \ref{Algebraic_MOND_disk} using the Newton-Raphson algorithm. This shows that $g_{_{N,r}}$ increases by 1.9\% compared to our previous expectation. As a result, $\nu_{ext}$ in the Solar neighbourhood falls from 1.5603 to 1.5525. Perhaps more meaningful is $\nu_{ext} \left( 1 + \frac{K_0}{3} \right)$, the angle-averaged ratio between Newtonian and MOND gravity (Equation \ref{eta_EFE}). This decreases from 1.423 to 1.417, implying that our results are hardly affected by the vertical component of the Galactic gravitational field.

\section{Conclusions}
\label{Conclusions}

We investigated the feasibility of testing MOND using the orbital velocities of wide binary stars whose very low orbital accelerations fall below the MOND $a_{_0}$ threshold. Several thousand such systems have recently been identified with a low contamination rate \citep{Andrews_2018} using Gaia observations \citep{GAIA_2018}. To keep the complexity manageable, we solved the MOND equations assuming all the mass in each binary was in one of the stars, though this assumption should not much affect forces in the Solar neighbourhood (Section \ref{Mass_ratio_effect}). We then self-consistently included the external field (EF) from the rest of the Galaxy, leading to an axisymmetric gravitational field. By using this to integrate systems with a range of total masses, semi-major axes, eccentricities and orbital planes over 20 revolutions, we showed that wide binary systems have relative velocities which can exceed Newtonian expectations by a significant amount. The excess is ${\approx 20\%}$ if we use the simple MOND interpolating function to transition between the Newtonian and modified regimes \citep{Famaey_Binney_2005}. This rather gradual transition works very well at explaining a variety of Galactic and extragalactic observations, which distinctly prefer this form over the sharper transition of the so-called `standard' function (Section \ref{Interpolating_function}). Once the interpolating function is fixed, the two major formulations of MOND yield numerically very similar results (Table \ref{Interpolating_function_effect}).

Consequently, Newtonian gravity should be distinguishable from MOND with ${\approx 500}$ well-observed systems with sky-projected separation $r_p = \left(1 - 20 \right)$ kAU, minimising contamination of the sample (Section \ref{Results}). Our results show that the test is best performed by considering the proportion of systems with $r_p > 3$ kAU and whose scaled velocity $\widetilde{v}$ (Equation \ref{v_tilde}) lies in the range 0.97$-$1.68.

The number of systems needed for the test could be halved if full 3D relative velocities were available, requiring follow-up measurements of the radial velocity accurate to within ${\approx 0.02}$ km/s. In this case, the best $\widetilde{v}$ range is 1.05$-$1.68. Additional improvements might come from allowing wider binaries, where MOND effects would be somewhat larger \citep[e.g. an upper limit of 40 kAU was used by][]{Andrews_2018}. However, wider systems would rotate slower and lead to a more contaminated sample.

It is very likely that a sufficient number of systems can be found within 150 pc \citep{Andrews_2018}. At this distance, the astrometric accuracy required is comparable to that of the most recent Gaia data release (Section \ref{Measurement_uncertainties}). Using stellar mass-luminosity relations \citep[e.g.][]{Mann_2015}, the observed magnitudes of the stars in each binary should be sufficient to accurately constrain their masses as their distances would be known very well (Section \ref{Mass_measurement}). It should also be feasible to obtain radial velocities of ${\approx 1000}$ nearby stars at the required accuracy \citep{Chubak_2012}.

Our results include how the EF from the rest of our Galaxy weakens the self-gravity of wide binary systems. However, some versions of MOND lack this external field effect (EFE). Such theories predict much larger deviations from Newtonian gravity that should become detectable with data from only ${\approx 100}$ systems, even without radial velocity measurements (Section \ref{MOND_no_EFE}). Although we do not consider it likely that MOND applies without the EFE, this would be strongly favoured by the discovery of systems with $\widetilde{v} > 2$ as the EFE limits the maximum value of $\widetilde{v}$ to ${\approx 1.7}$. Without the EFE, our results suggest that $\widetilde{v}$ could reach values as large as 3.2. Even larger values might arise if WB systems with apocentres beyond 100 kAU remain stable against encounters with passing stars.

We considered a number of systematic errors that could hamper our ability to test gravity using wide binary stars (Section \ref{Systematics}). The most serious issue would probably be a low-mass undetected companion to at least one of the stars (Section \ref{Hierarchical_systems}). There are several ways in which the effect of such companions could be minimised. Most straightforwardly, it can cause $\widetilde{v}$ to exceed the maximum possible in MOND. This allows such high-$\widetilde{v}$ systems to create a statistical model for the properties of low-mass binary companions. This model could be extended down to the critical $\widetilde{v}$ range of 0.9$-$1.7 that is most sensitive to the underlying law of gravity. In this way, we might be able to estimate the true fraction of systems whose $\widetilde{v}$ lies in this range.

To evade direct detection, any such companion must have a rather low mass. For such an object to noticeably affect $\widetilde{v}$, it would need to be quite close to one of the detected stars. This would lead to a much shorter orbital period (though still several centuries), creating a much larger acceleration than for the wide binary orbit (Equation \ref{g_close_wide_ratio}). Astrometric observations over a few years should be able to detect this by finding a parabolic sky trajectory, with a deviation from a linear trend of $\approx 100 \, \mu$as after 5 years. This might also be detectable as a radial velocity trend of ${\approx 4}$ m/s per year.

Wide binary stars are not very rare $-$ the nearest star to the Sun is in just such a system. Motivated by this, we considered the orbit of Proxima Cen about $\alpha$ Cen in some detail (Section \ref{Proxima_Centauri}). The orbital acceleration is ${\approx 40\%}$ higher in MOND, something that could be tested with the proposed Theia mission \citep{Theia_2017} if it achieves an astrometric precision of $\approx 1 \, \mu$as over 5 years (Figure \ref{Theia_test}).

To explore the long-term stability of the orbit, we integrated it backwards for 5 Gyr in an EF that varies with time due to the Galactic orbit of the system. We also considered rotated versions of the Proxima Cen-$\alpha$ Cen orbit that trace all possible orbital poles $\widehat{\bm{h}}$. Some $\widehat{\bm{h}}$ turn out to be unstable in the sense that there was a very close encounter at some time in the past 5 Gyr. These $\widehat{\bm{h}}$ are nearly orthogonal to the EF direction at the present time. This arises because the component of $\widehat{\bm{h}}$ along the EF direction is conserved in the limit that the EF is constant with time, as the potential is axisymmetric about the EF direction. Because changes in the EF arise only from the rather slow Galactic orbit of the system, we expect that systems with a lot of angular momentum along the EF direction would benefit from an angular momentum barrier that prevents them from undergoing an extremely close encounter.

Systems like this are indeed generally stable, with the minimum separation well correlated with the fraction of the angular momentum that presently lies along the EF direction (Figure \ref{Centauri_r_min}). Moreover, the angle between $\widehat{\bm{h}}$ and the EF is conserved rather well despite the systems completing many Galactic orbits that completely reorient the EF (Figure \ref{Centauri_adiabatic_test}). This suggests that we may gain a deeper analytic understanding of wide binaries by assuming that the binary orbit adjusts adiabatically to the much slower Galactic orbit. The dependence of orbital stability on $\widehat{\bm{h}}$ could also provide a basis for testing MOND using novel direction-dependent effects (Section \ref{Secular_effects}).

Wide binary stars promise to be an exciting new frontier in the quest to understand the law of gravity relevant to low-acceleration astrophysical systems like galaxies. Such understanding almost certainly requires independent non-galactic tests of the gravity law, an area that has received little attention so far. Wide binary systems promise to provide just such a test, perhaps using the catalogue of such systems identified by \citet{Andrews_2018} based on the Gaia mission \citep{Perryman_2001}. Although Gaia has already released data on more than a billion stars \citep{GAIA_2018}, precise data on only a minute fraction of them should be sufficient to answer fundamental questions about the laws governing our Universe.

\begin{appendix}

\section{Derivation of Equation \ref{EXTERIOR_PDM_ADJUSTMENT}}
\label{Potential_correction}

In this section, we determine the contribution to the potential at the point $\left( r, \theta \right)$ arising from phantom dark matter exterior to the radius $r_{_{out}}$ (itself $\geq r$) relative to a point mass $M$ embedded in a uniform external field $\bm{g}_{ext}$. Combining equations 24 and 28 of \citet{Banik_2015}, we get that
\begin{eqnarray}
	\nabla^2 \Phi ~=~ -\frac{GM\nu_{_{ext}}K_0}{r^3} \left(3 \cos^2 \theta - 1 \right)
	\label{rho_ph_equation}
\end{eqnarray}

We now consider the phantom dark matter within a thin shell of radius $r_{_{sh}}$ and thickness $dr_{_{sh}}$. Because $\nabla^2 \Phi = 0$ everywhere outside the thin shell and its surface density has the same angular dependence as the second Legendre polynomial, the solution must be $\Phi \propto r^n \left(3 \cos^2 \theta - 1 \right)$. The Laplacian of this is 0 if and only if $n = 2$ or $-3$. The choice is fixed by the need to avoid the potential diverging at small and large radii.
\begin{eqnarray}
	\Phi = 
\left\{
	\begin{array}{ll}
		u \left(3 \cos^2 \theta - 1 \right) \left(\frac{r}{r_{_{sh}}} \right)^2 & \mbox{if } r \leq r_{_{sh}} \\
		u \left(3 \cos^2 \theta - 1 \right) \left(\frac{r}{r_{_{sh}}} \right)^{-3} & \mbox{if } r \geq r_{_{sh}}
	\end{array}
\right.
\end{eqnarray}

To find the multiplicative constant $u$, we need to consider the discontinuity in $\frac{\partial \Phi}{\partial r}$ across the shell at $r = r_{_{sh}}$. This must equal the surface density of phantom dark matter in the shell (Equation \ref{rho_ph_equation}).
\begin{eqnarray}
	-5u\left(3 \cos^2 \theta - 1 \right) ~&=&~ -\frac{GM\nu_{_{ext}}K_0}{{r_{_{sh}}}^2} \left(3 \cos^2 \theta - 1 \right) dr_{_{sh}} \nonumber \\
	u ~&=&~ \frac{GM\nu_{_{ext}}K_0}{5{r_{_{sh}}}^2} dr_{_{sh}}
\end{eqnarray}

This lets us find the contribution to $\Phi$ from shells with $r_{_{sh}} \geq r_{_{out}}$.
\begin{eqnarray}
	\Delta \Phi ~&=&~ \int_{r_{_{out}}}^\infty \frac{GM\nu_{_{ext}}K_0 \left(3 \cos^2 \theta - 1 \right)}{5{r_{_{sh}}}^2} \left(\frac{r}{r_{_{sh}}} \right)^2 dr_{_{sh}} \nonumber \\
	 ~&=&~ \frac{GM\nu_{_{ext}}K_0r^2 \left(3 \cos^2 \theta - 1 \right)}{15{r_{_{out}}}^3}
	\label{Exterior_shell_contribution}
\end{eqnarray}

Differentiating this reproduces our result in Equation \ref{EXTERIOR_PDM_ADJUSTMENT}, something that is most easily verified using Cartesian co-ordinates. We can perform a similar calculation to get the contribution to $\Phi$ from all shells with $r_{_{sh}} \leq r_{_{in}}$ for some $r_{_{in}} \leq r$.
\begin{eqnarray}
	\Delta \Phi ~&=&~ \int_0^{r_{_{in}}} \frac{GM\nu_{_{ext}}K_0}{5{r_{_{sh}}}^2} \left(\frac{r}{r_{_{sh}}} \right)^{-3} dr_{_{sh}} \\
	 ~&=&~ \frac{GM\nu_{_{ext}}K_0 \left(3 \cos^2 \theta - 1 \right) {r_{_{in}}}^2}{10 r^3}
	\label{Interior_shell_contribution}
\end{eqnarray}

To find the total potential at some point $\left(r, \theta \right)$, we can combine the results of Equations \ref{Exterior_shell_contribution} and \ref{Interior_shell_contribution} evaluated for $r_{_{in}} = r_{_{out}} = r$.
\begin{eqnarray}
	\Phi ~&=&~ \frac{GM\nu_{_{ext}}K_0 \left(3 \cos^2 \theta - 1 \right)}{r} \left( \frac{1}{15} + \frac{1}{10} \right) \\
	~&=&~ \frac{GM\nu_{_{ext}}K_0 \left(3 \cos^2 \theta - 1 \right)}{6r}
\end{eqnarray}

This reproduces equation 29 of \citet{Banik_2015}, which was obtained using equation 2.95 of \citet{Galactic_Dynamics}. Another consistency check is that our solution for $\Phi$ satisfies Equation \ref{rho_ph_equation}.

\end{appendix}

\section*{Acknowledgements}

IB was partially supported by Science and Technology Facilities Council studentship 1506672. He is grateful to the Weizmann Institute of Science for hosting him during a week-long visit. During this visit, he had some very helpful discussions with Mordehai Milgrom, Boaz Katz and Doron Kushnir that greatly advanced this contribution. He also thanks Will Sutherland for providing an observational perspective and Duncan Forgan for comments on tides. The authors wish to thank the referee for comments which improved this manuscript.

\bibliographystyle{mnras}
\bibliography{CEN_bbl}
\bsp
\label{lastpage}
\end{document}